\documentclass[letter,10pt,reqno]{article}
\usepackage{palatino}
\usepackage{color}
\usepackage{amsmath}

\usepackage{tikz, pgfplots, ulem,mathrsfs}
\usetikzlibrary{decorations.markings,arrows, calc, calligraphy}
\definecolor{darkblue}{rgb}{0.1,0.1,0.45}
\usepackage{amssymb}
\usepackage[colorlinks=true, pdfstartview=FitV, linkcolor=darkblue, citecolor=black, urlcolor=blue]{hyperref}
\definecolor{shadecolor}{rgb}{0.9, 0.9, 0.81}

\def\xh{\hat{x}}
\def\Mcal {{\mathcal M}}
\def\Mcomb{\overline \Mcal _{g,n}[{\bf p}]}
\def \eqref #1{ (\ref{#1})}

\def \JS {\mathscr J_{{\bf p}}}

\def\Wfive{W_{5} }
\def\Wbdr{W_{1,1}}
\def\Wff{W_{2,2}}
\def \&{\hspace{-18pt}&}
\def\arg{{\rm arg}}

\def\QED{ {\hfill $\blacksquare$}\par \vskip 4pt}
\def\wt{\widetilde}
\newtheorem{theorem}{Theorem}[section]
\newtheorem{definition}[theorem]{Definition}
\newtheorem{proposition}[theorem]{Proposition}
\newtheorem{corollary}[theorem]{Corollary}
\newtheorem{remark}[theorem]{Remark}
\newtheorem{lemma}[theorem]{Lemma}

\def\nzh{{\widehat{\nz}}}
\def\be{\begin{equation}}
\def\ee{\end{equation}}
\def\ba{\begin{array}}
\def\ea{\end{array}}
\def\la{\label}
\def\p{\partial}
\def\nz{{\bf M}}
\def\db{{\bf d}}
\def\Pcal{{\mathcal P}}
\def\Hcal{{\mathcal H}}
\def\Ccalh{\widehat{{\mathcal C}}}
\def\zetah{\hat{\zeta}}
\def\rb{{\bf r}}
\def\sb{{\bf s}}
\def\Poss{{\mathcal P}}

\def\ah{\hat{a}}
\def\bh{\hat{b}}
\def\Ah{\widehat{A}}
\def\Bh{\hat{B}}
\def\Ch{\hat{C}}
\def\Dh{\widehat{D}}
\def\Tor{{\bf t}}
\def\kb{{\bf k}}

\def\qdt{\qd}
\def\Eh{{\widehat{E}}}
\def\Kh{{\widehat{K}}}

\def\Acalh{{\widehat{{\mathcal A}}}}
\makeatletter
\@addtoreset{equation}{section}
\makeatother

\def\gm{g_-}
\def\mo{{m_{odd}}}
\def\me{{m_{even}}}

\def\pb{{\bf p}}
\def\zh{\hat{z}}
\def\ph{\hat{q}}
\def\kh{\hat{\kappa}}
\def\dh{\hat{d}}
\def\C{{\mathbb C}}
\def\R{{\mathbb R}}
\def\Z{{\mathbb Z}}
\def\Q{{\mathbb Q}}

\def\O{\Omega}
\def\Oh{{\widehat{\Omega}}}

\def\Qcal{{\mathcal Q}}
\def\Mcal{{\mathcal M}}
\def\Lcal{{\mathcal L}}
\def\Mcal{{\mathcal M}}

\def\f{\frac}
\def\e{\epsilon} 
\def\d{\delta}
\def\a{\alpha}
\def\b{\beta}
\def\g{\gamma}
\def\s{\sigma}
\def\o{\omega}
\def\deg{{\rm deg}}
\def\qd{Q}
\def\CC{C}
\def\Ch{\widehat{C}}
\def\Ccal{{\mathcal C}}

\def\Bh{\widehat{B}}
\def\gh{\hat{g}}
\def\at{\tilde{a}}

\def\ka{\kappa}
\def\wh{\widehat}

\def\Lcal{{\mathcal L}}
\def\Acal{{\mathcal A}}
\def\bk{{\bf k}}
\def\bl{{\bf l}}
\def\l{\lambda}
\def\sh{\hat{\sigma}}
\def\tauh{\widehat{\tau}}

\def\dim{{\rm dim}}
\begin{document}
\baselineskip 14pt plus 1pt minus 1pt

\vspace{0.2cm}
\begin{center}
\begin{Large}
\fontfamily{cmss}
\fontsize{17pt}{27pt}
\selectfont
\textbf{Hodge and Prym tau functions,  Jenkins-Strebel differentials and  combinatorial model of $\Mcal_{g,n}$}
\end{Large}\\
\bigskip
M. Bertola$^{\dagger\ddagger}$\footnote{Marco.Bertola@concordia.ca, mbertola@sissa.it},  
D. Korotkin$^{\dagger}$ \footnote{Dmitry.Korotkin@concordia.ca},
\\
\bigskip
\begin{small}
$^{\dagger}$ {\it   Department of Mathematics and
Statistics, Concordia University\\ 1455 de Maisonneuve W., Montr\'eal, Qu\'ebec,
Canada H3G 1M8} \\
\smallskip
$^{\ddagger}$ {\it  SISSA/ISAS,  Area of Mathematics\\ via Bonomea 265, 34136 Trieste, Italy }\\
\end{small}
\vspace{0.5cm}
{\bf Abstract} \end{center}
%
%\maketitle

The  goal of the paper is to apply the  approach inspired by the theory of integrable systems to  construct explicit sections of line bundles over  the combinatorial model of the moduli space of pointed Riemann surfaces based on Jenkins-Strebel differentials. The line bundles are  tensor products of the determinants of the Hodge or Prym vector bundles with the standard tautological line bundles $\mathcal L_j$ and the sections are constructed in terms of tau functions. The combinatorial model is  interpreted as the real slice  of a  complex analytic moduli space of quadratic differentials where  the phase of each  tau-function provides a section of a circle bundle. 
The phase of the ratio of the Prym and Hodge tau functions gives a section of the $\kappa_1$-circle bundle.

By evaluating the increment of the phase around co-dimension $2$ sub-complexes, we identify the  Poincar\'e\ dual cycles to the Chern classes of the  corresponding line bundles: they are expressed explicitly as combination of Witten's cycle $\Wfive$ and  Kontsevich's boundary. This provides combinatorial analogues of Mumford's relations  on  $\Mcal_{g,n}$ and Penner's relations in the hyperbolic combinatorial model. The  free homotopy classes of loops around $\Wfive$ are  interpreted as pentagon moves while those of loops around Kontsevich's boundary as combinatorial Dehn twists.

Throughout the paper we exploit the classical description of the combinatorial model in terms of Jenkins--Strebel differentials, parametrized in terms of period, or {\it homological} coordinates; we  show that they provide Darboux coordinates for the symplectic structure introduced by Kontsevich. We also express the latter   as the intersection pairing in the odd homology of the  canonical double cover.

\tableofcontents

\section{Introduction}
Tau-functions play an important role in the theory of moduli spaces starting from the works   \cite{Witten} and \cite{Kontsevich} where it was shown that the tau-function of a  special solution of the Korteveg-de-Vries hierarchy generates  intersection numbers of $\psi$-classes over $\Mcal_{g,n}$.  
This led to numerous further important results. For example  intersection numbers of $\kappa$-classes with 
$\psi$-classes as well as Hodge integrals  turn out to be generated by  KdV and KP  tau-functions \cite{Kazarian,ELSV};
tau-functions of KP and Toda equation
define also Hurwitz numbers of various  kind \cite{Okounkov}. We refer to  \cite{LanZvon,Pandharipande} and references therein for detailed discussion
 of the subject.

Tau-functions of another type were introduced in  \cite{Annalen} while 
studying the Jimbo-Miwa isomonodromic deformations of  quasi-permutation monodromy groups. They 
are special cases of general isomonodromic Jimbo-Miwa tau-functions \cite{JM,Malgrange}; these tau-functions  
were named after  Bergman due to their relation to the Bergman projective connection.
The Bergman tau functions naturally appear   in various contexts,  including Dubrovin's  theory of Frobenius manifolds  \cite{Dubrovin,IMRN1}.

In \cite{Advances} it was shown that the Bergman tau-function  on Hurwitz spaces is essentially  a section of  the determinant line bundle of the  Hodge bundle. This observation was used  to compute the Hodge class over spaces of admissible covers via boundary divisors, generalizing a  result 
of Cornalba-Harris \cite{CorHar}. The theory of Bergman tau-functions was further applied
 to derive new relations in the Picard group  of the spaces of abelian,  quadratic and $n$--holomorphic  differentials over Riemann surfaces \cite{JDG,MRL,contemp, KSZ} and give an elementary proof of Farkas-Verra relations \cite{Farkas} in the Picard group of
  moduli spaces of spin curves \cite{Basok}.
 
In \cite{Leipzig,contemp} it was shown that there exist  two natural tau-functions associated to spaces of quadratic differentials
over Riemann surfaces. The first  tau-function is a holomorphic section of the determinant line bundle of the Hodge vector bundle $\Lambda_H$; the second tau-function 
is a section of   the determinant of the Prym vector bundle $\Lambda_P$. Therefore, these two tau-functions (as well as their analogs in
the case of Jenkins-Strebel differentials) are  naturally called the {\it Hodge} ($\tau_+$) and the {\it Prym} ($\tau_-$) tau-functions, respectively (see Sec. \ref{Hodge_sec} and Sec. \ref{prymsect}).
The study of the properties of $\tau_\pm$  gives in particular a direct analytical proof of  Mumford's formula
$\lambda_2-13 \lambda_1=-\delta$  (which holds in $Pic(\overline \Mcal_g,\Q)$), where $\delta$ is the class of Deligne-Mumford boundary \cite{contemp}. The Hodge and Prym  tau-functions turn out to be also relevant in the study of Lyapunov exponents of the Teichm\"uller flow 
\cite{EKZ}.  

The combinatorial  model   of $\Mcal_{g,n}$, denoted by   $\Mcal_{g,n}[{\bf p}]$,  for fixed $\pb\in \R_+^n$, is based on the theory of Jenkins-Strebel (JS) differentials. This model  was developed starting from ideas of  Harer, Mumford and Thurston (see \cite{HarMum}; a modern exposition is given in \cite{Mondello2}). 
We call it the {\it flat combinatorial model} as opposed to Penner's {\it hyperbolic combinatorial model} \cite{Penner}.
The  flat combinatorial model was central in Kontsevich's proof \cite{Kontsevich} of 
 Witten's conjecture \cite{Witten}.

Cells of $\Mcal_{g,n}[{\bf p}]$, with ${\bf p} = (p_1,\dots, p_n)\in \R^n_+$,   are labeled by  ribbon graphs (or fatgraphs) of given topology on a Riemann surface $\CC$ 
of genus $g$ punctured at $n$ points. The vertices of the ribbon graph are  zeros of the JS quadratic differential $\qd$ while the faces 
correspond to poles of $\qd$. The quadratic residues at the poles are given by $-p_i^2/4\pi^2$, where $p_1,\dots,p_n$ are the perimeters of the faces.
 The lengths of the edges in the metric $|\qd|$ coordinatize the cell.
 All cells with given multiplicities $k_1,\dots,k_m$ of the zeros of  $\qd$ form a stratum  of
$\Mcal_{g,n}[{\bf p}]$  labeled by the vector ${\bf k}$. 

There exists a compactification $\overline{\Mcal}_{g,n}[\pb]$ of the combinatorial model \cite{Mondello2, Looijenga} which provides a partial parametrization of the Deligne--Mumford compactification $\overline{\Mcal}_{g,n}$; points of the combinatorial (Kontsevich's) boundary correspond to quadratic differentials with simple poles, i.e.,  ribbon graphs with uni-valent vertices.
The main stratum of $\Mcal_{g,n}[{\bf p}]$  corresponds to quadratic differentials with all   simple zeros, so that the Strebel graph has only three-valent vertices; we denote this stratum by $W$.  The boundaries of real co-dimension $1$  of the cells of $W$ (the {\it facets})  correspond to JS differentials $\qd$ with one
double zero, while all other zeros remain simple;  the corresponding ribbon graphs have one 4-valent vertex while all other vertices are three-valent.
The union of cells of $W$ and their facets will be denoted by $\tilde{W}$. The complement, $\Mcomb\setminus \tilde{W}$,
contains cells of (real) co-dimension 2 and higher.

In  co-dimension 2 there exist two special sub-complexes of $\Mcomb$:
\begin{enumerate}
\item [--] The first, called "Witten's cycle" $\Wfive $, corresponds to ribbon graphs with at least one  
5-valent vertex. In the main stratum of $\Wfive$ the  differential $\qd$ has one zero of order 3 while all other zeros are simple.
\item [--]
The second,  $\Wbdr $ (also a cycle) is the  "Kontsevich's boundary" of $\Mcal_{g,n}[\pb]$. The main stratum corresponds to the collapse of two edges forming a homotopically non-trivial loop, and
further 
resolution of the arising two-valent vertex into two 1-valent ones.
Then the differential $\qd$ acquires  two simple poles at the nodal points in the normalization of the curve \cite{Mondello2}. The corresponding ribbon graph either remains connected (and has two 1-valent vertices) or consists of two components with one 1-valent vertex each.
The cycle $\Wbdr $ is the sum of several sub-cycles that correspond to  different topological types of the boundary of
$\Mcal_{g,n}$. 
\end{enumerate}
%See Fig. \ref{figw5deg}.
 While $\Mcal_{g,n}[\pb]$ is in one-to-one correspondence with $\Mcal_{g,n}$ itself,
this isomorphism does not extend to the boundary: some components of Deligne-Mumford boundary  $\overline{\Mcal}_{g,n}\setminus \Mcal_{g,n}$ (the ones corresponding 
to homologically trivial vanishing cycles which leave all punctures in  one of the connected components of the stable curve) are not represented by cells of highest dimension of $\Wbdr $.

On the main stratum of each Witten-Kontsevich cycle (see Def. \ref{defWK}) of $ \Mcomb$ there is an orientation that extends consistently from cell to cell. This orientation is
 defined by the top power of Kontsevich's symplectic form 
\be
\label{Kontintro}
\Omega   = \sum_{i=1}^n p_i^2 \omega _i
\ee
 where $\omega_i$ is  the representative of the  first Chern class $\psi_i$  of the $i$-th tautological line bundle $\Lcal_i$ \cite{Kontsevich}, see \eqref{defetaf}. 
 The integral of the top power of the form $\Omega$ over $W$
was used in \cite{Kontsevich} to generate the intersection indices of $\psi$-classes over $\Mcal_{g,n}$.
 \vskip 5pt

Denote by $\lambda$ and $\lambda^{(n)}_2$ the first Chern classes of the determinant line bundles of the Hodge vector bundle $\Lambda_H$ and the vector bundle $\Lambda_2^{(n)}$ of quadratic differentials with at most simple poles at punctures
over $\overline{\Mcal}_{g,n}$, respectively. Then Mumford formulas   (Th.7.6 of \cite{ArbCor_book}) in ${\rm Pic}(\overline{\Mcal}_{g,n},\Q)$ give the following relation between $\lambda$,  $\lambda^{(n)}_2$, the classes $\psi_i$ and the Deligne--Mumford boundary $\delta_{DM}$:
\be
\lambda^{(n)}_2-13\lambda=\sum_{i=1}^n\psi_i -\delta_{DM}.
\la{Mumint}
\ee
The formula (\ref{Mumint}) was established at the level of differential
forms in \cite{TZ} where the class $\sum_{i=1}^n \psi_i$ on $\Mcal_{g,n}$ was expressed via Eisenstein series.
%:

Another formula also due to Mumford expresses the   kappa--class $\kappa_1$    as follows (see Th.7.6 of \cite{ArbCor_book})
\be
\kappa_1=12\lambda+\sum_{i=1}^n \psi_i -\delta_{DM}.
\la{kaplam}
\ee

The formulas (\ref{Mumint}) and (\ref{kaplam}) imply the expression of  $\kappa_1$ via $\l$ and $\lambda^{(n)}_2$:
\be
\kappa_1=\l_2^{(n)}-\l\;,
\la{kall}
\ee
therefore $\kappa_1$ coincides with the first Chern class of the following line bundle over $\overline{\Mcal}_{g,n}$:
\be
\chi_\kappa=\frac{{\rm det}\Lambda_2^{(n)}}{{\rm det}\Lambda_H}.
\la{Lkappa}\ee

In the hyperbolic combinatorial model of $\overline{\Mcal}_{g,n}$ the  following relation was found by Penner \cite{Penner}: $12\kappa_1=\Wfive ^{hyp}+\Wbdr ^{hyp}$ where $\Wfive ^{hyp}$ is the analog of Witten's cycle and $\Wbdr ^{hyp}$ is the cycle corresponding to the boundary
of $\Mcal_{g,n}$ in the hyperbolic combinatorial model.
An analog of Penner's formula 
\be
12\kappa_1=\Wfive +\Wbdr 
\la{kaWW}
\ee
 in the flat  combinatorial model was  proved in 
\cite{ArbCor}; see 
\cite{Mondello1,Igusa1} for alternative proofs (notice that the sum of boundary divisors in the second formula of Corollary A.1  of \cite{Mondello1} coincides with $\Wbdr $).

\vskip 10pt

\subsection{Summary of results}
 The principal  goal of this paper is to express the Hodge class $\l$, the class $\l_2^{(n)}$ (and the closely related Prym class $\lambda_P=c_1(\det \Lambda_P)$, see details in   Sec. \ref{PrymonQ}) via $\psi$-classes and  combinatorial cycles $\Wfive $ and $\Wbdr $ using  the  Hodge and Prym tau functions $\tau_\pm$.  
As a corollary we provide a new proof of (\ref{kaWW}) and 
derive the analogs of Mumford's relations \eqref{Mumint} and  \eqref{kaplam} in the context of the  flat combinatorial model.

Denote by $S[\chi_\kappa]$ the circle bundle
associated  to the line bundle $\chi_\kappa$ over the flat combinatorial model. We prove that a section of  this ``kappa circle bundle'' is given by the argument of  the ratio $\frac {\tau_-}{\tau_+}$ (Corollary \ref{corsection}).

 We show that 
the periods of the square root of the $JS$ differential on the two-sheeted canonical cover give
 Darboux coordinates for the Kontsevich symplectic form \eqref{Kontintro} which allows to  effectively define the orientation of different cycles of the combinatorial model. 
\vskip0.3cm
Let us now describe the content of the paper in more detail.

\noindent {\bf Canonical cover defined by a meromorphic quadratic differential.} 
Let $\qd$ be a meromorphic quadratic differential on a curve $C$ of genus $g$ such that  the only even-order poles are of second order and denoted by $z_1,\dots,z_n$. It is convenient to group  the zeros of odd multiplicity together with the  poles of odd order, denoting the total number by $\mo$.  We denote by $\me$  the number of zeros of even multiplicity  and  write the divisor $(Q)$ (${\rm deg} (Q)=4g-4$) in form
\be
(\qd)=\sum_{i=1}^{\nz} d_i q_i\equiv\sum_{i=1}^\mo (2k_i+1)x_i+\sum_{i=\mo+1}^{m} 2l _i x_i -\sum_{i=1}^{n} 2 z_i\;.
\la{divqd1}
\ee
where $k_i\in \Z$, $l_i\in \mathbb N$, $m= \mo + \me$. 

The moduli space of pairs $(C,\qd)$ where divisor of $\qd$ is of the form 
(\ref{divqd1}) will be denoted by $\Qcal_{g,n}^{\bk,\bl}$.

The canonical covering $\Ch$ of $\CC$ is defined by the resolution of the nodes of the curve 
\be
v^2=Q
\la{cancov},
\ee 
 in $T^*\CC$.
In the case of holomorphic quadratic differentials the canonical covering appeared first in  Teichm\"uller theory (see \cite{Abikoff}); it also gives a simplest example of the {\it spectral curve} 
 of Hitchin's systems \cite{Hitchin} and  plays a role in the theory of supersymmetric Yang-Mills equations \cite{Seiberg-Witten} where the differential is allowed to be meromorphic.

The covering (\ref{cancov}) is branched at  poles and zeros of odd multiplicity $\{x_i\}_{i=1}^\mo$.
Note that $\mo\in 2\mathbb N$ because  $\deg (\qd) = 4g-4\in 2 \mathbb N$.
Therefore the  genus of $\Ch$ equals
 \be
\hat{g}=2g+\f{\mo}{2}-1.
\la{genush}
\ee
Denote by $\pi$ the canonical projection $\Ch\to \CC$ and by $\mu$ the involution of $\Ch$ interchanging the sheets. 

Each of  the zeros of even multiplicity $\{x_i\}_{i=\mo+1}^m$ has  two preimages on $\Ch$  which we denote by $\xh_i$ and 
$\xh_i^\mu$.
Similarly, the preimages of a pole $z_i$ are denoted  by $\zh_i$ and $\zh_i^\mu$. The pre-images on $\Ch$ of poles and zeros
of odd multiplicity $\{x_i\}$  are branch points of the projection $\Ch\to \CC$; therefore, we shall continue to denote them by the same letters, omitting the hat.

The holomorphic  involution $\mu:\Ch \to \Ch$ induces a linear map with eigenvalues $\pm 1$ on the cohomology of $\Ch$ and therefore the space $H^{(1,0)}(\Ch)$
 of holomorphic differentials on $\Ch$ can be decomposed into the direct sum of  two eigenspaces $H^\pm$
\be
H^{(1,0)}(\Ch) = H^+ \oplus H^- \ ,\qquad \dim \, H^+ = g\ ,\qquad \dim \, H^{-} =\gm \;.
\ee 
where
\be
\gm= g+\f{\mo}{2}-1
\la{gmd}
\ee
The space $H^+$ can be identified with the fiber of the {\it Hodge vector bundle} $\Lambda_H$ over 
$\Qcal_{g,n}^{\bk,\bl}$. 
The space $H^-$ is the space of holomorphic {\it Prym differentials}; it  is the fiber of the {\it Prym vector bundle}  $\Lambda_P$ over $\Qcal_{g,n}^{\bk,\bl}$.

The differential $v$ satisfies 
$\mu^*v= -v$;  it  is a  meromorphic Prym differential. The differential $v$   is holomorphic (i.e. $v\in H^-$) if $\qd$ does not have poles of order higher than 1.

The  homology group of $\Ch\setminus\{ \zh_j,\zh^\mu_j\}_{j=1}^n$, relative to $ \{\hat{x}_j,\hat{x}_j^\mu\}_{j=\mo+1}^{m}$, which we denote by 
\be
H_1(\Ch\setminus \{ \zh_j, \zh^\mu_j\}_{j=1}^n;
\{\hat{x}_j,\hat{x}_j^\mu\}_{j=\mo+1}^{m})
\la{homsplit}
\ee
(we consider it over $\R$)
 decomposes as  
$H_+\oplus H_-$,
where ${\rm dim} \, H_+=2g$, ${\rm dim}\,  H_-=2\gm+n+m_{even}$.

 Notice that
$\dim H_-=\dim \Qcal_{g,n}^{{\bk,\bl}}\;.$
Choosing some basis of cycles $\{s_j\}$  in $H_-$ we  define the {\it period}, or {\it homological coordinates} on the moduli space $\Qcal_{g,n}^{{\bk,\bl}}$ by
\be
\label{homcoordint}
{\mathcal P}_{s_j}  =\int_{s_j} v\;,\hskip0.7cm j=1,\dots,2 g_- + n+m_{even}.
\ee
When $\qd$ has at most double poles (therefore  $k_i\geq -1$)
the real slice $\{\Pcal_{s_j}\in \R, \ \ \forall j=1\dots \wh g\}$ of  $\Qcal_{g,n}^{\bk,\bl}$
corresponds to  Jenkins-Strebel differentials.

{\bf Hodge and Prym tau-functions.} The Hodge ($\tau_+$) and Prym ($\tau_-$) tau-functions on spaces of
quadratic differentials are the main analytical tools used in this paper; in their definition we follow
\cite{Leipzig,contemp,CMP}.
They satisfy a system of differential equations with respect to the homological coordinates $\Pcal_j$  \eqref{homcoordint}
which can be solved in terms of theta-functions and other canonical objects \cite{IMRN,JDG}. Here we write the formula for $\tau_+$ 
on the space $\Qcal_{g,n}^{\bk,\bl}$ referring to 
Section \ref{prymsect} for the formula for $\tau_-$. 

Denote by $E(x,y)$ the prime-form on $\CC$, by $\Acal_x$ the Abel map with  base-point $x$  and by $K^x$ the vector of Riemann constants. 
For $n\geq 1$ the fundamental polygon of $\CC$ can always be chosen so that (see Lemma 6 of \cite{JDG})
$
\f{1}{2}\Acal_x((\qd))+2K^x=0.
$

Introduce the following multi-valued $g(1-g)/2$ - differential $\Ccal(x)$ \cite{Fay92}:
\be
\Ccal(x)=\f{1}{W(x)}\left(\sum_{i=1}^g v_i(x)\f{\p}{\p w_i}\right)^g\theta(w,\O)\Big|_{w=K^x}
\ ,\qquad \ W(x):= \det \left[\frac {{\rm d}^{k-1}}{{\rm d} x^{k-1}} v_j\right]_{1\leq j,k\leq g}
\la{defCint}
\ee
where $\Omega$ is the period matrix of $\CC$, $\{v_j\}_{j=1}^{g}$ are holomorphic 1-forms on $\CC$ normalized by
$\oint_{a_i} v_j=\delta_{ij}$  and $\theta$ is the corresponding theta-function.

The Hodge tau function $\tau_+$  is then expressed in terms of the divisor $(\qd)=\sum d_i q_i$ as follows:
\be
\tau_+=
\Ccal^{2/3}(x)\left(\f{\qd(x)}{\prod_{i=1}^{\nz}E^{d_i}(x,q_i)}\right)^{(g-1)/6} 
\prod_{i<j} E(q_i,q_j)^{\f{d_i d_j}{24}}\, .
\la{taupint}
\ee
Although the formula \eqref{taupint} seems to depend on $x$, in fact it is constant with respect to it. 
The  prime-forms in \eqref{taupint}  are evaluated at the points $q_i$ in the so--called {\it distinguished}  
local coordinates $\zeta_i$ as follows:
\be
E(x,q_i)=\lim_{y\to q_i} E(x,y) \sqrt{d\zeta_i(y)},
\la{defEpi}
\ee
\be
 E(q_i,q_j)=\lim_{x\to q_j, y\to q_i} E(x,y) \sqrt{d\zeta_i(y)} \sqrt{d\zeta_j(x)}.
\la{defEppi}
\ee
 Near zeros  $x_j$ of odd/even multiplicities (see \eqref{divqd1})   the distinguished local coordinate are defined by, respectively, 
\be
\zeta_j(x)=\left[\int_{x_j}^x v\right]^{2/(2k_j+3)}\;,\hskip0.4cm j=1,\dots, \mo\;;
\hskip0.7cm
\zeta_j(x)=\left[\int_{x_j}^x v\right]^{1/(l_j+1)}\;,\hskip0.4cm j=\mo+1,\dots,m,
\ee
while, near the double poles $z_j$ they  are given by
\be
\xi_j(x)=\exp\left\{\f{2\pi i}{p_j}\int_{x_1}^x v\right\}
\la{xijint}
\ee
where $x_1$ is a chosen zero of $\qd$ and $p_k=2\pi i \, {\rm res}|_{z_k} v$. The expression (\ref{taupint}) depends on the choice of the ``first'' zero  $x_1$ and on the  
integration paths in \eqref{xijint}; different choices affect the coordinates $\{\xi_j\}$ by an $x$--independent factor.
Under the change of Torelli marking on $\CC$ with an $Sp(2g,\Z)$ matrix $\left(\ba{cc} A & B \\ C & D\ea\right)$ the function  $\tau_+$ 
is multiplied by ${\rm det}(C\Omega+D)$ up to a root of unity.

The function $\tau_+$  (\ref{taupint})  has degree of homogeneity equal to  $-1/12$ with respect to this factor  and hence a certain integer power  of the expression
\be
\tau_+^{12}\prod_{k=1}^n d\xi_k(z_k)\,,
\la{defTpmint}
\ee
is invariant under the choice of local parameters $\xi_j$ near $z_j$ and also under  the choice of signs in the definition of $v$, the prime-forms and local parameters $\zeta_j$.

The Prym  tau-function $\tau_-$ is similarly constructed on any $ \Qcal_{g,n}^{\bk,\bl} $ in Section \ref{prymsect}; under a change of canonical basis
in  $H_-$ by a symplectic matrix $\left(\ba{cc} A_- & B_- \\ C_- & D_-\ea\right)$ the function  $\tau_- $ 
is multiplied by ${\rm det} (C_-\Omega_-+D_-)$ (up to a root of unity) where $\Omega_-=2\Pi$ with $\Pi$ being the Prym matrix
of the canonical cover.

\paragraph{Homological symplectic structure on the combinatorial model.}
When all $k_i$ and $l_i$ in \eqref{divqd1} are equal to $0$ the space $ \Qcal_{g,n}^{\bk,\bl}  $ will be denoted by $\Qcal^0_{g,n}$: in this case all zeros of $\qd$ are simple and all poles are of second order at $z_1,\dots,z_n$
i.e. $\me=0$ and $m=\mo=4g-4+2n$.
Moreover  $g_-={\rm dim}\, H^-=3g-3+n$ and ${\rm dim}\,H_-={\rm dim}\,\Qcal^0_{g,n}=6g-6+3n = 2g_-+n$.

 On the other hand, the intersection pairing on $H_-$ has rank $6g-6+2n = \dim\,H_- -n$. Given any two cycles $s,\wt s\in H_-$ and their corresponding homological coordinates $\mathcal P_{s}$ and $ \mathcal P_{\wt s}$ we define the {\bf homological Poisson bracket} by the formula 
 \be
 \label{introPois}
 \left\{ \mathcal P_s ,\mathcal P_{\wt s}\right\} = s\circ \wt s.
 \ee 
 For $n=0$ (spaces of holomorphic quadratic differentials) the bracket \eqref{introPois} was introduced  in  \cite{BKN}. 

 The Casimir functions of  \eqref{introPois} are $\mathcal P_{\gamma_j}, \ \ j=1,\dots, n$,  where   $\gamma_j\in H_-$  is a half of the difference of two small positively oriented circles around the poles $z_j, z_j^\mu$ of $v$ on the cover 
 $\Ch$.  
  Denote now by  $\Qcal^0_{g,n}[{\bf p}],  \ \ {\bf p} = (p_1,\dots, p_n)$ the  symplectic leaves of \eqref{introPois} in 
$\Qcal_{g,n}^0$ where the quadratic residues at the poles  $z_j$'s   have the  fixed values  $-\frac {p_j^2}{4\pi^2}$.  
The symplectic form on the leaves $\Qcal^0_{g,n}[{\bf p}]$ is given by  
\be
\Omega_{hom}= \sum_{i=1}^{g_-} dA_i\wedge dB_i\;.
\la{omint}
\ee
where $A_j = \mathcal P_{a_j}, B_j = \mathcal P_{b_j}, \ j=1,\dots,g_- = 3g-3+n$ are the homological coordinates corresponding to a symplectic basis in $(H_-\!\!\mod \R\{\gamma_j\}_{j=1}^n)$.  The coordinates $A_j, B_j$ are thus defined up to addition of integer multiples of $p_\ell$'s, but the form \eqref{omint} is well--defined on the symplectic leaves. Similarly, the form (\ref{omint}) remains well-defined defined on symplectic leaves  $\Qcal^{\kb,0}_{g,n}[{\bf p}]$
when all zeros have odd multiplicity.

\paragraph{Real section of $\Qcal^0_{g,n}[{\bf p}]$ and combinatorial model.}
The real slice  of $\Qcal^0_{g,n}[{\bf p}]$ (where all homological coordinates including the perimeters $p_j$ are real) corresponds to the space of quadratic Jenkins--Strebel differentials with simple zeros, and hence it is in one-to-one correspondence with points of the largest stratum  of the flat combinatorial model $\Mcal_{g,n}[{\bf p}]$.

The complex symplectic form \eqref{omint} induces a real symplectic form on each maximal  cell of  the flat combinatorial model.  A parallel construction provides a symplectic structure on  every  cell  where the Jenkins--Strebel differential $\qd$ has only zeros of  odd multiplicity. The following theorem states the equality between Kontsevich's and homological symplectic forms in the combinatorial setting.

\vspace {10pt}

{\bf Theorem} [Thm. \ref{thmomega}]\ {\it 
The homological symplectic structure $\Omega_{hom}$ \eqref{omint} coincides with the Kontsevich symplectic form $\Omega$  \eqref{Kontintro} on each cell of  $\Mcal_{g,n}[{\bf p}]$ labelled by ribbon graphs with only odd-valent vertices.
Therefore the homological coordinates $\{A_j, B_j\}_{j=1}^{g_-}$ are Darboux coordinates of $\Omega$. }
\vspace {10pt}

The symplectic form allows us to define an orientation within each of the  cells mentioned in the theorem; analogously to  \cite{Kontsevich, Mondello2} we  prove in Section \ref{Orient} that this orientation propagates consistently between neighbouring cells of the Kontsevich-Witten cycles (see Def. \ref{defWK} and Prop. \ref{proporientation}). Moreover the orientation is also consistently propagating  between Kontsevich-Witten cycles of different dimensions.

\paragraph{Hodge and Prym  classes on $\overline \Mcal _{g,n}[{\bf p}]$ using $\tau_\pm$.}
The Prym bundle $\Lambda_P$ can be extended from  $\Mcal _{g,n}[{\bf p}]$ to   $\overline \Mcal _{g,n}[{\bf p}]$ 
following \cite{contemp,KSZ}. Over the largest stratum  of  $\Mcal _{g,n}[{\bf p}]$ the fiber of $\Lambda_P$ coincides with the space $H^-$ of holomorphic Prym differentials. Over smaller strata the fibers of $\Lambda_P$ contain also meromorphic Prym differentials.

In previous papers \cite{Advances,MRL,contemp, KSZ} the study of the divisor  of tau-functions allowed to  express the first Chern class of the  Hodge
(and Prym in \cite{contemp}) vector bundles via boundary divisors on various moduli spaces. In these cases the tau-functions are complex-analytic 
on the corresponding moduli spaces.
The straightforward application of this approach to the combinatorial model  is not possible because 
$\Mcal _{g,n}[{\bf p}]$ is not complex-analytic itself. 
On the other hand, since the combinatorial model  is embedded into the space  $\Qcal_{g,n}[{\bf p}]$ as its real slice, we can restrict $\tau_\pm$  from strata of $\Qcal_{g,n}[{\bf p}]$ to strata of $\Mcal _{g,n}[{\bf p}]$.
This gives   sections of (some integer powers of) the bundles $\det \Lambda_{(P,H)}^{12} \prod_{i=1}^n \mathcal L_i$, respectively. These sections are real--analytic within each maximal cell of the combinatorial model. 

 Therefore,  similarly to
\cite{Kontsevich} (see \cite{Zvonkine} for details) instead of working with line bundles it is more natural to consider the associated  circle (or $U(1)$) bundles. 

The computation of  the Poincar\'e\ duals of $c_1\left(\det \Lambda_{(P,H)} ^{12}\prod_{i=1}^n \mathcal L_i\right)$ 
in the combinatorial model  requires the evaluation of  the increment of the arguments  $\Phi_\pm = \arg\, \tau_\pm$ around cycles of co-dimension $2$, which are the Witten cycles $\Wfive$ and the Kontsevich boundary $\Wbdr $.  Closed paths around $\Wfive $ are represented  by pentagon moves, while  closed paths around  $\Wbdr $ are represented by  combinatorial Dehn's twists (see Section \ref{ChekFock}).
The absolute values  $|\tau_\pm|$ vanish on all facets of co-dimension $1$; on the other hand (Prop. \ref{propagation})  $\Phi_\pm$ can be extended continuously across the  facets of the complex allowing us to compute the increment around the co-dimension $2$ sub-complex. 

The computation of the  increment of $\Phi_\pm$  along closed paths can be performed by studying local models using results of   \cite{BK1} where special types  of genus one Boutroux curves were analyzed.
This leads to the following \par\vskip 5pt

{\bf Theorem [Thm. \ref {thmlambda1}]} {\it 
The following relations hold:
\be
\lambda+\f{1}{12}\sum\psi_i= \f{1}{144}\Wfive  +\f{13}{144} \Wbdr \la{112}\;\;\;,
\ee
\be
\lambda_P+\f{1}{12}\sum\psi_i= \f{13}{144}\Wfive  +\f{25}{144}\Wbdr \la{113}\;\;\;.
\ee
}
\par\vskip 5pt
We prove that the  Prym vector bundle $\Lambda_P$  over $\Mcomb$ is  isomorphic to the vector bundle $\Lambda_2^{(n)}$ of meromorphic quadratic 
differentials on $\CC$ with simple poles at punctures (an analog of this statement in the case of holomorphic quadratic differentials was established in \cite{contemp,KSZ}).
Therefore, taking into account that $\kappa_1 =\lambda_2^{(n)}-\lambda$  \eqref{kall} 
one can express the class $\ka_1$ via $\Wfive $ and $\Wbdr$ reproducing the Arbarello-Cornalba formula (\ref{kaWW}).

Denote  the circle bundle corresponding to the line bundle $\Lcal_\kappa$ (\ref{Lkappa}) by
$S[\Lcal_\kappa]$ (we recall that $c_1(\Lcal_\kappa)=\kappa_1$).
Then the relations (\ref{112}), (\ref{113})  imply the following 
\par\vskip 5pt
\noindent {\bf Corollary [Cor. \ref{corsection}]} {\it 
   A section of  the circle bundle $ S[(\Lcal_{\ka})^{48}]$  over $\Mcomb$ is given by $ \frac {\Theta_-}{\Theta_+}$ where $ \Theta_\pm = 
\left(\f{\tau_\pm} {|\tau_\pm|} \right)^{48}$.
}
\par\vskip 5pt

The elimination of 
the Witten cycle $\Wfive $ from \eqref{112}, \eqref{113} leads to the following corollary

\par\vskip 5pt
\noindent {\bf Corollary [Cor. \ref{corsection}]} {\it 
The following  relation holds:
\be
\lambda_2^{(n)}- 13 \lambda- \sum_{i=1}^n \psi_i= - \Wbdr \;.
\la{MumMgn}
\ee
}
\par\vskip 5pt
The formula (\ref{MumMgn}) is a combinatorial version of the classical Mumford's relation
 (\ref{Mumint}) to be understood in the same way as \eqref{112} and  \eqref{113}. 
 
 \par \vskip 5pt
 
The Arbarello--Cornalba formula \eqref{kaWW} and Mumford relations  \eqref{MumMgn} were known as statement about classes, with the latter relation actually being formulated on $\Mcal_{g,n}$ and not on the combinatorial model (recall that Kontsevich's boundary
$\Wbdr $ is "smaller" than the DM boundary of $\Mcal_{g,n}$).  In contrast, we obtain these formulas from the presentation of {\it explicit} sections  $\tau_{\pm}$ of the corresponding line--bundles. 
The explicit formulas for the circle bundles associated to $\kappa_1, \lambda, \lambda_P$ should be seen as far reaching analogies of the relationship between  sections of powers of  $\psi_1$ on $\Mcal_{1,1}$ in terms of modular form and Eisenstein series  (\cite{Zvonkine}, Sect. 2.2.2).

The advantage of explicit formulas for sections of line bundles is that, for example,  questions about the Poincar\'e\ cycles representing a given line bundle can be immediately addressed by analyzing the divisor of the section.  For example, a quite elementary question that remains unanswered is to identify the Poincar\'e\ dual of a $\psi$--class, and therefore also $\sum_{j=1}^n \psi_j$ (Takhtajan--Zograf class \cite{TZ}) in the combinatorial picture. In the $\Mcal_{1,1}$ case this question is elementary because of the identification   between sections and modular forms but in general the answer is not known.  Therefore  the construction of  explicit sections of line bundles has the twofold advantage of providing non-trivial special functions in the theory of integrable systems 
%generalizing Riemann theta functions, 
as well as providing tangible insight into the geometry of moduli spaces. 
\paragraph{\bf Organization of the paper.}
 In Section \ref{combmod} we discuss  the flat combinatorial model of $\Mcal_{g,n}$ based on Jenkins-Strebel differentials. Here we interpret pentagon moves and combinatorial Dehn's twist as paths around
the Witten-Kontsevich cycles $\Wfive $ and $\Wbdr $, respectively. We also prove that homological, or period, coordinates are Darboux coordinates for the 
Kontsevich's symplectic form used to orient the flat combinatorial model.
In Section \ref{2scan} we describe the geometry of canonical covering 
of a Riemann surface defined by a meromorphic quadratic differential and introduce homological coordinates on moduli spaces of quadratic differentials with given multiplicities of poles and zeros.
In Section \ref{HPsec} we define and study the Hodge and Prym tau-functions on moduli spaces of quadratic differentials with given  orders of poles and zeros.  We treat in details two main examples in genus zero which play a key role in application to flat combinatorial model of $\overline{\cal M}_{g,n}$.
In Section \ref{final_res} we compute the increments of the argument of $\tau_\pm$ with respect to pentagon move and combinatorial Dehn's twist (i.e. the monodromy around Witten's cycle $\Wfive $ and Kontsevich's boundary
$\Wbdr $) which leads to  the formulas (\ref{112}) and (\ref{113}) expressing   Hodge and Prym classes via Kontsevich-Witten cycles in the combinatorial setting. 
As a corollary we  prove  relation (\ref{MumMgn}) between classes $\lambda_1$ and $\lambda_2^{(n)}$, the formula (\ref{kaWW}) and construct an explicit section of circle bundle  $S[\Lcal_\kappa]$ such that $c_1(\Lcal_\kappa)$
is the first kappa-class
$\kappa_1$.

\section{Combinatorial model of $\Mcal_{g,n}$ via Jenkins-Strebel differentials}
\la{combmod}

To introduce  the  flat combinatorial model of $\Mcal_{g,n}$ we
start from the following definition (equivalent to the standard one \cite{Strebel}):
\begin{definition}
\label{defJS}
A quadratic differential $\qd$ on an $n$--marked smooth curve $\CC$ is called a {\rm Jenkins--Strebel} differential if it has   double poles at the marked points   and  all homological (or period) coordinates \eqref{homcoordint} are real:
\be
\Pcal_{s} \in \R,\qquad \hbox{  $ \forall \ s \in H_-$},
\ee
where $H_-$ is   the odd part of the homology of the canonical cover $\Ch$ defined by  \eqref{cancov}.
\end{definition}

The reality of the periods implies that the quadratic residues of $\qd$ at its double  poles $z_j$ are real and negative i.e. there exist  $p_j\in \R_+$ such
that in any local coordinate $\zeta$ near $z_j$ one has:
\be
\qd(\zeta) =\frac {-(p_j/2\pi) ^2}{\zeta^2}\left(1 + \mathcal O(\zeta)\right) ({\rm d} \zeta)^2
\la{locQ}
\ee

The Thm. 23.5 of Strebel's book \cite{Strebel} states that for each Riemann surface $\CC$ of genus $g$ with $n$ marked points $z_1,\dots,z_n$  and given ${\bf p}= (p_1,\dots, p_n)\in \R_+^n$ there exists a unique Jenkins-Strebel differential $Q$ with the expansion  (\ref{locQ}) near each $z_j$ and no other poles.

The flat combinatorial model $\Mcal_{g,n}[\pb]$ of $\Mcal_{g,n}$ is constructed as follows (see \cite{Zvonkine} for references).  
Given a JS differential $\qd$ the corresponding  {\it ribbon graph} $\Gamma$ is the  unoriented graph embedded in $\CC$ defined as follows.

The vertices of $\Gamma$ correspond to the zeroes and simple poles of $\qd$; zeros of multiplicity $k$ are vertices of valence $k+2$ and simple poles are uni-valent vertices.
 The edges of $\Gamma$ correspond to arcs of horizontal trajectories (in the flat metric $|\qd|$) connecting vertices.
 The faces of $\Gamma$  are the connected components of $\CC \setminus \Gamma$, which are  in one-to-one correspondence with the marked points.

Strebel's result  implies that 
there is a  one-to-one correspondence between interior points of  $\Mcal_{g,n}$ (i.e. smooth curves) and  metrized ribbon graphs $\Gamma$ with vertices of valence $\geq 3$,
where the  lengths $\ell_e$ of the  edges $e$ of $\Gamma$  in the metric $|\qd|$ provide local coordinates on $\Mcal_{g,n}[\pb]$.  
The topological types of the graphs $\Gamma$ label cells of $\Mcal_{g,n}[\pb]$. The  strata of $\Mcal_{g,n}[\pb]$ are  labeled by the valences of the vertices.
Namely, the stratum  $\Mcal_{g,n}^{\bk,\bl}[\pb]$ consists of punctured Riemann surfaces  such that the Jenkins-Strebel differential
$Q$ has zeros  of odd multiplicities $2k_j+1$, $j=1,\dots,m_{odd}$, $k_j\geq 1$ and even multiplicities $2l_j$ ($l_j\geq 1$),
$j=1,\dots,m_{even}$ \eqref{divqd1}. The (real) dimension of the stratum   $\Mcal_{g,n}^{\bk,\bl}[\pb]$ is given by
\be
\label{Mklgn}
\dim_\R \Mcal_{g,n}^{\bk,\bl}[\pb]= 2g-2+m
\ee
where $m=m_{even}+m_{odd}$.
The  largest stratum, $W$, corresponds to Jenkins-Strebel differentials with simple zeros (all vertices of $\Gamma$ are then tri-valent)  and has real dimension 
  $6g-6+2n$ (in this case $m=4g-4+2n$)  which coincides with the real dimension of $\Mcal_{g,n}\times \R_+^n$.

The length $\ell_e$ of an edge $e$ connecting two vertices $v_1, v_2$  is  equal to the absolute value of the integral of $v$ along the horizontal trajectory connecting the two vertices. In turn, up to  a factor of $\pm 1/2$,  this length coincides with the period of $v$ over the integer cycles in  $H_-$ consisting of the trajectory $e$ on one sheet of the projection  $\Ch \to \CC$ and the same trajectory in the opposite direction on the other sheet.
Fixing the vector $\pb\in \R_+^n$ one  imposes $n$ linear constraints on the  lengths of the edges: for  the  $j$--th face $f_j ,\ \ j=1,\dots, n$ we have $\sum_{e\in \partial f_j} \ell_e = p_j$. 

\begin{figure}
\begin{center}
\includegraphics[width=0.5\textwidth]{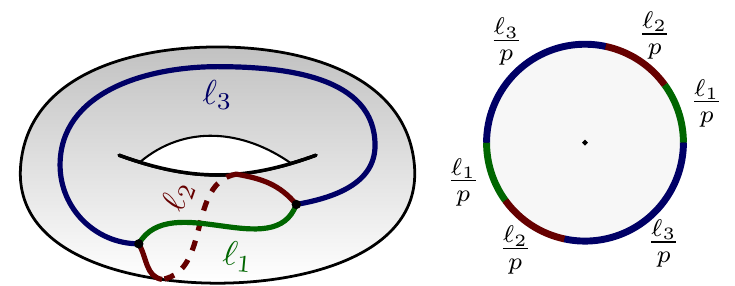}
\end{center}
\caption{Ribbon graph on a genus $1$ Riemann surface representing a point  in $\Mcal_{1,1}[\pb]$. Two simple zeros $x_1$ and $x_2$ of $Q$ are connected by 3 edges of lengths $\ell_1$, $\ell_2$ and $\ell_3$. The ribbon graph has only one face of perimeter $p=2(\ell_1+\ell_2+\ell_3)$. The lengths $\ell_1$ and $\ell_2$ can be used as coordinates on  $\Mcal_{1,1}[p]$.
}\label{M11}
\end{figure}
The faces  of the graph $\Gamma$ are embedded disks in $\CC$. The face $f_j$ containing the pole $z_j$ is uniformized  to the unit disk explicitly by  the map
\be
w_j(x)=\exp\left(\frac{2\pi i}{p_j} \int_{x^o_j}^x v\right)\;;\
\la{faceunit}
\ee
with $z_j$ being mapped to the origin where one chooses the branch of $v$ which has  residue $+\frac{p_j}{2\pi i} $ at $z_j$.
Here $x^o_j$ is one of the zeros of $\qd$ corresponding to a vertex on the boundary of $f_j$, arbitrarily   chosen as basepoint of integration; $w_j( x^o_j )=1$.

Within  the face $f_j$ the flat  metric $ds^2=|\qd|$ on $\CC$    is expressed as:
\be
ds^2= \frac{p_j^2}{4\pi^2} \left|\frac{dw_j}{w_j}\right|^2\;\qquad 0<|w_j|\leq 1.
\ee
For example, the ribbon graph corresponding to the stratum of highest dimension in the combinatorial model $\Mcal_{1,1}[p]$ 
has only one face ( Fig. \ref{M11} ). The constraint between the lengths of the edges  in this case reads $2(\ell_1+\ell_2+\ell_3)=p$.

Inverting this logic  it is possible to construct a polyhedral Riemann surface (i.e. Riemann surface with flat metric with conical singularities)
from a ribbon graph equipped with lengths of all edges by the  procedure of  "conformal welding'' \cite{Strebel}.

 The union of all strata  $\Mcal_{g,n}^{\bk,\bl}[\pb]$ with $k_j\geq 0, l_j\geq 1$  for fixed $\pb\in \R_+^n$ forms the combinatorial model  $\Mcal_{g,n}[\pb]$ of $\Mcal_{g,n}$, and it is  set-theoretically isomorphic to $\Mcal_{g,n}$.

\paragraph{Compactification of $\Mcal_{g,n}[\pb]$  and $\overline \Mcal_{g,n}$.} 

We discuss now a compactification of  $\Mcal_{g,n}[\pb]$  (see \cite{Looijenga, Zvonkine} for discussion and further references)  and its relationship to  the Deligne--Mumford compactification $\overline \Mcal_{g,n}$.  Denote by $\Delta_{DM}= \overline \Mcal_{g,n}\setminus  \Mcal_{g,n}$ the Deligne--Mumford boundary. 
Let  $\CC^{(0)} \in \Delta_{DM}$ be a stable curve  and let $\CC^{(0)} = \bigcup_j \CC^{(0)}_j$ be its decomposition into irreducible components. The 
normalization $\wt {\CC}^{(0)}_j$ of each of these components is an element of  $\Mcal_{g_j, n_j}$  for some $g_j, n_j$,
where $\sum n_j= n+2 \#\{\hbox{ nodes}\}$. In other words, the set of marked points on $\CC^{(0)}_j$ consists of the original $n$ marked points and also marked points arising from resolution of nodes of $\CC^{(0)}$.

Let  now $\CC^{(t)}\in \Mcal_{g,n}$ be a  smooth family of smooth curves for $t\neq 0$. Consider the corresponding family in the combinatorial model $\Mcal_{g,n}[{\bf p}]$; in general \cite{Zvonkine} 
the corresponding Jenkins--Strebel quadratic differentials   $\qd^{(t)}$ in the limit $|t|\to 0$ acquires {\it at most} a simple pole at the nodes of $\CC^{(0)}$.  

Therefore each component $\CC^{(0)}_j$ gets equipped with the Jenkins--Strebel quadratic differentials   $\qd^{(0)}_j$ with (generically) simple poles at resolutions of nodal points. 
This differential is identically zero on components where all marked points 
are the result of the resolution of nodes, see \cite{Zvonkine}.
On the  remaining components the differential $\qd^{(0)}_j$ defines the canonical covering $\Ch^{(0)}_j$ of $\CC^{(0)}_j$ by the  equation $v_j^2=\qd^{(0)}_j$.
The resolutions of nodal points of $\CC^{(0)}$   are (generically) branch points of  $\Ch^{(0)}_j$.

Following \cite{Zvonkine} we extend Def. \ref{defJS} to nodal stable curves. 
\begin{definition}
\label{deJSnode}
Let $\CC^{(0)} \in \Delta_{DM}$ be a stable nodal $n$--marked curve. A quadratic differential $\qd$ on  $\CC^{(0)} $ is called a {\rm Jenkins--Strebel} differential if it has   double poles at the marked points, at most simple poles at the nodes in the normalization of   $\CC^{(0)}$  and in the normalization of each component  all homological coordinates \eqref{homcoordint} are real. 
\end{definition}

The reality of homological coordinates immediately implies the vanishing of  $\qd$ on components of $\CC^{(0)} $ where no second order poles 
of $\qd$ left: on those components the Abelian differential $\sqrt{\qd}$ is holomorphic with real periods and therefore identically vanishing.

On those components $\CC_j^{(0)}$ where $Q_j^{(0)}$ has second order poles  all nodal
points are one-valent vertices of the corresponding ribbon graph (the valency 1 arises in generic case; if the length of the edge arriving to the one-valent vertex 
equals 0 then this vertex merges with another vertex of higher valency; this corresponds to node placed at zero of $Q_j^{(0)}$).
Therefore, we assume $k_j\geq -1$ in the definition of $\Mcal_{g,n}^{\bk, \bl}[\mathbf p]$.

The union of all strata $\Mcal_{g,n}^{\bk_j,\bl_j}[\pb_j]$ gives the  compactification  $ \Mcomb$ of the combinatorial model. The  corresponding Jenkins--Strebel differentials are now allowed to have simple poles and hence ribbon graphs may have univalent  vertices \cite{Zvonkine, Looijenga, Mondello2}.

Notice that the boundary $ \Mcomb\setminus \Mcal_{g,n}[{\bf p}]$ has real co-dimension $2$ because the degeneration of a single edge in the main stratum (where all vertices are tri-valent) cannot lead to a degenerate surface. 

%We  use henceforth the notation  $\Wbdr$ for this boundary (also known as the {\it Kontsevich boundary})  because in its main stratum the ribbon graphs have   exactly two uni-valent vertices  (one for each connected component if the graph is disconnected) and all other vertices are tri-valent.
%According to Prop. \ref{proporientation}, $\Wbdr$  is a cycle.

%The full  $\Wbdr$ is the complex of cells labelled by  possibly disconnected graphs where each component has at least  one vertex of arbitrary valence $\geq 1$ in correspondence with each of the nodes,  while all other vertices have valence $\geq 3$. 

Rephrasing  Strebel's result (Thm. 23.5 \cite{Strebel}) for each  $n$--pointed possibly singular, stable curve $\Ccal\in \overline {\Mcal}_{g,n}$  and $n$--tuple of positive numbers ${\bf p}=(p_1,\dots p_n)$ there is a unique JS quadratic differential  (according to Def. \ref{deJSnode}) $\qd$ on $\CC$ with the quadratic residues at the $n$--marked points given by $-p_i^2/(4\pi^2)$. In turns, $\qd$ defines a ribbon graph for each component of the normalization, thus giving an element of $\Wbdr$.  
This defines the  {\it Jenkins-Strebel map}
\be
\label{JS}
\JS: \overline \Mcal_{g,n} \to \Mcomb.
\ee

As we discussed above, the map $\JS$ is one-to-one on $\Mcal_{g,n} $. On the other hand, the part of Deligne-Mumford boundary $\Delta_{DM}$ which contains a stable component without marked points is mapped to a lower-dimensional component of $\overline \Mcal_{g,n}[\pb]\setminus \Mcal_{g,n}[\pb]$ since any component without marked points is blown--down to a point by $\JS$.

We are going to use the notation 
\be
W_{1,1}=\overline \Mcal_{g,n}[\pb]\setminus \Mcal_{g,n}[\pb] = \JS[\Delta_{DM}]
\la{defW11}
\ee
for the boundary of $\Mcal_{g,n}[\pb]$; the map $\JS:\Delta_{DM}\to W_{1,1}$ is surjective but not injective.

The set $W_{1,1}$ (which is also called the  "Kontsevich's boundary"  of $\Mcal_{g,n}[\pb]$) is a subcomplex of $\Mcal_{g,n}[\pb]$, and, moreover, is known to be a cycle according to Prop.\ref{proporientation} (see also \cite{Mondello2}).
The notation $W_{1,1}$ is motivated by valencies of non-standard  (i.e. non-trivalent) vertices in its largest cells: these graphs have exactly 2 one-valent vertices.

\paragraph{Co-dimension $2$ subcomplexes.}
In total there are only two subcomplexes of $\Mcomb$  in co-dimension $2$: $\Wfive$ and $\Wbdr$.
The largest cells of these two cycles  are obtained by the contraction of  two edges having at least one vertex in common: 
if the two edges have only {\it one} vertex in common, their contraction leads to the largest cells of  $\Wfive$ (ribbon graphs with  one  vertex of valence $\geq 5$  and all  other vertices of valence $\geq 3$). 

\begin{figure}
\begin{center}
\includegraphics[width=0.8\textwidth]{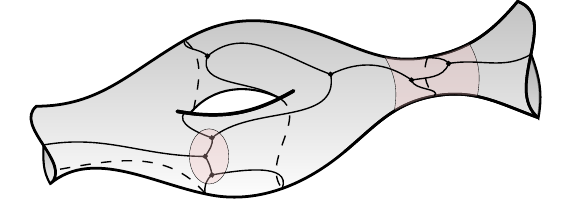}
\end{center}
\caption {A typical embedded Strebel graph. The two highlighted  regions are where the contraction of edges leads either to the
Witten's cycle $\Wfive  $ (left region) or to the Kontsevich's boundary $\Wbdr  $ (right region). }
\label{figw5deg}
\end{figure}

On the other hand, if  the two edges connect the  same pair of vertices then they necessarily form a loop that is homotopically non-trivial on the Riemann surface (if these two edges had formed a homotopically trivial loop, then they would  form the boundary of a face of the graph, but since the perimeters are kept fixed they cannot tend to zero simultaneously). 
Therefore, the simultaneous degeneration of two such edges  pinches the Riemann surface and  gives a point (of a cell of highest dimension of) the cycle $\Wbdr$.  
These two types of degenerations  are shown in Fig.\ref{figw5deg}.

\begin{figure}
\centering
\includegraphics[width=90mm]{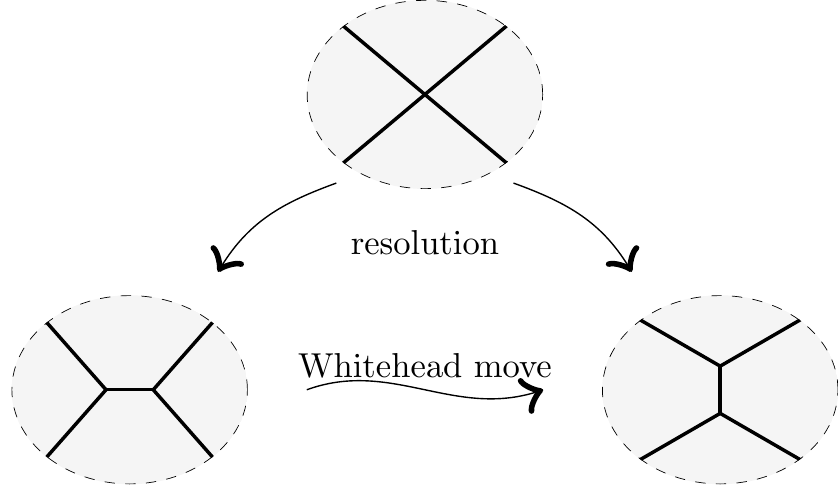}
\caption{Resolution of a 4-valent vertex and Whitehead move.}
\la{resolve4}
\end{figure}
\subsection{Orientation of $\mathcal M_{g,n}[{\bf p}]$ in homological coordinates and Witten--Kontsevich cycles}
\la{Orient}
\paragraph{Symplectic structure.}
Consider a point $(\CC, \qd)$ in the stratum $\mathcal M_{g,n}^{\bk,\bl}[{\mathbf p}]$ with $k_j\geq -1$.
% (at most simple poles besides the double poles at the punctures
%$z_1,\dots,z_n$ , see \eqref{divqd1}).
 If $\bl=0$ i.e. all multiplicities are {\it odd} (in which case we simply omit the superscript $\bl$ and denote the stratum by $\mathcal M_{g,n}^{\bk}[{\mathbf p}]$)  then the canonical cover $\Ch$ is branched at all the zeroes and at the simple poles.  The odd homology $H_-=H_-\left(\Ch\setminus \{\wh z_j, \wh z_j^\mu\}_{i=1}^n\right)$  is generated by $2g_-+n$  cycles and the rank of intersection form equals $2g_-$.

\begin{definition}[Homological Poisson Structure]
 Given two arbitrary cycles $\gamma, \wt \gamma \in H_-\left(\Ch\setminus \{\wh z_j, \wh z_j^\mu\}_{i=1}^n\right)$ the Poisson bracket  on $\mathcal M_{g,n}^{\bk}[{\mathbf p}]$ between  the corresponding homological coordinates  is defined by the formula
 \be
\left \{\oint_\gamma v , \oint_{\wt \gamma} v\right \}=  \gamma \circ \wt \gamma\ ,\hskip0.7cm  \forall \gamma, \wt \gamma\in H_-.
\label{Phom}
\ee
where $ \gamma \circ \wt \gamma$  denotes the intersection pairing. 
\end{definition}

The complex homological Poisson structure (\ref{Phom}) in the complex case  $n=0$ was introduced in  \cite{BKN}.

Let us now describe a suitable basis in $H_-$ (recall that ${\rm dim} \,H_-=2g_-+n$). For each point $z_j$, $j=1,\dots, n$ let $c_j$ be a small loop around  it on the base curve $\CC$ and denote $\g_j = \frac 1 2( \wh c_j - \wh c_j^\mu)\in H_-$ so that the associated homological coordinates are  $\pm  p_j$. 
Choose also   $2g_-$ cycles $\{a_\ell^-, b_\ell^-\}_{\ell=1}^{g_-}$ with intersection index 
 $a_\ell^-\circ b_\ell^-=1$ and all other intersection indices vanishing (the choice of these cycles is ambiguous: they are defined up to a linear combination of cycles $\g_j$).
Let $\{A_\ell=\int_{a_\ell^-}v, \;B_\ell=\int_{b_\ell^-}v\}_{\ell=1}^{g_-}$ be  the corresponding real homological  coordinates.
The intersection form in $H_-$ thus defines a {\it Poisson} structure if the perimeters $\{p_j\}_{j=1}^n$ are constants.

\begin{proposition}
On  each  cell of $\mathcal M_{g,n}^{\bk}[{\mathbf p}]$
the Poisson structure \eqref{Phom} is non-degenerate and admits an inverse given by  the symplectic form
 \be
\Omega_{\text {{ \tiny Hom}}} := \sum_{j=1}^{g_-}  {\mathrm d} A_j \wedge {\mathrm d} B_j.
\label{omegahom}
\ee
\label{sympl}

\end{proposition}

\noindent {\bf Proof.}
First, it is obvious that the form \eqref{omegahom} induces the Poisson structure \eqref{Phom}. 
The expression \eqref{omegahom}   is invariant under the choice of  cycles $\{a_\ell, b_\ell\}_{\ell=1}^{g_-}$ in $H_-$ having the intersection indices
$a_{\ell}\circ b_k=\delta_{\ell k}$, $a_{\ell}\circ a_k=b_{\ell}\circ b_k=0$ since any two such sets of cycles are related by a symplectic transformation up to an addition of a linear combination of 
the cycles $\gamma_j$. Moreover the periods over cycles $\gamma_j$ (which are equal to perimeters $p_j$'s) are constant on the space $\mathcal M_{g,n}^{\bk}[{\mathbf p}]$.
 \QED

The space  $\mathcal M_{g,n}^{\bk}[{\mathbf p}]$  is the union of several cells of (real) dimension $2g_-$. Each of these cells is a  polytope in $\R^{2g_-}$ with the symplectic form \eqref{omegahom} and   with  the induced volume form and orientation. 
The common boundaries of these cells are strata of real co-dimension $1$ (in fact, pieces of hyperplanes). 
\begin{remark}\rm
The form (\ref{omegahom}) can also be defined on each cell of a stratum  $\mathcal M_{g,n}^{\bk,\bl}[{\mathbf p}]$ with some $l_j\neq 0$ (i.e. with some zeros having even multiplicity). In that case the covering $\Ch$ 
has node at the corresponding zero $x_j$ of $\qd$ of even multiplicity. However, the dimension of such stratum is greater than the 
number of periods $(A_j,B_j)$ since the full set of homological coordinates in this case contains also integrals between zeros of $v$ related by the involution $\mu$ (these integrals are Casimirs of the homological Poisson structure on such strata.
Thus the form (\ref{omegahom})  can not be used to induce an orientation on  $\mathcal M_{g,n}^{\bk,\bl}[{\mathbf p}]$ for $\bl\neq 0$.
\end{remark}

Consider now the sub-complexes $W_{\bf r }$  of $\Mcomb$ whose cells of maximal dimension 
coincide with $\mathcal M_{g,n}^{\bk,\bl}[{\mathbf p}]$ (with   $r_j=2k_j+1$ for $k_j\neq 1$ or $r_j=2l_j$).
By convention, the trivalent vertices are not listed in the valence vector ${\bf r}$.

\begin{proposition}
\label{proporientation}
The sub-complexes  $W_{\bf r }$  with all odd $r_j$'s are orientable   cycles with the orientation induced  by the $g_-$th power of symplectic form \eqref{omegahom}.
\end{proposition}

Before proceeding with the proof, we need to define  how to ``propagate'' a symplectic basis in $H_-$ across a facet that corresponds to a Whitehead move (Fig. \ref{resolve4}). This will allow us to define the orientation of the whole cycle in a consistent way.
\begin{definition}[Propagation of symplectic bases under a Whitehead move]
\la{propbases}
Let $e$ be an edge that connects two trivalent vertices. 
Then the  contraction of $e$  leads to the formation of a vertex of valence $4$. 
Consider two graphs  corresponding to the two ways of resolving the vertex into two vertices of valence $3$, as depicted in Fig. \ref{Wh_prop2} and  denote the two corresponding canonical covers by $\Ch_1, \Ch_{2}$ respectively; the nodal covering corresponding to the boundary we denote by $\Ch_0$.  In $H_-(\Ch_1)$ the basis is chosen so that  $a_1^-, b_1^-$ are as in  Fig. \ref{Wh_prop2} (left) and the remaining cycles are chosen arbitrarily so as to complete to a full canonical set.

We then  choose the  canonical cycles in $H_-( \Ch_2)$ 
by continuous deformation  of all cycles  in $H_-( \Ch_1) $ except for $a_1^-, b_1^-$, which are transformed as shown in Fig. \ref{Wh_prop2}, right.
\end{definition}

\begin{figure}
\centering
\includegraphics[width=90mm]{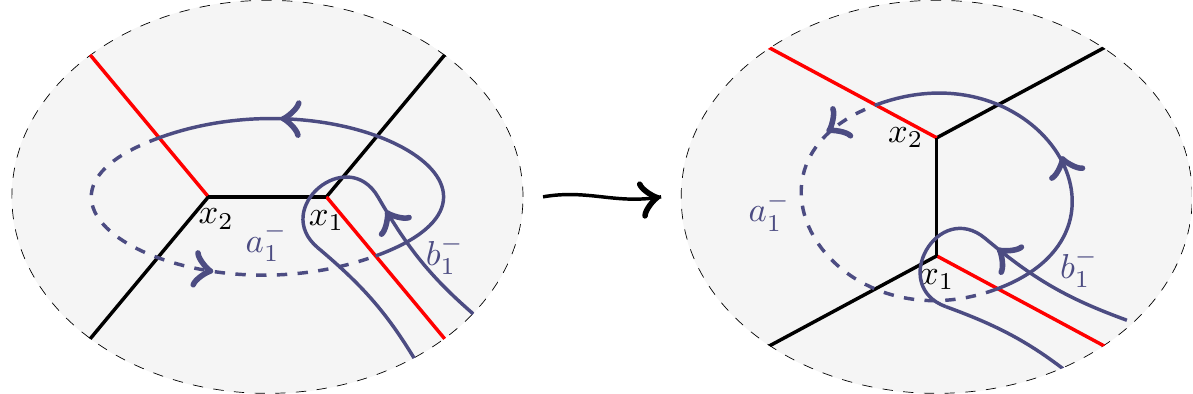}
\caption{
Canonical bases of cycles in $H_-(\Ch_1)$ and $H_-(\Ch_2)$ in two cells of $W$ which agree on their common boundary points. This gives a natural way to propagate the symplectic basis in $H_-$ into the neighbouring cell.
}
\la{Wh_prop2}
\end{figure}

{\bf Proof of Prop.  \ref{proporientation}.}
Using the definition \ref{propbases}  describing the propagation of the canonical basis on $\Ch$ under the Whitehead move we are going to  verify that the orientation induced on each cell by the homological symplectic structure \eqref{omegahom} defines a consistent orientation on    $W_{\bf r }$.   This consistency is a corollary of two facts:
\begin{enumerate}
\item each boundary of co-dimension $1$  is shared by an even number of cells (generically $2$).
\item The  orientation  on  each component of the  co-dimension $1$ boundary   induced from a half of the cells is the opposite to the orientation induced from the other half.  
\end{enumerate}

The co-dimension $1$ part of the  boundary of a maximal cell of $W_{\bf r }$ corresponds to one contraction of any  of the edges of the corresponding ribbon graph.  We start from analysis of the simplest case consisting of  the degeneration of an edge between two tri-valent 
vertices and then analyze the general case.

{\it Coalescence of two tri-valent vertices.} Consider first the case where the vanishing edge connects two vertices of valence $3$ each. 
We choose canonical symplectic bases in $H_-( \Ch_1)$ and  $H_-( \Ch_2) $ as explained in Def. \ref{propbases}.
Then the two cells can be mapped  locally to  neighbourhoods in the half--spaces in $(A_1,B_1, \dots, A_{g_-} B_{g-})\in \R^{2g_-}$ separated by the hyperplane $A_1=0$ (on this hyperplane  the cycle $a_1^-$ is a  vanishing cycle on $\Ch_{0}$).
The orientation extends through the boundary if $A_1$ has {\it opposite signs} in the two cells.

According to  Fig. \ref{Wh_prop2} the  orientation of $a_1^-$ before and after the Whitehead move is uniquely determined by the orientation of $b_1^-$. The sign of $A_1$ can be deduced from the following local considerations.
In a distinguished parameter $\zeta$  around the zero (denoted by  $ x_1$ on both $C_1$ and $C_{2}$)  
the differential $v$ reads $v = \frac 32 \sqrt{\zeta} {\rm d} \zeta$ and  the local flat coordinate near $x_1$ on $\Ch_{1,2}$ is  $ z =  \zeta^\frac 32$; we choose the branch-cuts (shown in red) on both $\CC_1$ and $\CC_{2}$ along the positive $\zeta$ axis.

The  period $A_1$  equals to $2 z(x_2)$ in both cases. 
 To determine the sign of $A_1$ in each case we notice that the flat coordinate $z$ maps the upper rim of the cut to the positive $z$--axis and the lower rim to the negative $z$--axis. The remaining two horizontal trajectories (black) moving counterclockwise from the red trajectory,  are mapped to the negative and positive $z$--axis, respectively.
 
Therefore, $A_1$ is
positive  in the case of $\Ch_1$ (left pane  of Fig. \ref{Wh_prop2}) ($x_2$ in that case lies on the trajectory corresponding to positive values of $z$). For analogous reasons, $A_1$ is negative in the case of $\Ch_{2}$ (right pane).
Therefore, indeed, the two sides of the Whitehead move are locally described by $A_1>0$ or $A_1<0$ and we conclude that the orientation propagates consistently from cell to cell.

\begin{figure}
\centering
\includegraphics[width=90mm]{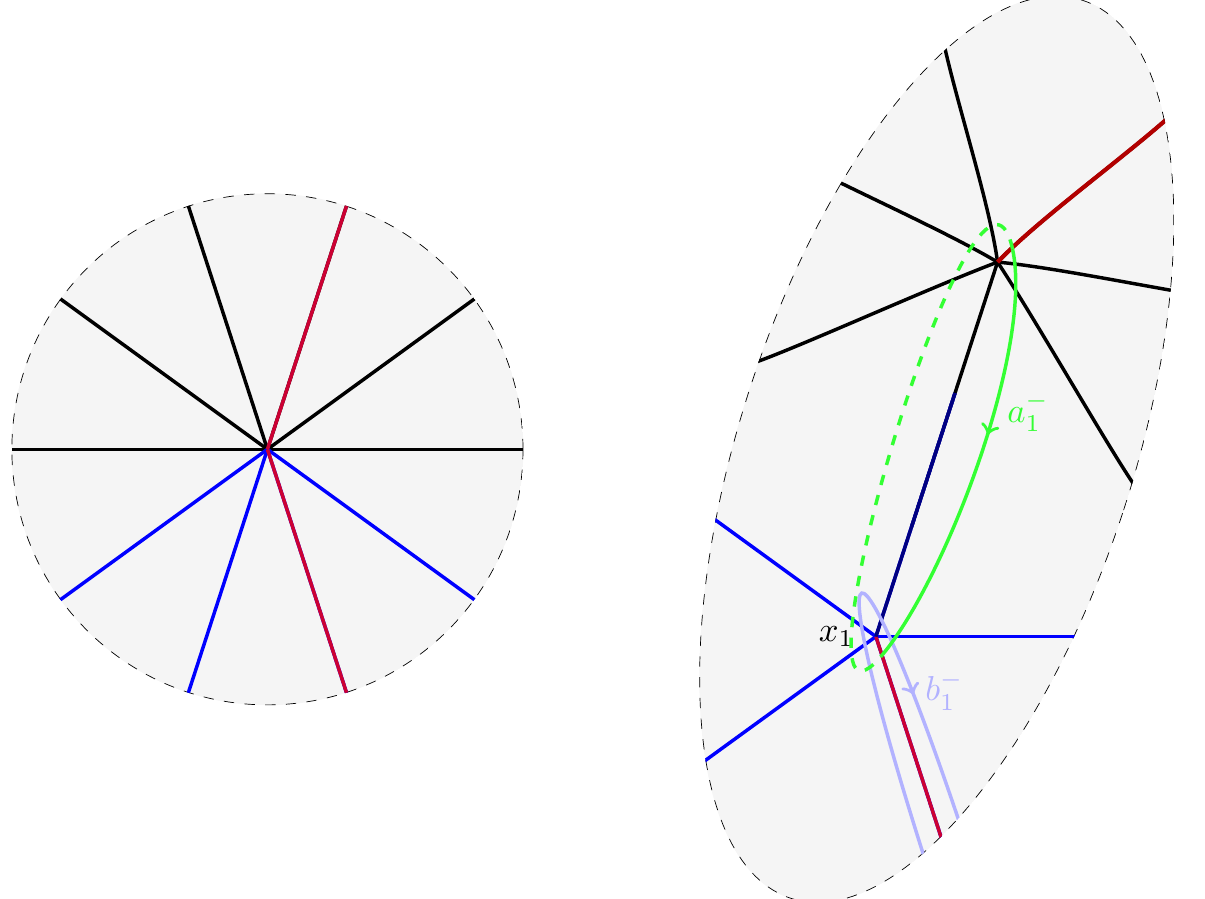}
\caption{Resolution of a vertex of valence $2(2+3)=10$ into two vertices of valence $2\cdot 2+1=5$ and $2\cdot 3+1=7$. Here $A_1B_1<0$. The branch cuts are shown in red (the  number of branch cuts issuing from each vertex is even or odd according to the parity of the vertex's valence).}
\la{Orientation2}
\end{figure}

{\it General case.}
Suppose now that the contracting edge connects two vertices of valences $2k+1$, $2r+1$,  i.e. two zeroes of $\qd$ of odd multiplicities $2k-1, 2r-1$, respectively. Then the merge produces a  zero of even multiplicity $2k+2r-2$ i.e. a  vertex of valence $2(k+r)$. This vertex can be  resolved  into two vertices of valences $2k-1, 2r-1$ in  $2(k+r)$ ways.  
As in the tri-valent case, we choose the symplectic bases in $H^-$ so that the vanishing cycle $a_1^-$ goes around the contracting edge and the dual cycle goes around one of the adjacent edges (see Fig. \ref{Orientation2}). 

In each resolution, the $b_1^-$ cycle goes around one of the $2(k+r)$ edges that are not contracted; as before we need to verify that the homological coordinate $A_1$ is positive in half of the cases and negative in the other half. 

We use again  the flat coordinate centered at  the vertex $x_1$  incident to the $b_1^-$ edge (which has valence either $2k+1$ or $2r+1$). An elementary local consideration along the same lines as in the previous case (paying attention to the arguments of the distinguished and flat coordinates) shows that
the sign of the product  $A_1 B_1$  equals $(-1)^\ell$ where $\ell$ is the number of edges between the $b_1^-$ edge and the contracting edge, counterclockwise.

Since the sign of $B_1$ is the same in every resolution, there are half cases with positive and half cases with the negative sign of $A_1$ and the theorem is proved.
 \QED
 
The above proposition justifies the following standard definition
\begin{definition}[Witten--Kontsevich cycles]
\label{defWK} 
The sub-complexes $W_{\bf r }$ with all $r_j$'s odd integers  are called   Witten--Kontsevich cycles.
\end{definition}
If some of the  $r_j$ is an even integer, the corresponding subcomplex  is not orientable and therefore it is not a cycle \cite{Mondello1}.

\subsubsection{Period coordinates as canonical coordinates for Kontsevich's form $\sum p_f^2\psi_f$}
 In \cite{Kontsevich} Kontsevich introduced the  two form    
$\Omega   = \sum_{f\in F(\Gamma)}p_f^2 \omega _f$   on  each cell of  $\mathcal M_{g,n}[{\bf p}]$ where 
$F(\Gamma)$ is the set of faces of $\Gamma$; 
$\omega_f$ 
 is the two--form representing the Chern class $\psi_f$ of the tautological line bundle at the marked point corresponding to the face $f$ of the ribbon graph  \cite{Kontsevich, Zvonkine}.

In terms of lengths of the edges the form $\Omega$ can be written as follows:
 \be
 \Omega=\sum_{f\in F(\Gamma)}  \eta_f
\la{Omega}
 \ee
 where 
 \be
 \eta_f=p^2_f \omega_f  = \sum_{e_j,e_k\in \p f\atop 1\leq j<k<n_f} {\rm d}  {\ell_j}\wedge {\rm d}  {\ell_k},
 \la{defetaf}
 \ee
 $e_1,\dots e_{n_f}$ are the edges bounding face $f$ ordered counterclockwise (the form is independent of the choice of the "first" edge on each face),
 and $\ell_j$ denotes the length of the edge $e_j$.
 
 Consider now the following Poisson bivector on each  cell of $\mathcal M_{g,n}[{\bf p}]$:
 \be
\Poss = \frac 1 4 \sum_{x\in V(\Gamma)} \sum_{e_j, e_k  \perp x \atop 1\leq j<k\leq n_x } (-1)^{k-j-1}\frac \p {\p \ell_j} \wedge \frac \p{\p \ell_{k}}
\label{bivect}
\ee
where $V(\Gamma)$ is the set of vertices of $\Gamma$ and $e_1,\dots,e_{n_x}$ are edges  incident to $x$ taken in counterclockwise order
(the bivector $\Poss $ also does not depend on the choice of the "initial" edge).

 In \cite{Kontsevich} it was shown that the two-form $\Omega$ is symplectic if all vertices of $\Gamma$ have odd valency.
 \begin{theorem}\cite{Kontsevich}
 The form $\Omega$ (\ref{Omega}) is symplectic on each top dimensional cell of any Witten-Kontsevich cycle 
 $W_{{\bf r} }$.
 \end{theorem}

 The proof of the next proposition (which in fact coincides with Lemma C.2 of \cite{Kontsevich} that was stated without proof there)  is essentially contained in the proof of Lemma 5.4 of \cite{Mondello2}. It shows that the bivector  $\Poss$ (\ref{bivect}) is the right  inverse to the 
 symplectic form $\Omega$ (\ref{Omega}) on each cell of top dimension of any Witten--Kontsevich cycle $W_{\bf r}$ (then it is also the left inverse since 
 $\Omega$ is non-degenerate  on those cells \cite{Kontsevich,Mondello2}).
 \begin{lemma}\la{MK}
 Let all vertices of $\Gamma$ have odd valence. Then for each edge $e$ of length $\ell$ the following relation holds:
 \be
 \Omega[\Poss(d\ell)]=d\ell
 \la{OP}
 \ee
  \end{lemma}
 {\it Proof.} 
 Let the edge $e$ connect vertices $x$ and $x'$ of valences $2k+1$ and $2k'+1$. Denote the remaining $2k$ edges emanating from $x$ by $e_1,\dots,e_{2k}$
 (enumerated counterclockwise starting from $\ell$). Similarly, the remaining $2k'$ edges emanating from $x'$ are denoted by 
 $e_{1'}\dots,e_{2k'}$ (enumerated counterclockwise).
 Then
 \be
 \la{Pdl}
 \Poss(d\ell)= \f{1}{4}\left(\sum_{j=1}^{2k} (-1)^{j-1} \partial_{\ell_j}+ \sum_{j=1}^{2k'} (-1)^{j'-1} \partial_{\ell'_j}\right)
 \ee
 (the formula (\ref{Pdl}) implies that the bivector $\Pcal$, up to factor $1/4$,  coincides with the bivector $\beta$ from Sect.C of \cite{Kontsevich}).
 Let us label the faces involved as shown in Fig.\ref{Faces}.
 \begin{figure}
\centering
\includegraphics[width=90mm]{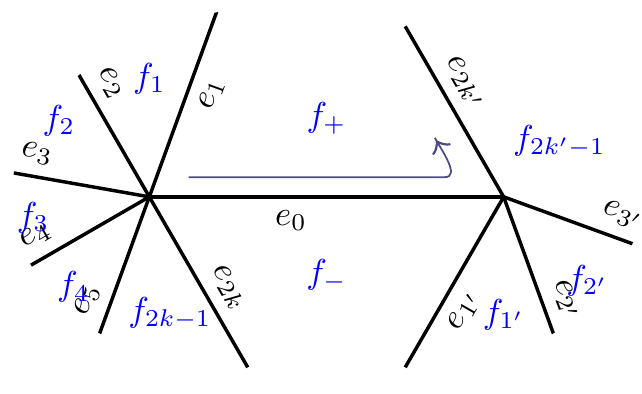}
\caption{ }
\la{Faces}
\end{figure}

 It is easy to verify that for any two consecutive edges $e$ and $ e'$ (such that $e$ precedes $ e'$ going counterclockwise) adjacent to the face $f$ we have
 \be
 \eta_f \left[\frac{\partial}{\partial \ell}-\frac {\partial} {\partial \ell'}\right]= d\ell+d \ell'\;.
 \la{etafdl}
 \ee
 We apply $\Omega$ in \eqref{Omega} to  \eqref{Pdl}; using \eqref{etafdl}, the separate  summands $\eta_f$ in $\Omega$ yield
 \begin{eqnarray}
 \eta_{f_+}(\Poss(d \ell) )= \frac 1 4 \left(d \ell_1 + 2 d\ell + d \ell_{2k'} \right)\ , \qquad
 \eta_{f_-} (\Poss(d \ell))= \frac 1 4 \left( d \ell_{2k} + 2 d\ell + d \ell_{1'} \right)\ ,
 \la{addemup1}
  \\[10pt]
 \eta_{f_j} (\Poss(d \ell) ) =\frac {(-1)^{j}}4 (d \ell_{j+1} + d \ell_j)\ ,\qquad 
 \eta_{f_{j'}} (\Poss(d \ell) ) = \frac{ (-1)^{j'} }4(d \ell_{j'+1} + d \ell_{j'})\;.
 \label{addemup}
  \end{eqnarray}
  Summation of  all terms in \eqref{addemup1} and \eqref{addemup} gives  $d \ell$. \QED

The next theorem shows that the form $\Omega$ coincides with the homological symplectic form $\Omega_{hom}$ (\ref{omegahom}) when all vertices of $\Gamma$ have odd valences.

\begin{theorem}
\label{thmomega}
The Kontsevich's two  form $\Omega$   \eqref{Omega} coincides with the  homological symplectic form $\Omega_{\text{\tiny Hom}}$ \eqref{omegahom}  on each  cell of maximal dimension of  the Witten--Kontsevich cycle $W_{{\bf r}}$ i.e.
\be
\sum_{f\in F(\Gamma)} p_f^2 \omega_f =\sum_{j=1}^{g_-} d A_j\wedge d B_j
\ee
\end{theorem}
{\bf Proof.}
Denote by $\Gamma$ the ribbon graph of a cell of maximal dimension in $W_{{\bf r}}$.
We start from showing that the  bi-vector representing the  Poisson bracket \eqref{Phom} is given by (\ref{bivect}) i.e.
 that 
 \be
 \{ \ell_e, \ell_{e'}\} =\Poss(d \ell_e, d \ell_{e'})
 \la{Pospos}
 \ee
  for any two edges $e,e'$ (where the Poisson bracket is defined in \eqref{Phom}).
 
We remind that for an edge $e$ the length is $\ell_e=\frac 12 \oint_{\gamma_e} v>0$  where $\gamma_e$ is the cycle in $H_-$  consisting of the edge $e$ on one sheet of $\Ch$  and the same edge on the other sheet in the opposite direction (with the overall orientation so that  $\oint_{\gamma_e} v$  is positive).
 If two edges $e,e'$ have no common vertex, then clearly the intersection number $\gamma_e\circ \gamma_{e'}=0$ and also $\Poss({\rm d}\ell_e, {\rm d}\ell_{e'})=0$. 

If $e,e'$ have one vertex $x_0$ of valence $2k+1$ in common then  $\Poss({\rm d}\ell_e, {\rm d}\ell_{e'}) = \frac 1 4 {c(e,e')}$ where $c(e,e')=\pm 1$ is the parity of the number of edges incident to $x_0$
between $e$ and $e'$ in counterclockwise direction (note that $c(e,e')=-c(e',e)$ since the valence of $x_0$ is odd).
To show that (\ref{Pospos}) holds in this case we need to show that $\gamma_e\circ\gamma_{e'}= c(e,e')$ (since $\{\int_{\gamma_e}v,\int_{\gamma_{e'}}v\} = \gamma_e\circ\gamma_{e'}$ for any two cycles). Clearly $\gamma_e\circ \gamma_{e'} = \pm 1$. To decide on the 
sign let $\zeta$ be the distinguished coordinate at $x$ so that $v = \frac {2}{2k+1} \zeta^{\frac {2k-1}2}{\rm d} \zeta$. The flat coordinate at $x$ is $z= \zeta^{\frac {2k+1}2}$ and the edges are arcs of the rays $\arg(\zeta)= j\frac {2\pi}{2k+1},\ j=0,\dots, 2k$. Placing the branch-cut along the $j=0$ - ray  an elementary analysis shows   that the orientations of the pair 
$(\gamma_e,\gamma_e')$ that makes both integrals $\int_{\gamma_e}v$  and $\int_{\gamma_{e'}}v$ positive yields an intersection number equal to $c(e,e')$; the remaining factor of $4$ comes from the fact that $\ell _e = \frac 1 2 \oint_{\gamma_e} v$. 
This proves (\ref{Pospos}).

To prove the statement of the theorem it remains to use  the relation between $\Poss$ and $\Omega$ given by Lemma \ref{MK} since, as a corollary of (\ref{Pospos}),  $\Poss=\sum_{j=1}^{g_-}\frac \p{\p A_j} \wedge \frac \p{\p B_j}$
in terms of homological coordinates. \QED
 \subsection{
 Pentagon moves and combinatorial Dehn twists}
 \la{sectionW5W11}
 
  The largest stratum of the cycle $\Wfive $ corresponds to JS differentials with
 one zero of multiplicity 3 and all remaining zeros of multiplicity 1. Equivalently, the corresponding ribbon graphs  have 
 one vertex of valence 5 while all other vertices are tri-valent.  
 
 The  largest stratum of the cycle $\Wbdr$ corresponds to differentials with two simple poles and simple zeros on nodal curves. The corresponding ribbon graphs are either connected graphs with two uni-valent vertices (nodal irreducible curve) or  union of two connected graphs each with one uni-valent vertex (two irreducible components).  Therefore the cycle $\Wbdr$ is the sum of a  cycle formed by irreducible components $\Wbdr^{irr}$  and several other cycles which correspond to degenerations where (in the largest stratum)  the genus of one of the component is $g-j$ and the genus of the other is $j$ ($0\leq j \leq \lfloor \frac g 2\rfloor $), while the $n$ marked points are distributed between the components in all possible ways (with at least one marked point in each component).

\subsubsection{ Cycle $\Wfive$ and pentagon moves }
\la{W5_pentamove}

 For each point of Witten's cycle $\Wfive $ the triple zero of $Q$ (i.e. the 5-valent vertex of the ribbon graph) can be split into 3 simple zeros (i.e.
three-valent vertices) in 5 different ways shown in Figure \ref{Wh_Penta}.
Therefore the {\it star} of   the given cell $K$ of $\Wfive$ consists of  five cells $K^{(0)},\dots, K^{(4)}$, cyclically ordered so that $K^{(j)}$ shares facets with $K^{(j+1)}$ and $K^{(j-1)}$ (indices taken modulo $5$). 

A  loop in a transversal cross--section around  a point of the cell $K$   consists of the union of five 
paths (one path in each $K^{(j)}$) that extend between different facets and form a continuous curve in $\tilde{W}$ (recall that $\tilde{W}$ denotes the  union of cells of $W$ and their facets). The ordering of the cells $K^{(j)}$'s determines an orientation of the loop. 

\begin{figure}[t]
\centering
\includegraphics[width=0.7\textwidth]{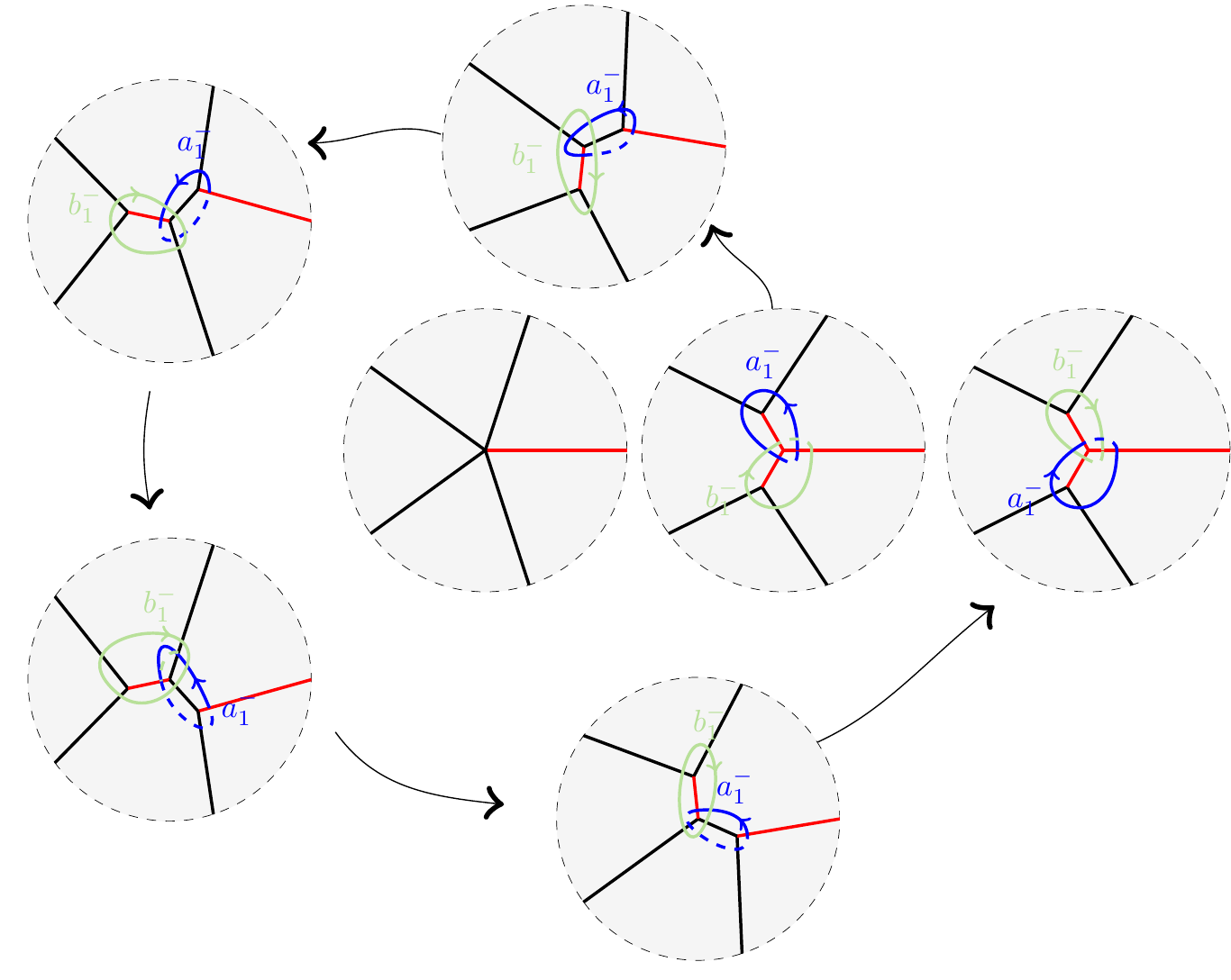}
\caption{Resolution of a five-valent vertex and transformation of the pair of canonical cycles $a_1^-$ and $b_1^-$ on $\Ch$ under pentagon move}
\la{Wh_Penta}
\end{figure}

The {\bf positive} orientation is, in turn, determined by the symplectic form \eqref{omegahom} as we now explain, referring to Fig. \ref{Wh_Penta}.
In each $K^{(j)}$ the homological coordinates are uniquely defined up to symplectic transformations. Let $A^{(j)}, B^{(j)} $ be the homological coordinates associated to the cycles $a_1^{(j)}, b_1^{(j)}$ in $H_-$  and chosen so that they are both positive. 
Then the orientation in the cone  $(A^{(j)},B^{(j)})\in \R_+^2$ is given by  ${\rm d} A^{(j)}\wedge {\rm d} B^{(j)}$;  the facets of $K^{(j)}$ correspond to the coordinate axes and the facet $A^{(j)}=0$ (``vertical axis'') is the facet that follows the facet $B^{(j)}=0$ (``horizontal axis'') in the  positive orientation. This allows to uniquely determine the order of the $K^{(j)}$'s by propagating the homology basis as explained in Def. \ref{propbases}.
Therefore, we get the following 
\begin{proposition}
\label{proppent}
The pentagon move, i.e. the sequence of five Whitehead moved depicted in Fig. \ref{Wh_Penta} represents  a path 
in $\tilde{W}$ which goes around Witten's cycle $\Wfive $ in the positive direction. 

\end{proposition}

 \subsubsection{Kontsevich's boundary $\Wbdr $ and combinatorial Dehn's twist}
 \label{combDehnsect}
 The Riemann surface $\CC$ corresponding to a point of the largest stratum of  $\Wbdr$  possesses a single nodal point $q_0$. The JS differential on such a curve has simple poles at the 
 two points of  the resolution of $q_0$ in the normalization of $\CC$ and $4g-6 + 2n$ simple zeros. 
 
 If the degeneration of the two edges as shown in Fig. \ref{figw5deg} (the region highlighted on the right) does not separate the Riemann surface (i.e. the shrinking loop is homologically non-trivial) then the corresponding ribbon graph is connected and has $4g-6+2n$ tri-valent and $2$ one-valent vertices. 
 Such ribbon graphs correspond to the irreducible component $\Wbdr^{irr}$ of $\Wbdr$.

 If the degeneration is separating,  i.e., the loop is homologically trivial,  and 
 both components contain at least one double  pole of $Q$, then  the corresponding ribbon graph %$\Gamma_0$ 
 consists of two connected components
 $\Gamma_1$ and $\Gamma_2$ of genera $g_1$ and $g_2$ (with $g_1,g_2\geq 0$ and $g=g_1+g_2$). The numbers of faces of the ribbon graphs $\Gamma_{1,2}$ equal $n_{1,2}$ ($n_{1,2}\geq 1$). Such ribbon graphs correspond to the reducible component $\Wbdr^{r}$ of $\Wbdr$.

\begin{figure}[htb]
\begin{center}\includegraphics[width=0.6\textwidth]{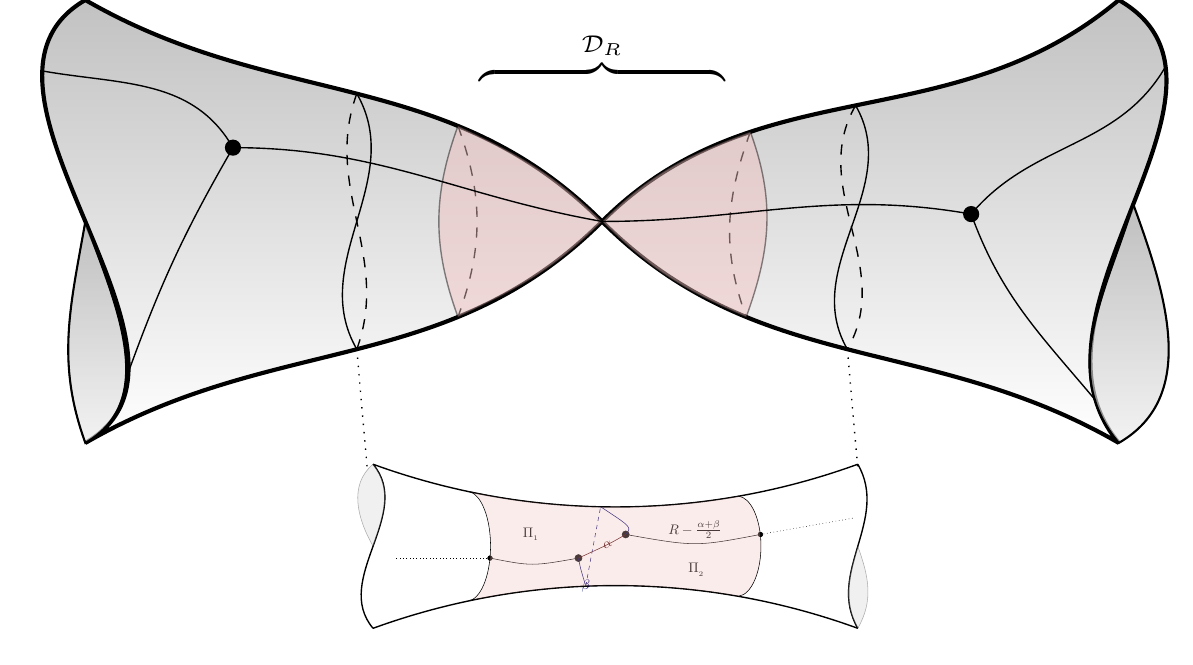}
\end{center}
\caption{Resolution of a double point $x_{1,2}^0$ on $\CC$ by insertion of an annulus with two simple zeros $x_1$ and $x_2$}
\label{W11plugin}
\end{figure}

We now discuss a neighbourhood of $\Wbdr$ in $\tilde W$. Starting from a point of  $\Wbdr $ (we denote the corresponding Riemann surface and quadratic differential by  $\CC_0$ and $\qd_0$)  one   replaces the two one-valent vertices of $\Gamma_0$ by two tri-valent vertices 
by inserting two small edges of lengths $\alpha = tA$ and $ \beta =tB$ with $A+B=1$ connecting these new vertices, as shown in Fig. \ref{W11plugin} (see \cite{BK1} for details;
the parameter $t\in R_+$ has the meaning of the length of the "vanishing cycle" going along horizontal trajectories which connect two arising simple zeros).

The resolution  is done by inserting the annulus domain instead of
two small circles around the simple poles of $\qd_0$ via procedure of "conformal welding", see Fig.\ref{W11plugin}.
In contrast to the resolution of a triple zero of $\qd$ discussed above, now there is only one cell of $W$ in a neighbourhood of a given point of $\Wbdr $.
The resolution can be done in countably  many ways  related by  Whitehead moves; two of these possible resolutions are shown in Fig. \ref{res_bound}.
\begin{figure}[htb]
\centering
\includegraphics[width=90mm]{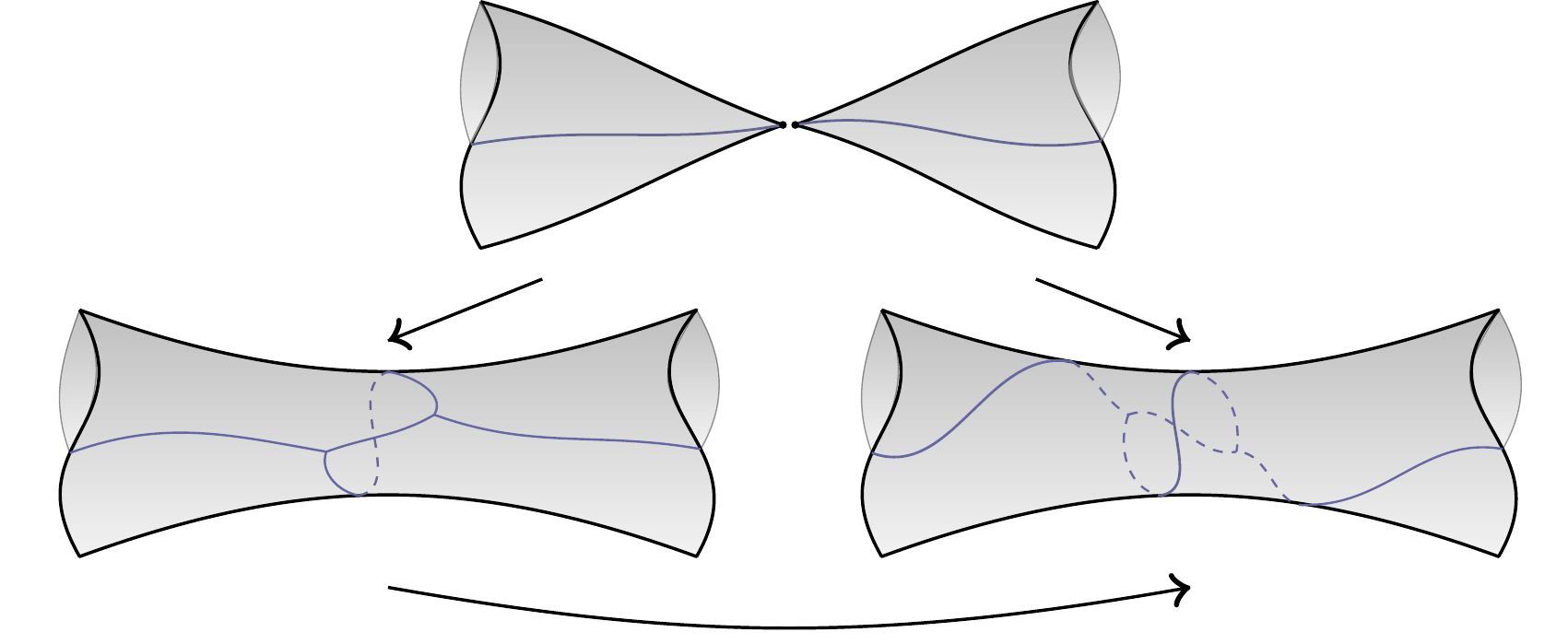}
\caption{Resolution of a point of $\Wbdr $. These two resolutions are related by a Whitehead move.}
%\blue{ corresponding to a combinatorial Dehn twist.}}
\la{res_bound}
\end{figure}

The transversal coordinates to $W_{1,1}$ are the homological coordinates $(B , \wt B)$ along the cycles 
$b_1^-, \wt b_1^-$ indicated in  Figures \ref{Dehnminus1}, \ref{Dehnminus2}; these cycles correspond to the two vanishing edges.

Let us denote by $K$ the cell in $ W$ obtained as a result of this resolution. This cell is
  "wrapped onto itself" in a neighbourhood of $\Wbdr$ in the sense that the  points on the facets of  $K$ that correspond to the limits $B\to 0$ and $\wt B\to 0$ are identified.
However, the path between points $(0,1)$ and $(1,0)$ in the $(B, \wt B)$-plane for fixed $t$ is topologically non-trivial in $\wt W$.

Let us  explain how the orientation induced by the symplectic form $\Omega_{_{Hom}}$ \eqref{omegahom}, together with the propagation of bases in Def. \ref{propbases}, translates to the $(B , \wt B)$ plane. 

The intersection numbers of $\wt b_1^-$ with  the elements of the basis $\{a_1^-, b_1^-, \dots,\}$ are $\wt b_1^- \circ a_1^-=-1, \wt b_1^-\circ  {b_1^-} = 2$ and all others are zero. 
In $H^-$ the intersection pairing has rank $2g^-$ and co-rank $n$, with the kernel being spanned by  $\{\g_j \}_{j=1}^n$. Therefore, the intersection is well--defined only as a pairing in the quotient space $H_- \mod \{\g_j \}_{j=1}^n$. The computation of intersection numbers shows  that  
\be
\la{bbt}
\wt b_1^- =  b_1^- - 2a_1^-   + \gamma
\ee
for some  $\gamma \in \Z\{\g_j\}_{j=1}^n$.  

%. 
\begin{lemma}
The homological coordinates $(B, \wt B)$ are transversal coordinates to a component of $\Wbdr$ within the adjacent cell of  $W$ and 
\begin{itemize}
\item the orientation induced by the symplectic structure $\Omega_{\text {{ \tiny Hom}}}$ \eqref{omegahom} in the $(B,\tilde{B})$-plane is  given by   $ {\rm d}   B \wedge {\rm d} \wt B$.
\item the homological coordinates $B,\wt  B$ have  {\bf opposite} signs.
\end{itemize}
\end{lemma}
{\bf Proof.}
From the relation   \eqref{bbt} it follows that   the orientation induced by $\Omega_{\text {{ \tiny Hom}}}$ \eqref{omegahom}   in the transversal $(B, \wt B)$--plane  is ${\rm d}   B \wedge {\rm d} \wt  B =   2 {\rm d}A \wedge {\rm d} B$.
The second item follows again from a local consideration  (similar to the ones in the proof of Prop. \ref{proporientation}) of the sign of $v = \sqrt{\qd}$ near the zero $x_1$ that lies within all three $a_1^-, b_1^-, \wt b_1^-$  see Figures \ref{Dehnminus1}, \ref{Dehnminus2}. 
In the distinguished coordinate $\zeta$ at $x_1$ we have $ \int_{x_1}^x v = \zeta^\frac 3 2$;  we choose the determination so that the cut is mapped to the positive $\zeta$--axis. Then $A_1=\oint_{a_1^-} v $ is positive,  $\wt B>0$ and $ B<0$. 
\QED
It is more convenient to use {\it positive} transversal coordinates; this can be done by defining $B' = -B/2$  and $A'=\wt B$.
Then the orientation in the $( A', B')$--plane  is ${\rm d} A' \wedge {\rm d}B'$ and we have: 
\begin{corollary}
\label{corW11}
The coordinates $ B'= - \frac 1 2  \oint_{b_1^-} v$ and $ A' =\oint_{\wt b_1^-} v$ are transversal local coordinates in a neighbourhood of $\Wbdr$ on $\wt W$.
The local model of the transversal manifold is a cone  $(\R_{\geq 0}^2\setminus \{(0,0)\})/\sim $ where the equivalence relation $\sim $ is the identification of the two axes $\{0\}\times \R_+$ and $\R_+\times \{0\}$.

The restriction of $\Omega_{\text {{ \tiny Hom}}}$ \eqref{omegahom} to the cone  is given by ${\rm d}A' \wedge {\rm d} B'$, which defines the induced orientation.
\end{corollary}

Recall  that the Dehn twist along a homotopically nontrivial loop $\ell$ of a Riemann surface corresponds to a closed path in  $\mathcal M_{g,n}$ around a component of the Deligne--Mumford boundary $\Delta_{DM}$. 
Now Cor. \ref{corW11} implies the following proposition.

\begin{proposition}
\la{PDehn}
Fix a cell of $W$ corresponding to a ribbon graph $\Gamma$ such that $x_1,x_2$ is a pair of vertices connected by two  edges $e, \wt e$ which are homotopically distinct on $\CC$. Consider the associated coordinates $(A',B')$ defined in Cor. \ref{corW11}. Then the 
path  in $\wt W$ starting from an interior point $A'>0<B'$ to the "wall" $A'=0$ followed by a path from the corresponding point on the other wall $B'=0$ back to the original point represents  the Dehn's twist along the path formed by the edges  $e, \wt e$.
 The orientation of this Dehn's twist  induced by the symplectic form (\ref{omegahom}) is positive.
\end{proposition}

On the transversal  cone $(\R_{\geq 0}^2\setminus \{(0,0)\})/\sim $ (see Cor. \ref{corW11}),  the Dehn twist is represented by a simple loop around the vertex of the cone. 
While traversing such path, the ribbon graph undergoes a single Whitehead move.

The next lemma describes the class of Dehn's twists which can be obtained via this construction.

\begin{lemma}
\label{lemmaDehn}
Let $\gamma$ be a loop on $\CC$ which is either non-separating or separates $\CC$ into 2 stable components
each of which contains at least one marked point. Let, moreover, the curve $\CC$ be sufficiently close to  the 
corresponding component $\Delta_{DM}^\gamma$ of the Deligne-Mumford boundary. Then the Dehn's twist based at the curve $\CC$ in $\mathcal M_{g,n}$  
along the loop $\gamma$ can be realized as a path in $\tilde{W}$ defined in  Proposition \ref{PDehn}.

\end{lemma}
{\bf Proof.} Let us start from an arbitrary curve $\CC_1$ with the contour $\gamma$ satisfying the condition of the lemma.
The free homotopy class of the  contour $\gamma$ identifies a component $\Delta_{DM}^\gamma$ of the Deligne--Mumford boundary $\Delta_{DM}$; the component $\Delta_{DM}^\gamma$ contains the stable nodal curves where $\gamma$ is shrunk to a point. Let $ (\CC_t,\gamma_t)$ be a smooth family parametrized by $t\in [0,1]$ so that  $ \CC_0 \in \Delta_{DM}^\gamma$ while $\gamma_0$  is collapsed to the node and $\gamma_1=\gamma$.
If $\CC_0$ is chosen generically in $ \Delta_{DM}^\gamma$ then the Jenkins--Strebel differential on the normalization of $\CC_0$ has  two  simple poles at the points obtained by resolving the node; the ribbon graph on the normalization of $\CC_0$ then has two uni-valent vertices as  discussed in  Section \ref{combDehnsect}.

Thus for a sufficiently small $t$ the curve $\CC_t$  falls within a neighbourhood which is parametrized in terms of the ``plumbing'' construction shown in Fig. \ref{W11plugin}. Then the  Dehn's twist along the loop $e\cup\wt e$ described in Proposition 
\ref{PDehn} (see  Fig. \ref{res_bound}) is homotopic to the Dehn's twist along the loop $\gamma=\gamma_t$.

\QED

\subsection{Combinatorial Dehn's twists in Chekhov-Fock modular groupoid}
\label{ChekFock}

\begin{figure}
\begin{center}
\includegraphics[width=0.5\textwidth]{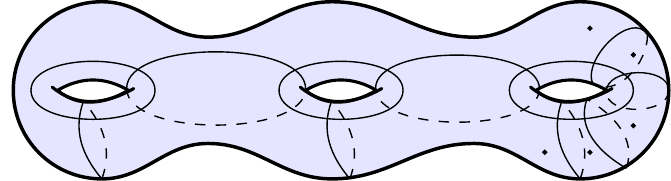}
\end{center}
\caption{The Dehn twist generators for $MPG_{g,n}$ (here illustrated for $g=3, n=5$).}
\label{figfarb}
\end{figure}

The fundamental group of the moduli space $\pi_1(\Mcal_{g,n},\CC_0)$  i.e. the mapping class group $MPG_{g,n}$
is known to be generated by a finite set of Dehn's twists; an example of  $2g+n$ Denh's twists $\{D_i\}_{i=1}^{2g+n}$ 
generating the full $MPG_{g,n}$
is shown in Fig. 4.10 of \cite{Farbmarg}, see Fig. \ref{figfarb}. Notice that all Dehn's twists  $\{D_i\}_{i=1}^{2g+n}$ act along closed {\it non-separating}  loops on the Riemann surface $\CC$ with $n$ puncture. The Dehn's twists represent paths around the  components of the Deligne-Mumford boundary $\Delta_{DM}$ obtained by pinching the loops.
Any other  Dehn twist (including those  along  separating loops)    can therefore be represented as product  of $\{D_i\}_{i=1}^{2g+n}$. 

This classical picture is naturally translated to the context of the combinatorial model $\mathcal M_{g,n}[{\bf p}]$.
We start from the following
\begin{proposition}
\label{propWW}
Let $\CC_0$ be mapped to a cell of maximal dimension by the Jenkins-Strebel map \eqref{JS} $\JS: \Mcal_{g,n} \to \mathcal M_{g,n}[{\bf p}]$. Then $\JS$  induces an isomorphism of the following fundamental groups:
\be
\pi_1(\Mcal_{g,n}, \CC_0) \simeq \pi_1(\mathcal M_{g,n}[{\bf p}], \JS(\CC_0)) \simeq \pi_1(\tilde{W}\cup W_5^0\cup W_{4,4}^0, \JS(\CC_0))\;.
\label{221}
\ee
\end{proposition}
{\bf Proof.}
Since the Jenkins--Strebel  map  $\JS$ \eqref{JS} from $\Mcal_{g,n}$ to $\mathcal M_{g,n}[{\bf p}]$ is a homeomorphism, the fundamental group 
$\pi_1(\Mcal_{g,n}, \CC_0)$ is isomorphic to $\pi_1(\mathcal M_{g,n}[{\bf p}], \JS(\CC_0))$. To prove the second isomorphism in \eqref{221} we observe that  the fundamental group of the complex $\mathcal M_{g,n}[{\bf p}]$ is the same as the one of  the sub-complex of all  cells  of codimension up to $2$. The union of cells of highest dimension 
forms the stratum $W$; the cells of codimension $1$ form the stratum $W^0_4$ whose cells correspond to
ribbon graphs with exactly one vertex of valence $4$ and all other vertices of valence $3$; recall that we denoted $\tilde{W}=W\cup W^0_4$.
Finally, there are two strata of codimension $2$:  $W_5^0$ (the stratum of highest dimension of the Witten's cycle $W_5$)
and $W_{4,4}^0$; the cells of $W_{4,4}^0$ correspond to ribbon graphs which have two vertices of valence $4$ and while other vertices have valence $3$. \QED

Now we  define a graph $G$ associated to the sub-complex $\tilde W$, which naturally encodes its fundamental group. 

Denote by $K_0$ the top-dimensional cell containing  $\JS(\CC_0)$.  We can represent an element of  $\pi_1(\mathcal M_{g,n}[{\bf p}], \JS(\CC_0))$ by a loop based at $ \JS(\CC_0)$ that remains within $\tilde W$. Such a path  traverses  several cells of $W$ crossing the facets between them.
Transitions between cells of $W$ correspond to the Whitehead moves on the corresponding edges.

Consider now the graph $G$ whose vertices $\{K_i\}$ are represented by cells of $W$.
Two vertices $K_1$ and $K_2$ are connected by an edge if they have a common facet i.e. if one can go from $K_1$ to $K_2$ via the Whitehead move. The edge then corresponds to a cell of $W_4^0$.
Since each cell in $\tilde W$ is homeomorphic to a ball, 
the fundamental group $\pi_1(\tilde W, \JS(\CC_0))$  is then isomorphic to  $\pi_1(G,K_0)$.

A path on the graph $G$ can be thought of as  a sequence of several Whitehead moves; the set of all Whitehead moves carries the  name of
{\it modular groupoid}  introduced by Chekhov and Fock \cite{Fock}. Two Whitehead moves $M_1$ and $M_2$ can be
multiplied if the end-cell of the edge of $G$ representing $M_1$ coincides with the initial cell of the edge representing $M_2$.

We now define a combinatorial complex $\wh G$ obtained from $G$ by adding ``faces'' to it; 
the  faces of $\wh G$ are  defined to be in one-to-one correspondence with cells of co-dimension $2$ in $\mathcal M_{g,n}[{\bf p}]$.  For a face $F$ of $\wh G$ (corresponding to a cell denoted by the same letter) the {\it boundary} is defined to be the union of  vertices and edges of $G$ corresponding to cells of co-dimensions zero and one in the {\it star} of $F$ (the star of a cell $F$  is the union of all cells whose closure intersects $F$).

 Since all cells of co-dimension $2$ belong either to $W_5^0$ or $W_{4,4}^0$, there exist two types of faces in $\wh G$: quadrilaterals (for cells of   $W_{4,4}^0$) and pentagons (for cells of   $W_{5}^0$).

Within $\wh G$ there exist  non-trivial homotopy  deformations defined by the condition that all boundaries of faces are homotopically trivial
(i.e. paths around $W_5^0$ and $W_{4,4}^0$ become trivial in $\pi_1(\wh G)$).

Thus we have the following 
\begin{proposition}
\label{propWTF}
The following  fundamental groups are  isomorphic
\be
\pi_1(\tilde{W},\JS(\CC_0))\simeq \pi_1(G,K_0)
\la{WG}
\ee
and
\be
\pi_1(\Mcal_{g,n}, \CC_0 )
\simeq \pi_1(\wh G, K_0)
\la{MCC}
\ee
where $K_0$ is the cell of $W$ such that $\JS(\CC_0)\in K_0$.
\end{proposition}
{\bf Proof.}
The isomorphism (\ref{WG}) is an immediate corollary of the definition of graph $G$ and the fact that all cells of $W$ and
$W_4$ are
homeomorphic to balls of corresponding dimensions.

To prove (\ref{MCC}) we notice that the
Prop. \ref{propWW}  establishes the isomorphism between $\pi_1(\Mcal_{g,n}, \CC_0 )$ and $\pi_1(\tilde{W}\cup W_5^0\cup W_{4,4}^0, \JS(\CC_0))$. 
The isomorphism between  the latter  fundamental group and $ \pi_1(\wh G, K_0)$  is constructed as follows. For each class $[\gamma]\in \pi_1(\tilde{W}\cup W_5^0\cup W_{4,4}^0, \JS(\CC_0))$   choose a representative $\gamma$ whose image $\JS(\gamma)$ only traverses cells of $\tilde W$ (avoiding strata of  codimension greater than $1$). This defines a loop in $G$ (and $\wh G$) along vertices and edges. Its representative in   $ \pi_1(\wh G, K_0)$ (with the homotopy induced by quadrilateral and pentagon moves) gives the image of the class $[\gamma]$. This image is independent of the choice of
the loop within the class $[\gamma]$ since all cells in the combinatorial model are homeomorphic to balls.
Conversely, let us fix a representative curve $\CC_j$ in each cell $K_j$ of $W$. For  the class $[\ell]\in  \pi_1(\wh G, K_0)$
we select a representative that stays on edges of the graph $G$ and map it to the loop in $ \tilde{W}$ obtained by connecting the representatives $\CC_j$ of the cell traversed by $\ell$. Images of different representatives of the same class $[\ell]$ then also lie in the same equivalence class of loops in $\tilde{W}\cup W_5^0\cup W_{4,4}^0$, and, therefore, represent the same element of
$\pi_1(\tilde{W}\cup W_5^0\cup W_{4,4}^0, \JS(\CC_0))$.
 \QED

\begin{figure}[t]
\centering
\includegraphics[width=0.7\textwidth]{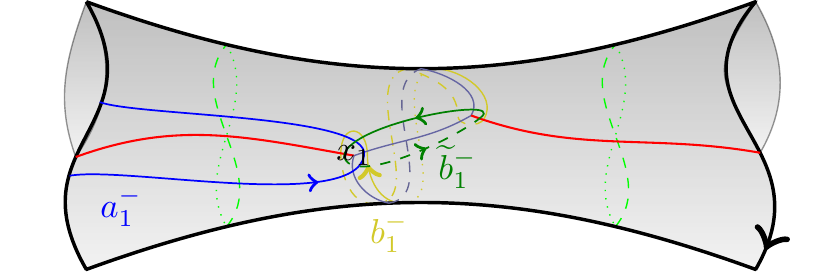}
\caption{The pinching region for a curve (and ribbon graph) in the neighbourhood of $\Wbdr$. Indicated the two vanishing cycles $b_1^-$ and $\wt b_1^-$.
}
\la{Dehnminus1}
\end{figure}
  The fundamental group of $\wh G$ can be generated by
the combinatorial Dehn's twists in the same way as  $\pi_1(\Mcal_{g,n})$ is generated  by the ordinary Dehn's twists.
\begin{definition}
For each component $\Wbdr^\gamma$ of $\Wbdr$ (corresponding to the collapse of a non-separating loop $\gamma$) denote by $K_\gamma$ the adjacent cell of $W$. 
The (local) combinatorial Dehn twist $D_\g^{comb}$ is defined to be the path in $G$ starting and ending at  the  vertex  $K_\gamma$.  This path follows the edge of $G$ which  represents the facet of $K_\gamma$ obtained by the collapse of one of two edges of  the corresponding ribbon graph.
%which for the loop $\gamma$
(see Cor. \ref{corW11}). 
\end{definition}
Such local combinatorial Dehn's twists can not be multiplied in the based fundamental group of $G$. 
To define the  Dehn's twist around $\Wbdr^\gamma$ based at any 
 vertex $K_0$ of $G$ we choose a path $\ell_\gamma$ on $G$ which connects $K_0$ to the cell $K_\gamma$ and  define
\be
D_\gamma^{comb}[K_0]= \ell_\gamma D_\gamma^{comb} \ell_\gamma^{-1}\;.
\la{defDjcomb}
\ee
The based combinatorial Dehn's twist defined in this way, being  considered as an element of $\pi_1(G,K_0)$  depends of course on the choice of the path $\ell_\gamma$. However, the equivalence class of $D_\gamma^{comb}[K_0]$ in $\pi_1(\wh G,K_0)$ does not depend on the choice of $\ell_\gamma$.

Combining Prop. \ref{propWTF} with the fact that $MPG_{g,n}$ is generated by Dehn twists along nonseparating loops $\gamma_1,\dots, \gamma_{2g+n}$ shown in Fig.\ref{figfarb} (Fig. 4.10 of \cite{Farbmarg})  we have proved
\begin{proposition}
The fundamental group $\pi_1(\wh G,K_0)$  is generated by (equivalence classes of) combinatorial Dehn's twists $D_{\g_j}^{comb}[K_0], \ \ j=1,\dots, 2g+n$ given by (\ref{defDjcomb}).
 \end{proposition}

This proposition implies in particular that the based Dehn's twists around {\it any}  part of the  DM boundary 
can be expressed as products of based Dehn twists around only those components of $W_{1,1}$ which are isomorphic to the corresponding components of $\Delta_{DM}$.

In our framework, Theorem 1 of \cite{Fock} can then be reformulated as follows:
\begin{proposition}
The fundamental group $\pi_1(G,K_0)\simeq \pi_1(\tilde{W},\JS(\CC_0))$ is generated by combinatorial Dehn's twists and paths around all pentagon and quadrilateral faces of $G$ which start and end at the same vertex $K_0$ of $G$.
\end{proposition}

\begin{figure}[t]
\centering
\includegraphics[width=0.45\textwidth]{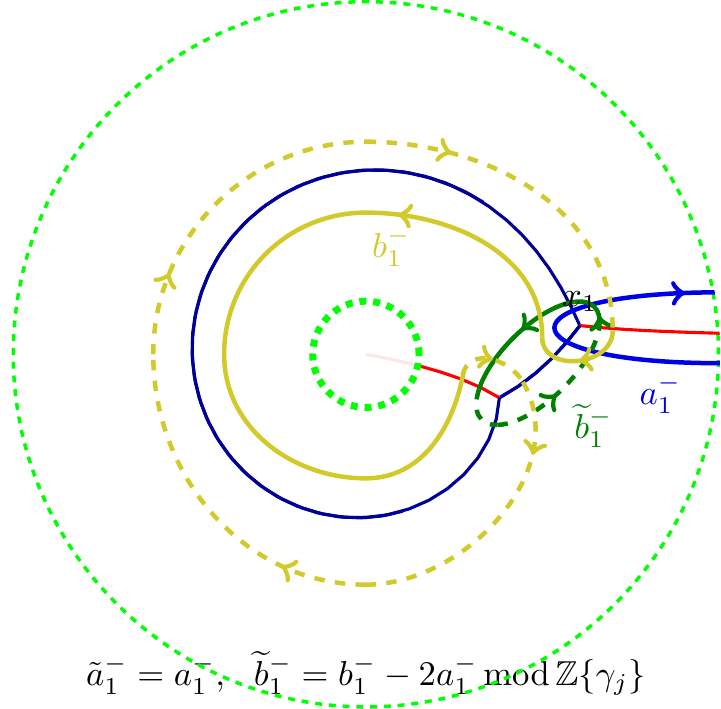}
\caption{ 
The same region as in Fig. \ref{Dehnminus1} can be mapped to an annulus with the metric induced by the quadratic differential $\qd$. The quadratic differential on this annulus is given by the restriction of a JS quadratic differential on $\C \mathbb P^1$ with two poles of order $3$ at $z=0, \ z=\infty$ (more details are given in Section 4  of \cite{BK1}). }
\la{Dehnminus2}
\end{figure}

\section{Meromorphic quadratic differential and canonical covering}
\la{2scan}
\subsection{Canonical covering}
Here we give more details about the geometry of the canonical covering $v^2=\qd$ (\ref{cancov}) defined by a meromorphic quadratic differential $\qd$ with divisor (\ref{divqd1}) (i.e. that the pair $(\CC,\qd)$
is in the space $\Qcal_{g,n}^{\bk,\bl}$).
The divisor of the  Abelian differential $v$ on $\Ch$  takes the form
\be
(v)=\sum_{i=1}^{\nzh}
\dh_i\ph_i\equiv\sum_{i=1}^\mo (2k_i+2)x_i+\sum_{i=\mo+1}^{m} l_i (\xh_i+\xh_i^\mu) -\sum_{i=1}^{n} (\zh_i+\zh_i^\mu)\;.
\la{divv}
\ee
 To derive the expression (\ref{divv}) for the divisor of the abelian differential $v$ on $\Ch$ from the expression (\ref{divqd1})  
 for the divisor of quadratic differential $\qd$ on $\CC$ one should  use the fact that the local coordinates  on $\Ch$ near poles and zeros of even multiplicity coincide with the local coordinates on $\CC$, while near the poles and zeros of odd multiplicity they are square roots of the local coordinate  on $\CC$.

The genus of $\Ch$ equals $\gh= 2g+\frac{m_{odd}}{2}-1$, or, equivalently,
$\gh=g+g_-$ with 
\be
g_-=g+\frac{m_{odd}}{2}-1\;.
\ee

For holomorphic $\qd$ the canonical cover $\Ch$ is a classical ingredient of Teichm\"uller theory.
In the case of meromorphic $\qd$ the covering $\Ch$  appears in the theory of
generalized $SL(2)$ Hitchin's systems \cite{Bottacin,Markman} under the name of "spectral cover".

Denote by $(a_i,b_i)$ a canonical basis in $H_1(\CC,\Z)$. The dual basis in $H^{(1,0)}(\CC,\C)$ will be denoted by $(v_1,\dots,v_g)$
with the normalization $\oint_{a_i}v_j=\delta_{ij}$.  Denote the period matrix of $\CC$ by $\O_{ij}=\oint_{b_i}v_j$.

\begin{figure}
\centerline{
\includegraphics[width=0.5\textwidth]{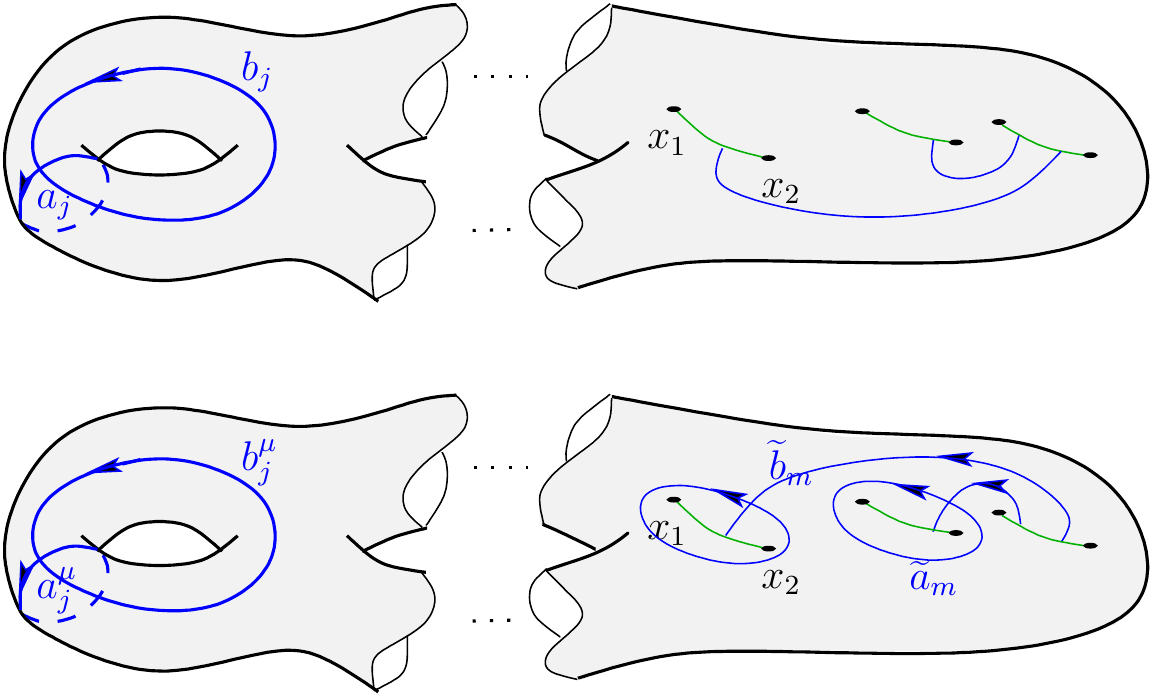}}
\caption{The canonical covering $\Ch$ and the symplectic  basis of cycles (\ref{mainbasis}). }
\label{canbasCh}
\end{figure}
A canonical basis of $H_1(\Ch,\Z)$ can be chosen as shown in Fig.\ref{canbasCh} 
\cite{Fay73,contemp,BKN}:
\be
\{a_j,  a_j^\mu, \at_k,\,b_j,  b_j^\mu\,,\wt b_k\}\,,\quad j=1,\dots,g,\;\;\;k=1,\dots,\gm-g\;,
\label{mainbasis}
\ee
where the subset $\{a_j,b_j,a_j^\mu,b_j^\mu\}_{j=1}^g$ is obtained by the  lift  of the canonical basis 
of cycles $\{a_j,b_j\}_{j=1}^g$ from $\CC$ to $\Ch$,
such that 
\be 
\mu_* a_j = a_j^\mu,\quad\mu_* b_j = b_j^\mu,
\quad\mu_* \at_k + \at_k= \mu_* \wt b_k + \wt b_k  = 0\;.
\la{canCh}
\ee
We shall denote by 
\be
\label{basisCH}
 {\{\wh{v}_j, \wh{v}_j^\mu, \wh{w}_k\}} 
 \ee
  the  basis of normalized Abelian differentials on $\Ch$ dual to the basis of cycles (\ref{canCh}). 
The differentials $ v_j^+=\wh v_j+\wh v_j^\mu,\; j=1,\dots,g$,
form a basis in the space $H^+$; these differentials are naturally identified with the normalized holomorphic differentials $v_j$ on 
$\CC$. Therefore, to simplify the notation, they shall often be denoted by $v_j$ instead of $v_j^+$.
 A basis in $H^-$ is given by the $\gm$ Prym differentials $v_l^-$, where
\be
 v_l^-=\begin{cases} \wh v_{l}- \wh v_{l}^\mu,\quad l=1,\dots, g\;,\\
\,\wh w_{l-g}, \qquad\; l=g+1,\dots, \gm \;.\end{cases}
\label{Prym}
\ee
The following  classes in $H_1(\Ch,\R)$ form a symplectic basis in $H_+$
\begin{eqnarray}
a_j^+ = \f{1}{2}(a_j+ a_j^\mu)\;,\quad
b_j^+ =  b_j+ b_j^\mu\;,\quad j=1,\dots,g\; ,
\label{abp}
\,\qquad
a_j^+\circ b^+_k=\delta_{jk}\ ,
\end{eqnarray}
 whereas the  classes
\be
a_l^- =  \f{1}{2}(a_{l}- a_{l}^\mu)\, ,\hskip0.5cm
b_l^- =  b_{l}- b_{l}^\mu,\quad l=1,\dots,g\;,
\la{abm1}
\ee
\be
a^-_{l}=\at_{l-g},\;\;\; b^-_{l}=\tilde b_{l-g}\;,\hskip0.5cm\, l=g+1,\dots,\gm
\label{abm}
\ee
form a symplectic basis in $H_-$.

The basis $\{a_j^+,a_l^-,b_j^+,b_l^-\}$, $j=1,\dots,g$, $l=1,\dots,\gm$, is related 
to the canonical basis (\ref{mainbasis}) as follows
\be
 \left(\ba{c} {\bf b_+}\\ {\bf b_-}  \ea\right)=T
 \left(\ba{c} {\bf b}\\ {\bf b}^\mu\\ {\bf \tilde{b}} \ea\right)\;,\hskip0.7cm
 \left(\ba{c} {\bf a_+}\\ {\bf a_-}  \ea\right)=T^{-1}
 \left(\ba{c} {\bf a}\\ {\bf a}^\mu\\ {\bf \tilde{a}} \ea\right) 
\ee 
 with symmetric matrix
\be
T=  \left(\ba{ccc} I_g & I_g& 0     \\
                     I_g & -I_g  & 0 \\
                    0& 0 & I_{\gm-g}  \ea\right)\;\qquad.\;,\hskip0.7cm
            T^{-1}=  \left(\ba{ccc} I_g/2 & I_g/2& 0     \\
                     I_g/2 & -I_g/2  & 0 \\
                    0& 0 & I_{\gm-g}  \ea\right)        
\la{MaMb}
\ee

Integration of the differentials $v_j^+$ over the cycles $\{a_k^+\}$ gives $\oint_{a_k^+} v_j^+=\delta_{jk}$. The integrals of  $v_j^+$ over the cycles
$\{b_k^+\}$ give twice the period matrix of $\CC$: 
\be
\O^+_{jk}\equiv \int_{b_k^+} {v}_j^+=2\O_{jk}\;.
\la{OOp}\ee
 Similarly, integration of  the Prym differentials (\ref{Prym}) over the cycles $\{a_j^-\}$
(\ref{abm1}), (\ref{abm}) yields the $\gm\times\gm$ unit matrix, while their integrals over the cycles $\{b_j^-\}$ give a $\gm \times \gm $
symmetric  matrix  $\O^-$ which equals twice the  {\it Prym} matrix $\Pi$ (as defined in \cite{Fay73}, p.86):
\be
\O^-_{jk}\equiv \int_{b_k^-} {v}_j^-=2\,\Pi_{jk}\ ,\ \ 1\leq j,k\leq \gm\;.
\la{Prymmatdef}
\ee
The Prym matrix can be  written in block form as follows; 
\be
\Pi= \left(\ba{cc} \Pi_1 & \Pi_2\\
                \Pi_2^t  & \Pi_3  \ea\right)
\la{Prymmat}
\ee
where $\Pi_1$ is $g\times g$ matrix; $\Pi_2$ is a $g\times (\gm-g)$ matrix and $\Pi_3$ is a $(\gm-g) \times (\gm-g)$ matrix.
Then the period matrix $\widehat{\O}$ of the double cover $\Ch$ in the basis (\ref{mainbasis}), (\ref{basisCH})  can be expressed in terms of $\Omega^+$ and $\Omega^-$ as follows:
\be
\widehat{\Omega}=
T^{-1}
 \left(\ba{cc} \Omega^+ & 0\\
                0  & \Omega^-  \ea\right) T^{-1}=2 T^{-1}
 \left(\ba{cc} \Omega & 0\\
                0  & \Pi  \ea\right) T^{-1} 
 \la{shs}
\ee
which gives
\be
\widehat{\Omega}= \left(\ba{ccc} \f{\Omega+\Pi_1}{2} & \f{\Omega-\Pi_1}{2} & \Pi_2\\
 \f{\Omega-\Pi_1}{2} & \f{\Omega+\Pi_1}{2} & - \Pi_2\\
   \Pi_2^t     & - \Pi_2^t & \Pi_3  \ea\right)
\la{OhOP}
\ee
The expression (\ref{OhOP}) coincides with (91) of \cite{Fay73} up to some signs and
 interchange of blocks. This difference with \cite{Fay73} is due to a  different choice of the canonical basis of cycles
(\ref{mainbasis}) made in this paper (our cycles $a_j^\mu$ and $b_j^\mu$  differ by sign from the ones used by Fay; also the
 ordering of the canonical cycles used by Fay is different from ours).
\begin{remark}\rm
The structure  of the canonical cover is also extensively discussed in the recent papers \cite{contemp,BKN}. Conventions used in \cite{BKN}  differ from the ones used here due to a different normalization of the canonical basis (\ref{abm1}) and (\ref{abm}) in $H_-$ (and the dual basis (\ref{Prym}) in 
$H^-$) used here and in \cite{contemp}. In particular, the Prym matrix $\tilde{\Pi}$ from \cite{BKN} is related to the matrix (\ref{Prymmat}) by the following transformation:
\be
\tilde{\Pi}= \left(\ba{cc} I_g & 0 \\
                           0  &  \sqrt{2}\, I_{\gm} \ea \right)\Pi  \left(\ba{cc} I_g & 0 \\
                           0  &   \sqrt{2}\, I_{\gm} \ea \right)\;.
\ee
\end{remark}

\subsection{Distinguished local coordinates on $\CC$ and $\Ch$}

A given meromorphic quadratic differential $\qd$ with divisor as in 
(\ref{divqd1}) defines a set of distinguished local parameters 
 on $\CC$ and on the canonical covering $\Ch$.  
The distinguished local parameters $\zeta_i$ on $\CC$ and $\zetah_i$ on $\Ch$ coincide near all points except the branch points of the covering (the points $\{x_i\}$ and  $\{y_i\}$
of (\ref{divqd1})); near the branch points we have $\zeta_i(x)=\zetah_i^2(x)$.

\begin{itemize}
\item
Near any point $x_0\in\Ch$ such that $\pi(x_0)\not\in (\qd)$   the local coordinates on $\CC$ and $\Ch$  can be chosen to be $\int_{x_0}^x v$. 
\item
Near a point $x_i$, $1=1,\dots,\mo$  
the distinguished local parameters $\zetah_i$ on $\Ch$ and $\zeta_i(x)$ on $\CC$ are  given by  
\be
\zetah_i(x)=\left[\int_{x_i}^x v\right]^{1/(2k_i+3)}\;,\hskip0.7cm \zeta_i(x)=\zetah^2_i(x)\;.
\la{dist20}
\ee

\item
Near  points $x_i$ and $x_i^\mu$ of $\Ch$ for $i=\mo+1,\dots,m$
we define
\be
\zetah_i(x)=\left[\int_{x_i}^x v\right]^{1/(l_i+1)}\;\hskip0.7cm
\zetah^\mu_i(x)=\left[\int_{x_i^\mu}^x v\right]^{1/(l_i+1)}
\ee 
which in fact coincide (up to a sign)  with the local parameter 
\be
\zeta_i(x)=\left[\int_{x_i}^x v\right]^{1/(l_i+1)}
\ee
near $x_i$ on $\CC$ for $i=\mo+1,\dots,m$.

\item
Near a second order pole $z_i\in \CC$ and corresponding points $\zh_i,\zh_i^\mu\in \Ch$  
the distinguished local coordinates $\zeta_i$ and $\zetah_i$  are defined as follows.
Denote the quadratic residue of $\qd$ at $z_i$ by $-p_i^2/4\pi^2$. Then the residues of $v$ at $ \zh_i,\zh_i^\mu$ are equal to $p_i/2\pi i$ and $-p_i/2\pi i$, respectively.
The local parameters near $ \zh_i$ and $\zh_i^\mu$ are given by 
\be
\hat{\xi}_i^\pm(x)=\exp\left\{\f{\pm 2\pi i}{p_i}\int_{x_1}^x v\right\}
\la{dist3}
\ee
where $x_1$ is a chosen "first"  zero of $\qd$. These parameters  depend on the choice of the zero $x_1$ and the paths of integration. 
To fix the integrals in (\ref{dist3}) uniquely we consider the fundamental polygon of $\CC_0$ of $\CC$ and cut it along contours $\gamma_1,\dots,\gamma_n$
such that $\gamma_i$ connects the zero $x_1$ with the pole $z_i$. The corresponding  cuts  $\hat{\gamma}_i$ and $\hat{\gamma}_i^\mu$ on 
the fundamental polygon $\Ch'$ of $\Ch$
are obtained by lifting $\gamma_i$ from
$\CC$ to $\Ch$ i.e. $\pi^{-1}(\gamma_i)=\{\hat{\gamma}_i,\hat{\gamma}_i^\mu\}$. The paths of integration in (\ref{dist3}) should then lie entirely in this 
fundamental polygon with these additional cuts.

The local parameters $\zeta_i(x)$  on $\CC$ near $z_i$ can be chosen to coincide with any  of $\zetah_i^\pm$ and have  the same ambiguity.
\end{itemize}

\subsection{Homological coordinates on $\Qcal_{g,n}^{\bk,\bl}$}
\la{varform}

As in the introduction, decompose the 
homology group of $\Ch\setminus  \{\pi^{-1}(z_j)\}_{j=1}^n $, relative to zeros of $v$ of even multiplicity,  into even and odd parts under the involution $\mu$:
\be
H_1(\Ch\setminus \{ \pi^{-1}(z_j)\}_{j=1}^n;
\{\pi^{-1}({x}_j)\}_{j=\mo+1}^{m})= H_+\oplus H_-\;.
\la{decom}
\ee
Then
\be
{\rm dim}\, H_-={\rm dim}\, \Qcal_{g,n}^{\bk,\bl}=2g-2+m+n
\ee
which, since $m=\me+\mo$, can also be written as ${\rm dim}\, H_-=2g_-+n+\me$ with $g_-=g+\mo/2 -1$.
For any basis of cycles $\{s_i\}$ in $H_-$ one defines {\it homological coordinates} on $\Qcal_{g,n}^{\bk,\bl}$ as follows:
\be
\Pcal_{s_i}=\int_{s_i} v\;\hskip0.7cm i=1,\dots,{\rm dim}\, H_-\;.
\la{homcoord}
\ee
We notice that $H_+$ can be identified with the relative homology group 
$H_1(\CC\setminus\{z_i\}_{i=1}^{n},\{x_i\}_{i=\mo+1}^{m})$.

The  homology group dual to (\ref{decom}) which can also be decomposed into corresponding even and odd parts is
\be
H_1(\Ch\setminus \{\pi^{-1}({x}_j)\}_{j=\mo+1}^{m} ;  \{ \pi^{-1}(z_j)\}_{j=1}^n)= H^*_+\oplus H^*_-\;.
\la{decom1}
\ee

The subspaces $H_\pm^*$ are dual to $H_\pm$  respectively. The basis in $H_-^*$ dual to the basis $\{s_i\}$ in $H_-$ 
is denoted by $\{s_i^*\}$ (the intersection index is $s_i\circ s_j^*=\delta_{ij}$).

\section{Hodge and Prym tau-functions on  $\Qcal_{g,n}^{\bk,\bl}$}

\la{HPsec}

The  tau-functions $\tau_+$ and $\tau_-$ on the moduli spaces of holomorphic quadratic differentials were defined in \cite{Leipzig,contemp} by restriction
of the Bergman tau-function  on an appropriate stratum of moduli spaces of Abelian differentials \cite{JDG}. 
For general theory of the Bergman tau-functions and its applications to the the theory of isomonodromic deformations, spectral geometry and theory of matrix models and Frobenius manifolds we refer to  \cite{Annalen,IMRN1,JDG,EKK}. For applications 
of Bergman tau-functions to geometry     of various moduli spaces and the theory of Teichm\"uller flow we refer to \cite{Advances,MRL,contemp,Basok,KSZ,EKZ}.

Under a change of Torelli marking of the base Riemann surface the tau-function $\tau_+$ transforms as a section of determinant line bundle of the Hodge vector bundle; thus we call it {\it Hodge tau-function}. Similarly, $\tau_-$ transforms as a section of
determinant line bundle of the Prym vector bundle over the space of quadratic differentials \cite{contemp}; thus we call it  {\it Prym tau-function}.

In our present setting of meromorphic  quadratic differentials  with  second order poles  we use the formalism of 
Bergman 
tau-functions on spaces of Abelian differentials of  third kind  \cite{CMP}.
We start from defining two tau-functions, $\tau_+$ and $\tauh$ by explicit formulas 
using the framework of \cite{CMP} and then define the Prym tau-function $\tau_-$ as the ratio $\tauh/\tau_+$.

\subsection{Hodge tau-function $\tau_+$}
\la{Hodge_sec}

Introduce two  vectors $\rb,\sb\in\f{1}{2}\Z^g$ such that 
\be
\f{1}{2}\Acal_x((\qd))+2K^x+\O\rb+\sb=0
\la{defZh}
\ee
and also the following notations \cite{JDG,contemp}:  
\be
E(x,q_i)=\lim_{y\to q_i} E(x,y) \sqrt{d\zeta_i(y)}
\la{defEp}
\ee
\be
 E(q_i,q_j)=\lim_{x\to q_j, y\to q_i} E(x,y) \sqrt{d\zeta_i(y)} \sqrt{d\zeta_j(x)}
\la{defEpp}
\ee
where $\zeta_i$ is the distinguished local parameter on $\CC$ near $q_i$.

Consider the  multi-valued $g(1-g)/2$-differential $\Ccal(x)$ (\ref{defCint}) on $\CC$ where   $K^x \in \C^g$ is the vector of Riemann constants corresponding to the basepoint $x$ \cite{Fay92}.

Let us choose a  system of cuts $\{\g_i\}_{i=1}^n$  on $\CC$  by connecting one zero (say, $x_1$) with $z_i$ by the cut $\g_i$, see Fig.\ref{Cancover}.
We assume that these cuts lie entirely in the fundamental polygon $\CC'$  of $\CC$. The integration contours in the definition of 
distinguished local parameters (\ref{dist3}) near $z_i$ are assumed to not intersect $\{\gamma_i\}_{i=1}^n$.

\begin{figure}[t]
\centering
\includegraphics[width=90mm]{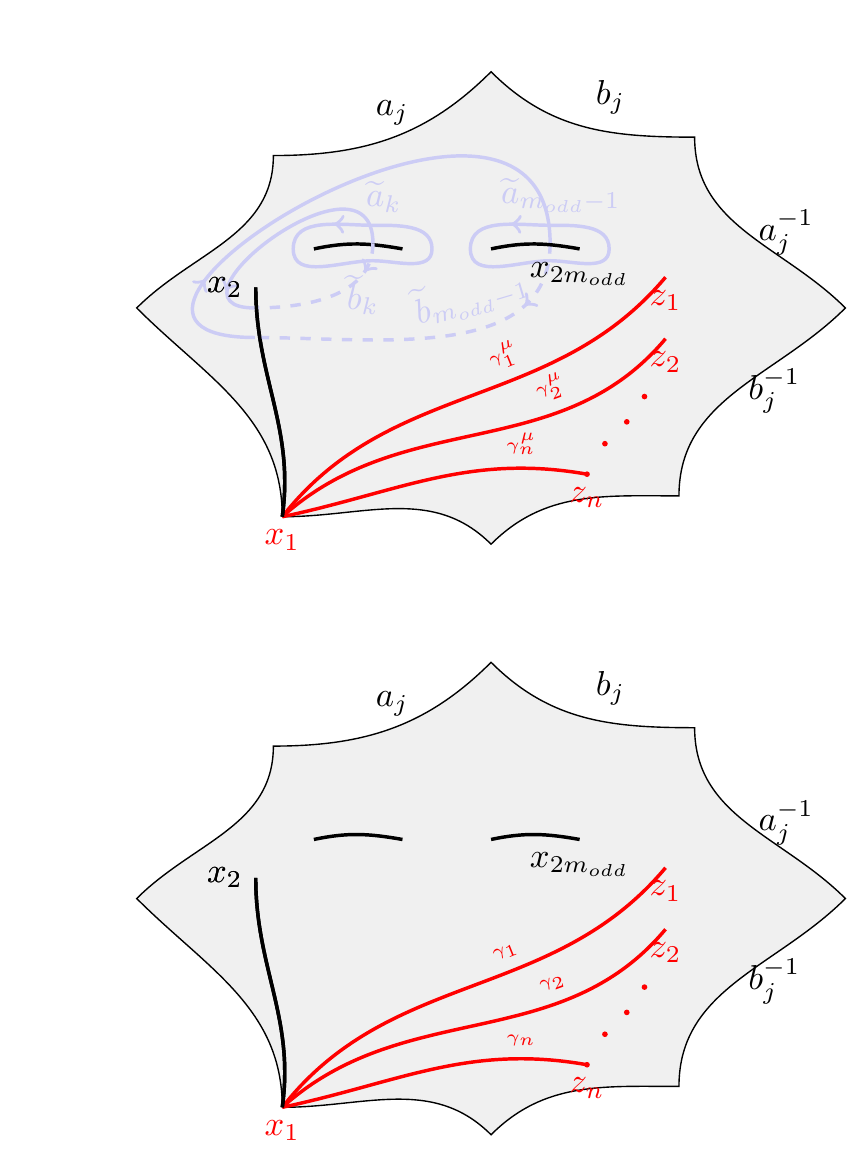}
\caption{Branch cuts and cuts $\{\gamma_i, \gamma_i^\mu\}$ connecting $x_1$  with poles of $v=\sqrt{\qd}$ on the canonical cover $\Ch$.}
\la{Cancover}
\end{figure}

\begin{definition}
For a given choice of Torelli marking ${\bf t}$ and cuts $\{\gamma_i\}$ the tau-function $\tau_+$  is given by the following expression which is independent of $x\in\CC$:
\be
\tau_+(\CC,\qd,{\bf t},\{\gamma_i\})=\Ccal^{2/3}(x)\left(\f{\qd(x)}{\prod_{i=1}^{\nz}E^{d_i}(x,q_i)}\right)^{(g-1)/6} 
\left(\prod_{i<j} E(q_i,q_j)^{d_i d_j}\right)^{1/24}\,e^{-\f{\pi i}{6} \langle\O {\rb},{\rb}\rangle-\f{2\pi i}{3}\langle{\rb},K^x\rangle }
\la{taupfor}
\ee
where $\sum_{i=1}^{\nz} d_i q_i=(\qd)$ is the divisor of quadratic differential $\qd$ on $\CC$.
\end{definition}

In analogy to Sec.3 of \cite{JDG} one can show that the expression (\ref{taupfor}) is independent of $x$.
Namely, since the product of prime-forms in the denominator of (\ref{taupfor}) compensates all poles and zeros of $\qd$,
it is sufficient to verify that all holonomy factors of $\tau_+$ are powers of unity. This can be checked using (\ref{defZh}) 
and transformation of   $E(x,y)$ (see p.4 of \cite{Fay92})  and  $\Ccal$ (see p.9  of \cite{Fay92}) between the opposite sides of the fundamental polygon:
\be
E^2(x+a_j,y)= E^2(x,y)\;\hskip0.7cm
E^2(x+b_j,y)= E^2(x,y)e^{-2\pi i \Omega_{jj}-4\pi i (\Acal_j)|_x^y}\;,
\ee
\be
\Ccal(x+a_j)=\Ccal(x),\;\hskip0.7cm
\Ccal(x+b_j)=\Ccal(x)e^{-\pi i \Omega_{jj}-2\pi i (g-1) K_j^x}\;.
\ee

The following properties of $\tau_+$ can be proved in parallel  to properties of the tau-function on spaces of 
holomorphic \cite{JDG} and meromorphic Abelian differentials.

\begin{proposition}\la{homplus}
The tau-function $\tau_+$ has the quasi-homogeneity property $\tau_+(\CC,\e \qd)=\e^{\ka_{+}} \tau_+(\CC, \qd)$ 
where
\be
\ka_+= \f{1}{48}\sum_{i=1}^{\nz}\f{d_i(d_i+4)}{d_i+2}   
\la{kapl}
\ee
\end{proposition}
{\it Proof.} The proof is based on the definition (\ref{taupfor}).  The only terms which change under the transformation $\qd\to \e\qd$ are the distinguished
local parameters $\zeta_i$ with exception of the local parameters (\ref{dist3}) near poles $z_j$ which are homogeneous in $\e$. Collecting together all powers of $\e$ we get the formula (\ref{kapl}).
An alternative proof of this proposition can be obtained using variational formulas for $\tau_+$ with
respect to homological coordinates, following \cite{Advances,JDG,contemp}.
\QED

Next we discuss the dependence of $\tau_+$ on the choice of local parameters at poles $z_j$.
Introduce the least common multiple of $d_i+2$:
\be
\alpha=LCM(d_1+2,\dots,d_{\nz}+2)\;.
\la{LCD}\ee

\begin{proposition}\la{proptauptaut}
The ($48\a$)th power of the expression
\be
\tau_+ (\CC,\qd)\left(\prod_{i=1}^n d \xi_i(z_i)\right)^{1/12}\;.
\la{defTcalp}
\ee
is independent of the choice of local parameters $\xi_i$ near $z_i$. 
\end{proposition}
{\it Proof.} The degree of differential $d\xi_i(z_i)$ in (\ref{taupfor}) can be computed as follows. 
The contribution of multiplier containing $\qd(x)$ equals $(g-1)/6$ multiplied with $-1/2$ (since $E(x,z_i)$ is in the denominator) and $-2$ (since $z_i$ enters $(\qd)$ with coefficient $-2$), which gives $(g-1)/6$.

The contribution of remaining product of prime-forms can be computed as 
$(-2)(1/2)(1/24)\sum_{(\qd)\setminus z_i} d_i$ which gives $(-1/24)(4g-2)$ (since ${\rm deg}(\qd)=4g-4$).
The sum of these two numbers equals $-1/12$ which proves the invariance of ($48\a$)th power of (\ref{defTcalp}) under a change of local coordinates near $z_j$ (i.e. it is a $4\a$-differential with respect to $\xi_i$).
\QED

\begin{proposition}\la{sympplus}
Let $\s$ be a symplectic transformation of $H_1(\CC,\Z)$  acting on canonical basis of cycles as follows:
 \be
\left(\ba{c} {\bf b}^\s \\ {\bf a}^\s\ea\right)=\s \left(\ba{c} {\bf b} \\ {\bf a} \ea\right)\;,\hskip0.7cm
\s=\left(\ba{cc} A & B  \\ C & D \ea\right)
\la{Shat}
\ee
Then the tau-function $\tau_+$ transforms under the action of $\s$ as follows:
\be
\f{\tau_+^\s}{\tau_+}=\gamma (\s) \det(C \O + D)
\la{tausigma}
\ee
where $\gamma^{48\alpha}(\s)=1$.
\end{proposition} 
{\it Proof.} Recall the transformation of the vector ${\bf v}$ of normalized differentials, the period matrix, the vector of Riemann constants , the differential $\Ccal(x)$ and the prime-form under symplectic transformation
(\cite{Fay92},  Lemma1.5 and formulas (1.19), (1.20), (1.23)). Denote $M={}^t\!(C\Omega+D)$;
then ${\bf v}^\s=M^{-1} {\bf v}$ and
$$\Omega^\s= (A\Omega+B)(C\Omega+D)^{-1}\;,$$
\be 
{E^{\s}}^2(x,y)=E^2(x,y)\exp\left\{2\pi i \langle(C\Omega+D)^{-1}C \Acal_x^y,\Acal_x^y\rangle\right\}\;.
\la{transE}
\ee
Introduce  two vectors whose entries equal to $0$ or $1/2$:
\be
\a_0=\f{1}{2}(C^tD)_0\;,\hskip0.7cm 
\b_0= \f{1}{2}(A^tB)_0
\la{ab0}
\ee
where index $0$ denotes the vector consisting of  diagonal entries of the matrix module $\Z^g$. Then
\be
(K^x)^\s=M^{-1} K^x+ \O^\s\a_0+\b_0\;;
\la{transK}
\ee
$$
\Ccal^{\s}(x)=\g'{\rm det}^{3/2}(C\O+D) \Ccal(x)
$$
\be
\times{\rm exp}\left\{-\pi i \,^t\alpha_0\Omega^\s\alpha_0 +\pi i ^t K^x(C\O+D)^{-1}C  K^x -2\pi i \,^t \a_0(C\O+D)^{-1} K^x\right\}
\la{transC}
\ee
where $(\g')^8=1$.

Substituting these transformations into (\ref{taupfor}) we get (\ref{tausigma}). The root of unity of degree $48\a$
appears due to ambiguity in the definition of distinguished local parameters at points of $\qd$ not 
coinciding with $\{z_j\}$. In the process of computation it is convenient to assume that the fundamental polygon is
chosen such that $\rb={\bf s}=0$ (the proof that this choice is always possible for $n\geq 1$ is identical to the proof of Lemma 6 
of \cite{JDG}); then $\rb^\s=-2\a_0$ and ${\bf s}^\s=2\b_0$.

Notice that the moduli-dependent term ${\rm det}(C\Omega+D)$ in (\ref{tausigma}) can also be obtained using variational formulas for $\tau_+$, similarly to (4.23)-(4.25) of \cite{CMP}. However, to prove that the constant multiplier $\gamma$ is a root of unity of degree $48\a$ one needs to use the explicit formula (\ref{taupfor}). \QED

The transformation (\ref{tausigma}) can be equivalently rewritten in terms of the basis $\{a_j^+,b_j^+\}$  in $H_+$ given by (\ref{abp}).
The period matrix $\O^+=2\O$ according to (\ref{OOp}) while the symplectic ($\Z/2$) matrix $\sigma_+$ acting in this basis is
related to the matrix $\s$ (\ref{Shat}) by
$$\s_+=\left(\ba{cc} A_+ & B_+  \\ C_+ & D_+ \ea\right)= \left(\ba{cc} A & 2B  \\ C/2 & D \ea\right)\;.$$
Thus $\det(C\O+D)= \det(C_+\O_++D_+)$ and transformation (\ref{tausigma}) can also be written as
\be
\f{\tau_+^\s}{\tau_+}=\gamma (\s) \det(C_+ \O_+ + D_+)\;.
\la{tausigmaplus}
\ee

\subsection{Tau-function $\tauh$}

 Denote by  $\Ccalh$ the differential
(\ref{defCint}) corresponding to the Riemann surface $\Ch$. Denote the Abel map on $\Ch$ corresponding to initial point $x\in \Ch$ by $\Acalh$, the prime-form on $\Ch$ by $\Eh$ and introduce two vectors  $\hat{{\rb}},\hat{{\sb}}\in \Z^{\gh}$ via relation
\be
\Acalh_x((v))+2\Kh^x+\Oh\hat{\rb}+\hat{\sb}=0
\la{defZh1}
\ee

In analogy to (\ref{defEp}), (\ref{defEpp}) we also define
\be
\Eh(x, \ph_i)=\lim_{y\to \ph_i} \Eh(x,y) \sqrt{d\zetah_i(y)}
\la{defEph}
\ee
\be
 \Eh(\ph_i,\ph_j)=\lim_{x\to \ph_j, y\to \ph_i} E(x,y) \sqrt{d\zetah_i(y)} \sqrt{d\zetah_j(x)}
\la{defEpph}
\ee
where $\zetah_i$ is the distinguished local parameter on $\Ch$ near $\ph_i$ given by (\ref{dist20}).

\begin{definition}
For a given Torelli marking $\hat{\bf t}=({\bf \ah}, {\bf \bh})$ of $\Ch$ and the set of contours $\hat{\Lcal}$ connecting $x_1$ with 
points $\{\zh_i,\zh_i^\mu\}$ 
the tau-function $\tauh$ is defined by the following expression independent of $x\in\Ch$:
\be
\tauh(\CC,\qd,\hat{\bf t},\{\gamma_i\}_{i=1}^n)=\Ccalh^{2/3}(x)\left(\f{v(x)}{\prod_{i=1}^{\nzh}\Eh^{\dh_i}(x,\ph_i)}\right)^{(\gh-1)/3} 
\left(\prod_{i<j} \Eh(\ph_i,\ph_j)^{\dh_i \dh_j}\right)^{1/6}\,e^{-\f{\pi}{6} \langle\Oh \hat{{\rb}}, \hat{{\sb}}\rangle-\f{2\pi i}{3}\langle\hat{{\rb}},\Kh^x\rangle }
\la{tauhfor}
\ee
where $\sum_{i=1}^{\nzh} \dh_i \ph_i=(v)$ is the divisor of abelian differential $v$ on $\Ch$. 
\end{definition}

\begin{proposition}
%\la{homhat}
The tau-function $\tauh$ has the quasi-homogeneity property $\tauh(\CC,\e \qd)=\e^{\kh} \tauh(\CC, \qd)$ 
where
\be
\kh= \f{1}{24}\sum_{i=1}^{\nzh}\f{\dh_i(\dh_i+2)}{\dh_i+1} 
\la{kaph}
\ee
\end{proposition}
{\it Proof}  is a corollary of the definition   (\ref{tauhfor}). The only terms which change under the transformation $\qd\to \e\qd$ are the distinguished
local parameters $\zetah_i$. Collecting together all powers of $\e$ we get the formula (\ref{kaph}). \QED

\begin{proposition}\la{propthtavt}
The $48\a$\,th power of the expression
\be
\tauh (\CC,\qd)\left(\prod_{i=1}^n d \xi_i(z_i)\right)^{1/6}\;.
\la{defTcalh}
\ee
is  invariant under the choice  of local parameters $\xi_i$ near $z_i$.
\end{proposition}
{\it Proof.}
 The differential $d\xi_i(z_i)$ enters the expression (\ref{taupfor}) with  degree 
which can be computed as follows. The degree $\f{\gh-1}{3}$ arises as contribution of the  multiplier containing $v(x)$:
there are two prime-forms in the denominator which contribute $d\xi_i(z_i)$ each with power $(-1)(1/2)$ (since corresponding $\dh_i=-1$).

The degree $(1/6)[1+\sum_{(v)\setminus \{\zh_i,\zh_i^\mu\}} \dh_i]$ comes from the last product in (\ref{tauhfor}).
Here $1$ is the contribution of $E(\zh_i,\zh_i^\mu)^{(-1)(-1)}$. The remaining sum is the contribution of other products 
of prime-forms where one of the arguments coincides with $\zh_i$ or $\zh_i^\mu$. Since ${\rm deg}(v)=2\gh-2$, this gives 
$1/6(1-2\gh)$.

In total the degree of $d\xi_i$ equals $-1/6$ which shows that $48\a$th power of (\ref{defTcalp}) is invariant under the change of local coordinates near $z_j$ (i.e. it is $8\a$-differential with respect to each $\xi_i$).\QED

\begin{proposition}\la{symphat}
Let $\sh$ be a symplectic transformation of $H_1(\CC,\Z)$  acting on canonical basis of cycles as follows:
 \be
\left(\ba{c} {\bf \bh}^\s \\ {\bf \ah}^\s\ea\right)=\sh \left(\ba{c} {\bf \bh} \\ {\bf \ah} \ea\right)\;,\hskip0.7cm
\sh=\left(\ba{cc} \Ah & \Bh  \\ \Ch & \Dh \ea\right)
%\la{Shat}
\ee
Then the tau-function $\tauh$ transforms under the action of $\sh$ as follows:
\be
\f{\tauh^{\sh}}{\tauh}=\gamma (\sh) \det(\Ch\Oh+ \Dh)
\la{tausigmahat}
\ee
where $\gamma^{48\alpha}(\sh)=1$.
\end{proposition} 
{\it Proof.} The transformation (\ref{tausigma}) is derived directly from (\ref{tauhfor}) using transformation formulas (\ref{transE}), (\ref{transK}), (\ref{transC}) in the case of canonical covering $\Ch$. The computation repeats the computation required to prove Prop.\ref{sympplus}.
\QED

\subsection{Prym tau-function $\tau_-$}
\label{prymsect}

\begin{definition}
Let $\{a_j,b_j\}_{j=1}^g$ be a canonical basis of cycles on $\CC$ and $\{\ah_j\bh_j\}_{j=1}^{\gh}$ be an associate basis
of cycles (\ref{mainbasis}) on $\Ch$. 
Then the Prym tau-function $\tau_-$ on the space $\Qcal_{g,n}^{\bk,\bl}$ 
is defined as ratio of $\tauh$ (\ref{tauhfor}) and $\tau_+$ (\ref{taupfor}):
\be
\tau_-=\frac{\tauh(\CC,\qd,\hat{{\bf t}},\{\gamma_i\}_{i=1}^n)}{\tau_+(\CC,\qd,{\bf t},\{\gamma_i\}_{i=1}^n)}\;.
\la{taum}
\ee
\end{definition}
The following properties of $\tau_-$ can be derived from the properties of tau-functions $\tau_+$ and $\tauh$.

\begin{proposition}\la{sympmin}
Let $\sh$ be a symplectic transformation of $H_1(\Ch,\Z)$ commuting with $\mu_*$, and let $\s_-$
be the matrix representing $\sh$ in the subspace $H_-$ such that
\be
  \left(\ba{c} {\bf b_-} \\ {\bf a_-}\ea\right)^\s =\s_- \left(\ba{c} {\bf b_-} \\ {\bf a_-} \ea\right)\;,\hskip0.7cm
\s_-=\left(\ba{cc} A_- & B_-  \\ C_- & D_- \ea\right)\;.
 \ee 
Then the tau-function $\tau_-$ transforms under the action of $\s$ as follows:
\be
\f{\tau^{\s^-}_-}{\tau_-}=\gamma_- \det(C_- \O_- + D_-)
\la{tausigmam}
\ee
(where according to (\ref{Prymmat}) $\O_-=2\Pi$ with $\Pi$ being the Prym matrix) and $\gamma_-^{48\alpha}=1$.

\end{proposition} 

{\it Proof.} 
The transformation (\ref{tausigmam}) of $\tau_-$ can be deduced from
transformation (\ref{tausigma}) of $\tau_+$ and (\ref{tausigmahat}) of $\tauh$.
Namely, any symplectic transformation $\hat{\sigma}$, commuting with $\mu_*$,
has in the basis (\ref{mainbasis}) the form 
\be
\left(\ba{c} \hat{{\bf b}}\\ \hat{{\bf a}}  \ea\right)^{\hat{\sigma}}= \hat{\sigma}
\left(\ba{c} {\bf b}\\ {\bf a}\ea\right)\;,\hskip0.7cm
\hat{\sigma}= \left(\ba{cc} T^{-1} & 0\\ 0 & T \ea\right)
\left(\ba{cccc} A_+ & 0 & B_+ & 0 \\ 
0 & A_- & 0 & B_- \\
 C_+ & 0 & D_+ & 0 \\
 0 & C_- & 0 & D_-
 \ea\right)
\left(\ba{cc} T & 0\\ 0 & T^{-1} \ea\right)
\ee
where 
\be
\sigma_\pm = \left(\ba{cc} A_\pm & B_\pm \\ C_\pm & D_\pm \ea\right)
\ee
are symplectic matrices acting on bases $({\bf b}_\pm, {\bf a}_\pm)$  given by (\ref{abp}),  (\ref{abm1}), (\ref{abm}).
 Then, since due to (\ref{shs}), 
 $$
 \widehat{\Omega}=
%\frac{1}{2}???? 
T^{-1}
 \left(\ba{cc} \Omega^+ & 0\\
                0  & \Omega^-  \ea\right) T^{-1}\;,
 $$
we get
\be
{\rm det}(\Ch\Oh+ \Dh)= {\rm det}(C_+\O_+ + D_+)\,
{\rm det}(C_-\O_- + D_-)
\ee 
which gives (\ref{tausigmam}) as a corollary of (\ref{tausigma}) and  (\ref{tausigmahat}).
\QED

Therefore, the Prym tau-function defined by (\ref{taum}) depends in fact on the following data:  $(\CC,\qd,{\bf t}_-,\{\gamma_i\}_{i=1}^n)$ where ${\bf t}_-$ is a choice of canonical basis $(a_j^-,b_j^-)$ in the space $H^-$ of 
holomorphic Prym differentials.

The following two properties of $\tau_-$ are parallel to the properties of $\tau_+$ and $\tauh$.

\begin{proposition}
%\la{homhat}
The tau-function $\tau_-$ has the quasi-homogeneity property 
$$\tau_-(\CC,\e \qd,\hat{{\bf t}},\{\gamma_i\}_{i=1}^n)=\e^{\kappa_-} \tau_-(\CC, \qd,\hat{{\bf t}},\{\gamma_i\}_{i=1}^n)$$
where
\be
\kappa_-= \ka_+
+\f{1}{8}\sum_{i=1}^\mo \f{1}{d_i+2}
\la{kam}
\ee
and $\kappa_+$ is given by (\ref{kapl}).
\end{proposition}

{\it Proof.} We have $\kappa_-=\kh-\kappa_+$. Using expressions (\ref{kaph}) and (\ref{kapl})  for $\kappa_+$ and $\kh$ we get (\ref{kam}).\QED

\begin{proposition} \la{proptaum}
The $48\a$\,th power of the expression
\be
\tau_- (\CC,\qd)\left(\prod_{i=1}^n d \xi_i(z_i)\right)^{1/12}\;.
\la{defTcalmin}
\ee
is independent of the choice of local parameters $\xi_i$ near $z_i$.
\end{proposition}

{\it Proof.} This proposition is an immediate corollary of Props. \ref{proptauptaut}  and \ref{sympplus}. \QED

Finally, we discuss the dependence of $\tau_\pm$ on the choice of cuts $\{\g_i\}_{i=1}^n$ used to define the local parameters $\xi_i$ near $z_i$.
\begin{proposition}
Under a change of integration contours  used to define distinguished local coordinates $\xi_i$ near $z_i$ 
(i.e. under a change of cuts defining the fundamental polygon $\CC'$ and cuts $\gamma_i$ between $x_1$ and $z_i$)
the tau-functions $\tau_\pm$ transform as follows:
\be
\tau_\pm(\CC,\qd,{\bf t_\pm},\{\gamma'_i\}_{i=1}^n) = \tau_\pm(\CC,\qd,{\bf t_\pm},\{\gamma_i\}_{i=1}^n) \,\exp\left\{  -\f{\pi i}{6}\sum_{j=1}^n\sum_{i=1}^{g_-+n}A_{ij}\f{\Pcal_{s_i}}{p_j}\right\}
\la{transtau}
\ee
where the set $\{s_i\}$ contains cycles $\{ a_l^-,  b_l^-/2,\;\;l=1,\dots,g$ and
$a_l^-,b_l^-, \;l=g+1,\dots,g_-\}$ 
(notations here follow (\ref{abm1}),(\ref{abm})) as well as $\f{1}{2}(l_j-l_j^\mu)$ where $l_j$ is a small loop around $z_j$;
 $A_{ij}$ is a matrix with half-integer entries. 
\end{proposition}
{\it Proof.} 
A change of integration  contours $\g_i$ in the formulas (\ref{taupfor}), (\ref{taum}),
(\ref{tauhfor}) for $\tau_\pm$
obviously leads to appearance of an exponential factor of the form  (\ref{transtau}).

 If the "first" zero is not changed then the corresponding  matrix $A_{ij}$ has integer entries. If $x_1$ is changed to some other zero $x_2$ then the integrals get an additional contribution of the form $\int_{x_1}^{x_2} v$ with can be expressed as a half-integer combination of the cycles $s_i$. Therefore in general the matrix $A$ can have half-integer entries.
\QED

\begin{remark}\rm
Relation (\ref{kam}) is analogous to the formula (2.4) of Theorem 2 of \cite{EKZ} (for the case of holomorphic $\qd$) which relates sums of Lyapunov exponents corresponding to Hodge and Prym vector bundles. This indicates that the tau-function $\tau_-$ has a close relationship to the action of the Teichm\"uller flow on Prym vector bundle, similarly to the role of $\tau_+$ in the study of the 
Teichm\"uller flow on Hodge vector bundle, as described in Sect. 5.2 of \cite{EKZ}.
\end{remark}

\begin{remark}\rm
In \cite{contemp} it was considered the case when all $d_i=1$ (i.e. $\qd$ is a holomorphic quadratic differential with simple zeros). Then $\nz=4g-4$ and the formulas (\ref{kapl}), (\ref{kam}) give
 $\ka_+=\f{5}{36}(g-1)$ and $\ka_-=\f{11}{36}(g-1)$. These numbers differ from $p_+$ and $p_-$ from Theorem 2 of \cite{contemp} by a factor of $48$; this is due to the fact that  the tau-functions $\tau_\pm$ from \cite{contemp}
 equal to the 48th power of $\tau_\pm$ used in this paper. The function $\tauh$ from \cite{contemp}   corresponds to 24th power of $\tauh$ used here.
\end{remark}

\subsection{Hodge and Prym classes over $\Qcal_{g,n}$ and $\tau_\pm$}
\la{PrymonQ}

The Hodge vector bundle $\Lambda_H$ over  $\Qcal_{g,n}$ is defined by the pullback of the 
Hodge vector bundle over $\Mcal_{g,n}$ (the fiber of $\Lambda_H$ is the space of holomorphic abelian differentials on $\CC$).

To define the Prym vector bundle $\Lambda_P$ over $\Qcal_{g,n}$ (following \cite{contemp,KSZ}) we consider first  the subspace $\Qcal_{g,n}^{0}$
of $\Qcal_{g,n}$ such that 
all zeros of $\qd$ are simple; this is the largest stratum of $\Qcal_{g,n}$.
The fiber of  $\Lambda_P$  over $\Qcal_{g,n}^0$ coincides with $H_-$.
 At those boundary components of $\Qcal_{g,n}^{0}$ where some zeros of $\qd$ become
multiple, the fiber of $\Lambda_P$ is obtained by decomposing the fiber of the Hodge vector  bundle
over $\overline{\Mcal}_{\gh}$ over the nodal curve $\Ch$ into $H_+\oplus H_-$. As a result $H_-$ consists of meromorphic 
differentials $u$ with poles of order $k_i$ at the branch point $x_i$ for odd $k_i$ and poles of order $k_i/2$ at
 points $(\hat{x}_i,\hat{x}_i^\mu)$ for a zero $x_i$ of even order $k_i$. 
 In this way one gets a well-defined vector bundle $\Lambda_P$ over the whole space $\Qcal_{g,n}$ \cite{KSZ}.
 \\[5pt]

Let us denote by $\Lambda_2^{(n)}$ the bundle over $\Mcal_{g,n}$ whose fibers consist of meromorphic quadratic differentials  with at most simple poles at $z_1,\dots,z_n$ ($\Lambda_2^{(n)}$ is, in fact, the cotangent bundle over $\Mcal_{g,n}$).

The following theorem is similar to the analogous statement from \cite{contemp,KSZ}:
\begin{proposition}\la{PrymQuad}
The fiber of the  Prym vector bundle $\Lambda_P$ over $\Qcal_{g,n}^0$ above the point $(\CC,\qd)$  is isomorphic to the fiber of vector bundle $\Lambda_2^{(n)}$ above the point $\CC$. The isomorphism is given by
\be
\pi^\star\tilde{\qd}=u v, \ \ \ 
\la{isom}
\ee
where $\pi:\Ch\to\CC$ is the canonical projection and $\tilde{\qd}$ is a meromorphic quadratic differential on $\CC$  with simple poles and $u\in H^-$ is a Prym differential.
\end{proposition}

{\it Proof.} Both $u,v$ are antisymmetric with respect to the involution $\mu$ and therefore they have odd--multiplicity zeroes at the branch-points, with $v$ having only simple zeros.  The product has a zero of even multiplicity on $\Ch$ at each branch-point. Denote by $\hat \zeta$ a local parameter on $\Ch$ near a branch-point, so that $v \sim \hat \zeta d\hat \zeta$ and $u \sim \hat \zeta^{2k+1} d\hat \zeta$.

Then $uv \sim  \hat \zeta^{2k+2} d\hat \zeta^2 \sim \zeta^k d \zeta^2$ with $\zeta = \hat \zeta^2$. Thus $uv$ 
is invariant under canonical involution on $\Ch$ and hence corresponds to  a  quadratic differential on $\CC$
with at most simple poles at punctures. \QED

The isomorphism can be appropriately extended to all strata of $\mathcal Q_{g,n}$ as discussed in  \cite{KSZ}.

 The following proposition is an immediate corollary of Props. \ref{proptauptaut}, \ref{sympplus}, \ref{sympmin}  and \ref{proptaum}:
\begin{proposition}\la{propTpm}
 $\tau_+^{48}(\CC,\qd)$  is a section of the line bundle ${\rm det}^{48}\Lambda_H\otimes \prod_{i=1}^n \Lcal _i^4 $ and 
 $\tau_-^{48}(\CC,\qd)$  is a section of the line bundle ${\rm det}^{48}\Lambda_P\otimes \prod_{i=1}^n \Lcal _i^4 $ 
over the space $\Qcal_{g,n}^0$.
\end{proposition}

Following \cite{Advances,MRL,contemp,KSZ} this proposition can be used to get relations between various classes
in the Picard group of $\Qcal_{g,n}$. However, in this paper we don't pursue this goal; instead we are going to 
use similar ideas in the context of combinatorial model of $\Mcal_{g,n}$.

\subsection{Differential equations for $\tau_\pm$ on $\Qcal_{g,n}^{\bk,\bl}$}

Differential equations for $\tau_\pm$ can be derived in parallel to \cite{JDG,contemp,CMP}. However, the presence of quadratic poles of $\qd$ introduces some new features. 

Consider the system of cuts $\{\gamma_i\}_{i=1}^n$ on $\CC$ and the associate system of cuts $\{\hat{\gamma}_i\}_{i=1}^n=\pi^{-1}(\{\gamma_i\}_{i=1}^n)$ on $\Ch$ used to define tau-functions $\tau_\pm$. Choose a basis $\{s_i\}_{i=1}^{g_-+m_{even}}$ in the odd subspace 
$H_-'$ of
$H_1(\Ch\setminus  \{\hat{\gamma}_i\}_{i=1}^n,\;\{\pi^{-1}(x_i)\}_{i=m_{odd}+1}^{m})$ and denote by $\{s_i^*\}_{i=1}^{g_-+m_{even}}$ 
the dual basis in ${H_-'}^*$ which is the odd subspace in $H_1(\Ch\setminus\{\hat{\gamma}_i\}_{i=1}^n\cup\{\pi^{-1}(x_i)\}_{i=m_{odd}+1}^{m}\})$.

Consider the meromorphic bidifferential $\Bh(x,y)=d_xd_y\log \Eh(x,y)$, where $\Eh$ corresponds to the same Torelli marking of $\Ch$ as in the formula (\ref{tauhfor}) for $\tauh$. 

Now put
\be
B_+ (x,y)=\Bh(x,y)+\mu_y^*\Bh(x,y)\,, \quad B_-(x,y)=\Bh(x,y)-\mu_y^*\Bh(x,y)\,,
\la{defBpm}
\ee 
(the subscript $y$ at $\mu^*$ means that we take the pullback with respect to the involution on the second factor in $\Ch\times\Ch$).
The bidifferential $B_+(x,y)$ is just the pullback of the canonical bidifferential $B(x,y)$ on $\CC\times \CC$. The bidifferential  $B_-$ was first introduced in \cite{contemp}. It is called the {\it Prym bidifferential}.

Consider the regularization of the bidifferentials $B_\pm$ near  the diagonal:
\be
B_\pm^{reg}(x,x)=\left(B_\pm(x,y)-\f{v(x)v(y)}{(\int_x^y v)^2}\right)\Big|_{y=x}
\la{Breg}
\ee
and introduce meromorphic Abelian differentials $w_v^\pm$ on $\Ch$ (anti-symmetric with respect to $\mu$) by
\be
w_v^\pm(x)=\f{B_\pm^{reg}(x,x)}{v(x)}
\la{Qvpm}
\ee
%In terms of  the  projective connections $B_\pm$ these differentials are expressed by:
%\be
%w_v^\pm(x)=\f{1}{6}\f{S_{B_\pm}-S_v}{v}
%\la{defwv}
%\ee
%where the meromorphic projective connection $S_v$ is given by Schwarzian derivative $S_v(\cdot)=\{\int^x v, \cdot\}$.

The complete set of local coordinates on $\Qcal_{g,n}^{\bk,\bl}$, besides $\{\Pcal_{s_i}\}_{i=1}^{g_-+m_{even}}$,
contains the residues $\{p_i\}_{i=1}^n$. However, we are not going to consider derivatives of $\tau_\pm$ with respect to $p_i$ since they will be kept fixed in the context of combinatorial model of $\Mcal_{g,n}$. The derivatives with respect to $\Pcal_{s_i}$ are given by the following proposition.

\begin{proposition}
The tau-functions $\tau_\pm$ satisfy the following equations:
\be
\f{\p\log\tau_\pm}{\p \Pcal_i}= -\f{1}{4\pi i} \int_{s_i^*} w_v^\pm\;,\hskip0.7cm i=1,\dots,g_-+m_{even}
\la{deftaupm}
\ee
\end{proposition}

{\it Proof.} The formula (\ref{deftaupm}) is a corollary of two facts.
The first is the variational formula for the Bergman tau-function on spaces 
of third kind abelian differentials given by  formula (4.4) of \cite{CMP}
(this formula is proved in parallel to the case of spaces of holomorphic abelian differentials treated in detail in \cite{JDG}). The second fact is the relationship between 
variational formulas for tau-functions on spaces of Abelian and quadratic differentials
which was established in Section 4.1 of \cite{contemp}.
\QED

\subsection{Examples.}

The following  two examples play an important role in the sequel since they allow to analyze the local behaviour of the tau-functions near different boundary components of the moduli spaces.

\subsubsection{Space $\Qcal_{0,0}^{1,1,1,-7}$}
\label{Q01117}
Here we consider  the moduli space of quadratic differentials  on Riemann sphere and $n=0$ 
with the divisor
$$(\qd)=x_1+x_2+x_3-7 x_4\;$$
(this notation is motivated by application to the combinatorial model). Then $\gh=1$.
The pole $x_4$ of multiplicity $7$ will be 
put to infinity of complex plane with coordinate $x$ while  $x_1$, $x_2$ and $x_3$ are positions of zeros in $x$-plane.
Using the rescaling and shift of coordinate $x$ we can then write
\be
\qd(x)=(x-x_1)(x-x_2)(x-x_3)(dx)^2
\la{qex1}\ee
for $x_1+x_2+x_3=0$
and
\be
v(x)=[(x-x_1)(x-x_2)(x-x_3)]^{1/2} dx
\la{vex1}\ee
such that on $\Ch$ we have $(v)=2x_1+2 x_2+2x_3-6\infty$.

{\bf Hodge tau function $\tau_+$.}
The distinguished local parameters on $\C P^1$ near points $x_i$ are given by $\zeta_i(x)=\left[\int_{x_i}^x v\right]^{2/3}$ while
the distinguished local parameter near $y_1=\infty$ equals $\zeta_\infty(x)=\left[\int_{x_1}^x v\right]^{-2/5}$.
Then, say, for $\zeta_1$ we have
\be
\f{d\zeta_1(x)}{dx}\big|_{x=x_1}=(2/3)^{-1/3}[(x_1-x_2)(x_1-x_3)]^{1/3}
\la{zet1x}
\ee
while $\f{\d\zeta_\infty(x)}{d(1/x)}\big|_{x=\infty}=(2/5)^{2/5}$. In the sequel we shall omit the unessential multiplicative
 constants  since all tau-functions are  defined up to  moduli-independent multiplicative  factors. 
For the prime-form we have in genus zero: $E(x,y)=\f{x-y}{\sqrt{dx}\sqrt{dx}}$.
Then, say,
\be
E(x,x_1)=const\,\f{(x-x_1)}{\sqrt{dx}} [(x_1-x_2)(x_1-x_3)]^{1/6}\;\hskip0.7cm
E(x,\infty)= const\,\f{1/x}{\sqrt{d(1/x)}}
\ee
and similar expressions for $E(x_i,x_j)$ and $E(x_i,\infty)$. 
The formula (\ref{taupfor}) for the function $\tau_+$ takes the form:
\be
\tau_+=\f{\qd^{-1/6}(x)\prod_{i=1}^3 E^{1/6}(x_i,x)}{ E^{7/6}(\infty,x)} \f{\prod_{i<j} E^{1/24}(x_i,x_j)}{\prod_{i=1}^3 E^{7/24}(x_i,\infty)}\;.
\ee
Substituting the expressions for the prime-forms computed in distinguished local parameters into this formula  we get
\be
\tau_+= [(x_1-x_2)(x_2-x_3)(x_3-x_1)]^{1/36}\;.
\la{tau111m7}
\ee

{\bf Prym tau function $\tau_-$.}
To get an expression for $\tau_-$ we shall first write down the formula for $\tauh$ and then find $\tau_-$ from relation $\tauh=\tau_+\tau_-$. The function $\tauh$ is in this case coincides with the Bergman
 tau-function on the space of Abelian differentials of the form (\ref{vex1}) i.e. on the stratum $\Hcal_1^{[2,2,2,-6]}$ of meromorphic Abelian differentials of second kind \cite{CMP}. The formula for $\tauh$ is given by (\ref{tauhfor}). 
 
 Equivalently and more conveniently,  we can use
 the formula (3.29) 
(and the next formula after that) of \cite{JDG} which describe dependence of $\tau_+$ on the choice of Abelian differential while keeping the Riemann surface fixed..
If we adjust this formula to the present situation we choose 
\be
v_0=\f{dx}{\sqrt{(x-x_1)(x-x_2)(x-x_3)}}
\la{vt}
\ee
which is a differential without zeros or poles. In that case $\tauh(\Ch,v_0)=\eta^2(\o_2/\o_1)$ \cite{JDG} where $\o_1$ and $\o_2$ are periods of the differential $v_0$:
\be
\o_1=\int_a v_0\;,\hskip0.7cm \o_2=\int_b v_0
\la{o12}\ee
and $\eta$ is the Dedekind's eta-function.
Then for an arbitrary differential $v$ of second kind with divisor $(v)=\sum_{i} m_i x_i$ an appropriate analog of  the formula (3.29) of
\cite{JDG} looks as follows:
\be
\tauh(\Ch,v) = \eta^2(\o_2/\o_1) \prod_{i} \left(\f{v_0}{d\zetah_i}\right)^{-m_i/12}
\la{Polyakov}
\ee
where,  $\zetah_i(x)=\left[\int_{x_i}^x v\right]^{1/(d_i+1)}$ is the {\it distinguished local parameter on $\Ch$}  near $x_i$ defined by $v$.

If $v$ is given by (\ref{vex1}) then 
\be
\f{v_0}{d\zetah_1}\Big|_{x=x_1}=const\, [(x_1-x_2)(x_1-x_3)]^{-2/3}
\la{v0xi1}
\ee
and 
\be
\f{v_0}{d\zetah_\infty}\Big|_{x=\infty}=const \;.
\la{v0xiinf}
\ee
Substituting (\ref{v0xi1}), (\ref{v0xiinf}) (as well as similar formulas at $x_{2,3}$) into (\ref{Polyakov}) we get
\be
\tauh=\eta^2(\o_2/\o_1) [(x_1-x_2)(x_1-x_3)(x_2-x_3)]^{2/9}
\la{tauh3}
\ee
and, dividing this expression by $\tau_+$ (\ref{tau111m7}), 
\be
\tau_-= \eta^2(\o_2/\o_1)[(x_1-x_2)(x_1-x_3)(x_2-x_3)]^{7/36}\;.
\la{taum3}
\ee
Notice that the homogeneity coefficients $\hat{\kappa}= 2/15$ and $\ka_-=7/60$ are in agreement with the ones given by the general formulas (\ref{kaph}), (\ref{kam}).

Finally, using the standard expression for the $\eta$-function via $w_1$ and the discriminant:
\be
\eta^{24}(\o_2/\o_1)=\frac{\o^{12}_1}{(2\pi)^{12}}[(x_1-x_2)(x_1-x_3)(x_2-x_3)]^{2}
\la{etadelta}
\ee
we get, up to a non-essential constant
\be
\tau_-= \o_1\,[(x_1-x_2)(x_1-x_3)(x_2-x_3)]^{13/36}
\la{taum33}
\ee
or, equivalently,
$$
\tau_-=\omega_1\,\tau_+^{13}\;.
$$

\subsubsection{Space $\Qcal^{1,1,-3,-3}_{0,0}$}
\label{Q001133}
Here we put $g=0$, $n=0$ and choose the divisor $(\qd)=x_1+x_2-3 x_3-3 x_4$.
Assuming $x_3=0$ and $x_4=\infty$, we have
\be
\qd=\f{(x-x_1)(x-x_2)}{x^3}(dx)^2
\la{q2z2p}
\ee
and 
\be
v(x)=x^{-3/2}[(x-x_1)(x-x_2)]^{1/2} dx\;.
\la{v1133}\ee
The poles $y_1$ and $y_2$ of order 3 are located at $x=0$ and $x=\infty$.

{\bf Hodge tau-function $\tau_+$.}
The distinguished local parameters on $\C P^1$ at $x_1$ and $x_2$ are given by  $\xi_i(x)=\left[\int_{x_i}^x v\right]^{2/3}$ and
$$
\f{d\xi_1}{dx}\Big|_{x=x_1}=const\,\mu_0^{1/3}\f{(x_1-x_2)^{1/3}}{x_1}\;,\hskip0.7cm
\f{d\xi_2}{dx}\Big|_{x=x_2}=const\,\mu_0^{1/3}\f{(x_1-x_2)^{1/3}}{x_2}\;.
$$
The distinguished local parameter at $x=0$ is given by $\zeta_0(x)=[\int_{x_1}^x v]^{-2}$ such that 
$\f{d\zeta_0(x)}{dx}\Big|_{x=0}=const (\mu_0 x_1 x_2)^{-1}$. Finally, the  distinguished local parameter at $x=\infty$ is
$\zeta_\infty(x)= [\int_{x_1}^x v]^{-2}$ such that $\f{d\zeta_\infty(x)}{dx}\Big|_{x=\infty}=const \mu_0^{-1}$.

The expression (\ref{taupfor}) for $\tau_+$ takes the form:
\be
\tau_+=\qd^{-1/6}(x)\f{[E(x_1,x)E(x_2,x)]^{1/6}}{[E(0,x)E(\infty,x)]^{1/2}}
\f{ E^{1/24}(x_1,x_2)E^{3/8}(0,\infty)}{[E(x_1,0)E(x_1,\infty)E(x_2,0)E(x_2,\infty)]^{1/8}}
\la{tau2pm3}
\ee
which gives
\be
\tau_+=(x_1 x_2)^{1/12} (x_1-x_2)^{1/36}\;.
\la{taup1133}
\ee

{\bf Prym tau-function  $\tau_-$.} Here we are going to use again the formula (\ref{Polyakov}) with $v_0$ given by the standard holomorphic differential (\ref{vt}) to compute $\tauh(v,\CC)$, that is, the Bergman tau-function on the stratum $\Hcal_1^{[2,2,-2,-2]}$ of the space of Abelian meromorphic differentials. The distinguished local coordinates near the points $x=x_{1,2}$, $x=0$ and $x=\infty$ of the divisor $(v)$ are: 
\be
\zetah_i(x)=\left[\int_{x_i}^x v\right]^{1/3}\;,\hskip0.7cm 
\zetah_0(x)=\left[\int_{x_1}^x v\right]^{-1}\;,\hskip0.7cm 
\zetah_\infty(x)=\left[\int_{x_1}^x v\right]^{-1}\;\hskip0.7cm 
\ee
so that
\be
\f{v_0}{d\zetah_i}\Big|_{x=x_i}=const\;  (x_1-x_2)^{-2/3}\;.
\la{v0xi11}
\ee
Verification  of formula (\ref{v0xi11}) goes as follows: as $x\to x_1$ we have $v_0\sim [x_1(x_1-x_2)]^{-1/2}\frac{dx}{(x-x_1)^{1/2}}$  and
$$d\zetah_1(x)\sim \left(\frac{(x_1-x_2)^{1/2}}{x_1^{3/2}}\right)^{3/2}((x-x_1)^{2/3})^{3/2}\frac{(x-x_1)^{1/2} dx}{x_1^{3/2} (x_1-x_2)^{-1/2}}
=(x_1-x_2)^{1/6}x_1^{-1/2}\frac{dx}{(x-x_1)^{1/2}}$$
with the ratio leading to (\ref{v0xi11}).

Similar computation gives
\be
\f{v_0}{d\zetah_0}\Big|_{x=0}=const\, ; \;\hskip0.7cm 
\f{v_0}{d\zetah_\infty}\Big|_{x=\infty}=const
\la{v0xi12}
\ee
such that the formula  (\ref{Polyakov}) with $m_1=m_2=2$ and $m_0=m_\infty=-2$ gives
\be
\tauh= \eta^2(\o_2/\o_1) (x_1-x_2)^{2/9}
\la{tauh113}
\ee
where, as before, $\o_1$ and $\o_2$ are $a$- and $b$- periods of the differential $v_0$, respectively, and
\be
\tau_-=\frac{\tauh}{\tau_+}= \eta^2(\o_2/\o_1) (x_1-x_2)^{7/36} (x_1x_2)^{-1/12}
\la{tauh1133}
\ee
In this case  the formula expressing the Dedekind's eta-function in terms of branch points $0,x_1,x_2$ and the Abelian integral $\o_1$ looks as follows:
\be
\eta^{24}(\o_2/\o_1)=\frac{\o^{12}_1}{(2\pi)^{12}}[x_1 x_2(x_1-x_2)]^{2}
\ee
and (\ref{tauh1133}) becomes
\be
\tau_-= \o_1\, (x_1 x_2)^{1/12} (x_1-x_2)^{13/36}
\la{tauh1134}
\ee
or
$$
\tau_-=\frac{ \omega_1}{x_1 x_2} \tau_+^{13} \;.
$$
 
\section{Hodge and Prym tau-functions in the combinatorial model of $\Mcal_{g,n}$ }
\label{final_res}

\subsection{Monodromy of circle bundle and Poincare dual of the first Chern class}

We  recall a few well-known facts (see for example \cite{GrifHar} for details). An effective way to compute the Poincar\'e dual to the first Chern class of a 
complex line bundle $\Lcal$ over a {\it smooth (oriented) manifold} $M$ of dimension $N$ goes as follows. Let $f$  be a  section of 
$\Lcal$ which is smooth and non-vanishing outside of several oriented connected  submanifolds $M_1,\dots,M_k$ of (real) co-dimension 2.  
Denote the circle (or $U(1)$) bundle over $M$ associated to $\Lcal$ by $S[\Lcal]$. Then $\Phi=f/|f|$ is a section of 
$S[\Lcal]$ over $M\setminus\{M_j\}_{j=1}^k$.  Denote variation of ${\rm arg}\Phi$ along a small positively oriented loop 
around $M_j$ by $2\pi  k_j$ with $k_j\in\Z$.
Then the Poincar\'e\ dual to $c_1(\Lcal)$ is given by a linear combination
$\sum_{j=1}^k k_j M_j$. The relation $c_1(\Lcal)=\sum_{j=1}^k k_j M_j$ holds in the Picard group ${\rm Pic}(M,\Z)$.

Notice that the smooth section $f$ may also vanish on a non-orientable submanifold $M_0$ of co-dimension $2$. However,
the variation of ${\rm arg} f$ around $M_0$ always vanishes since a small loop $l_0$ around $M_0$ can be smoothly
deformed to $l_0^{-1}$. 
(This observation  is consistent with the
standard fact that the Poincar\'e\ dual of a given form is always orientable). 

The above construction needs to be adjusted if $M$ is an orbifold (we assume that $M$ is a "good" orbifold according to Thurston's classification) i.e. it is a quotient
of a smooth manifold $\widehat{M}$ over a discrete group. In this case $M_j$ can also be suborbifolds of $M$.
The variation of ${\rm arg}\Phi$ around $M_j$ is then equal to $2\pi i q_j$ where $q_j\in \Q$.
Then we have the relation $c_1(\Lcal)=\sum_{j=1}^k q_j M_j$ which this time holds in ${\rm Pic}(M,\Q)$ (in the case of orbifold the group ${\rm Pic}(M,\Z)$ can have torsion while the group ${\rm Pic}(M,\Q)$ is torsion-free).

This description can be appropriately adjusted and  applied to the combinatorial model $\mathcal M_{g,n}[{\bf p}]$.

We consider the restrictions of  the power  $\tau^{48}_\pm$ of the  tau-functions $\tau_\pm$ of Section \ref{HPsec} 
to each cell of the
largest stratum $W$ of $\mathcal M_{g,n}[{\bf p}]$ ($W$ is the ``real locus" of the main stratum $\mathcal Q_{g,n}^0[{\bf p}]$ of $\mathcal Q_{g,n}[{\bf p}]$).

On each of those cells $\tau^{48}_\pm$
 are sections of the line bundles $({\rm det}\Lambda_H)^{48} \otimes\prod_{i=1}^n\Lcal^{4}_i$ and $({\rm det}\Lambda_P)^{48} \otimes\prod_{i=1}^n\Lcal^{4}_i$, respectively.
Introduce the sections $\Phi_\pm={\rm arg}\tau_\pm$ of these circle bundles on each cell of $W$.

In  Lemma \ref{tauW4} below it is shown that both sections $\Phi_\pm$  admit  continuous extension through the facets and provide continuous sections of the circle bundles over $\tilde{W}$ (we recall that $\tilde{W}$  denotes the union of $W$ and facets between its cells).
Recall also that the complement of $\tilde{W}$ in $\mathcal M_{g,n}[{\bf p}]$ is the union of
three co-dimension 2 subcomplexes: the Witten's cycle $\Wfive $, Kontsevich's boundary (also a cycle) $\Wbdr $ and the non-orientable 
subcomplex $\Wff$, which is not a cycle (we refer to \cite{Mondello1,Mondello2} for the proof of non-orientability of $\Wff$). Due to its non-orientability, $\Wff$ can not contribute to the Poincar\'e duals of 
$c_1({\rm det}^{48}\Lambda_H \otimes \prod_{i=1}^n \Lcal_i^4)$ and $c_1({\rm det}^{48}\Lambda_P \otimes \prod_{i=1}^n \Lcal_i^4)$.
Therefore, in the Picard group ${\rm Pic}(\tilde{W},\Q)$  (which coincides with   ${\rm Pic}(\mathcal M_{g,n}[{\bf p}],\Q)$) these Chern classes are linear combinations of  $\Wfive $ and $\Wbdr $ with rational coefficients which 
are obtained by computing the monodromies of $\tau_\pm/|\tau_\pm|$  (i.e. the increments of $\Phi_\pm={\rm arg}\tau_\pm$) around these cycles.

\subsection{Tau-functions $\tau_\pm$ and sections of circle  bundles over  $\tilde{W}$}
\label{sectcircle}

Since $\Mcal_{g,n}[\pb]$ is a slice of $\Qcal_{g,n}[\pb]$ defined by the  reality  of all periods of the  Abelian differential $v=\sqrt{Q}$ on $\Ch$, the Bergman tau-functions $\tau_\pm$ can be defined   on each cell of  $\Mcal_{g,n}[\pb]$ by the same formulas (\ref{taupfor}), (\ref{tauhfor}), (\ref{taum}) 
as in the case  of  $\Qcal_{g,n}[\pb]$.

Let us summarize the data which define $\tau_\pm$ on a given cell of $\Mcal_{g,n}[\pb]$ (we are mainly interested in the case when the cell is in the largest stratum $W$):
\begin{enumerate}
\item
Torelli markings of $\CC$ and of $\Ch$: the tau-functions $\tau_\pm$ depend on the choice of the Lagrangian subspaces
of $a$-cycles in $H_\pm$.

\item
The choice of one of the zeros (say, $x_1$) of $\qd$ which is used as an initial point of integration to define the distinguished local coordinates 
in neighbourhoods of poles $z_i$ via
\be
\xi_i(x)=\exp\left\{\frac{2\pi i}{p_i}\int_{x_1}^x v\right\}
\la{defxicom}\ee
\item
The choice of  cuts $\g_i$  between $x_1$ and  $z_i$; the integration contours  in 
(\ref{defxicom}) lie in $\Ch_0\setminus\{\pi^{-1}\gamma_i\}_{i=1}^n$  where $\Ch_0$ is the fundamental polygon of $\Ch$  defined above.
\end{enumerate}
The tau-functions $\tau_\pm$ are real-analytic within each cell of $\Mcal_{g,n}[\pb]$.

If the initial point of integration $x_1$ and/or the paths connecting $x_1$ with $z_1,\dots,z_n$  are chosen differently  then the tau-functions $\tau_\pm$ are  multiplied 
by the  factor of the form (\ref{transtau}),  which on $\Mcal_{g,n}[\pb]$ is expressed as
\be
\tau_\pm \to \tau_\pm \,\exp\left\{  -\f{\pi i}{3}\sum_{j=1}^n\sum_{k=1}^{2g+2m-2}A_{kj}\f{\ell_k}{p_j}   \right\}.
\la{transtau3}
\ee
%\todo{Verify factor}
The matrix $A_{kj}$ is a matrix with integer or half-integer entries and $\ell_k$ are the Strebel lengths of the edges of the ribbon graph; notice that this multiplier is always unitary.  The extra factor of 2 in the exponent of (\ref{transtau3}) comparing with
(\ref{transtau}) appears since the length of an edge is equal to $1/2$ of
an integer combination of periods $\Pcal_{s_i}$ from (\ref{transtau}).

 Following Prop. \ref{proptauptaut}, we  introduce the following expressions on each stratum $\Mcal_{g,n}^\db[\pb]$ of the combinatorial model $\Mcal_{g,n}[\pb]$:
\be
\tau_\pm \left(\prod_{i=1}^n d\xi_i(z_i)\right)^{1/12}
\la{defTcalpm}
\ee

Similarly to Prop. \ref{propTpm}, defining $\a$ via (\ref{LCD}) on a given stratum, we conclude that the 
$48\alpha$th power of  (\ref{defTcalpm}) is independent of the choice
of the base-point  $x_1$ and the cuts $\gamma_i$.

This allows to  interpret  $\tau_\pm^{48\a}$ as sections of the line bundles $\left({\rm det}^{12}\Lambda_{(H,P)}\otimes \prod_{i=1}^n \Lcal _i\right)^{4\a} $ over the stratum  of the combinatorial model $\Mcal_{g,n}^\db[\pb]$. 

%First we use $\tau_\pm$ to construct sections of the corresponding  line and circle bundles over the 
For the main stratum $W$  $\a=1$.
Since $W$ consists of disconnected topologically trivial cells, we consider the topologically non-trivial space $\tilde{W}$ which 
contains also the  facets of the cells of $W$ (as discussed above these facets correspond to ribbon graphs with one 4-valent vertex and the crossing from 
one cell of $W$ to another corresponds to a Whitehead move on one of edges of the ribbon graph).

The absolute values of $\tau_\pm$ vanish on  the facets of cells of $W$ i.e. when two vertices of $\Gamma$ glue together.
 However,  the following sections of the circle bundles over $\tilde{W}$:
\be
\Theta_\pm = \left(\frac{\tau_\pm}{|\tau_\pm |} \right)^{48}=e^{48 i\Phi_\pm}\;,
 \la{Tpm}
 \ee
 can be extended continuously through the facets.
 
 The formalism of circle bundles (instead of line bundles) is adequate in
the analysis of the intersection theory on moduli spaces using the JS combinatorial model; in \cite{Zvonkine}  it was used to discuss various subtle points in Kontsevich's proof of Witten's conjecture \cite{Kontsevich}.
To find the  Poincar\'e dual to first Chern class of ${\rm det}^{48}\Lambda_{(H,P)} \otimes \prod_{i=1}^n \Lcal_i^4$ 
one needs to
compute the increments of the argument $\Phi_\pm$  along closed paths in $\tilde{W}$ around cycles of co-dimension 2 in $\Mcomb$ (i.e. around $W_5$ and $W_{1,1}$). These increments (divided by $2\pi$) give coefficients in the linear combination of the cycles $W_5$ and $W_{1,1,}$ (while $W_{1,1}$ is the sum of several cycles, all of them turn out to contribute with the same coefficient) when computing the Poincar\'e duals.

To continue ${\rm arg}\tau_\pm$ from one cell of $W$ to another via a  Whitehead move one needs to study the asymptotics of $\tau_\pm$
as the length of one of the edges of the ribbon graph  $\Gamma$ tends to zero while perimeters of all faces remain the same.

Denote the length of the "vanishing edge" by $t$ and the lengths of the edges of the limiting ribbon graph $\Gamma_0$ by $\ell_i^0$ (while  keeping the perimeters constant). One can then assume that the limit $t\to 0$ of the graph $\Gamma$ is taken in such a way that all lengths
$\ell_i$ of the edges of $\Gamma$ depend linearly on $t$ i.e. $\ell_i=\ell_i^0+\alpha_i t$.

Denote the JS differential arising in the limit $t\to 0$  form by $\qd_0$;
multiplicities of its zeros are $(2,1,\dots,1)$; the double zero of corresponding JS differential coincides with the four-valent vertex of $\Gamma$. The genus of the canonical cover given by  $v_0^2=\qd_0$ 
(we denote it by  $\Ch_0$) equals $\gh-1$
since one of the branch cuts on $\Ch$ degenerates as $t\to 0$.
The set of homological coordinates on the facet includes the integral of $v_0=\sqrt{\qd_0}$ on $\Ch_0$ between the double zeros of $\qd_0$ (i.e. simple zeros of $v_0$) on the different sheets of $\Ch_0$.

Tau-functions $\tau_\pm$ on each facet of $W$ i.e. on a cell of $\tilde{W}\setminus W$ are  defined by the 
 formulas (\ref{taupfor}), (\ref{tauhfor}), (\ref{taum}) with multiplicities of zeros of $\qd_0$ having the above form.
 Such definition also depends on the choice of Torelli marking ${\bf t}_0$ on $\Ch_0$ 
 and the choice of cuts $\{\g^0_j\}_{j=1}^n$ connecting a chosen zero
 (which we denote by $x_1$ and assume that it remains simple as $t\to 0$)  of $\qd_0$ with its poles.

Then the  behaviour of $\tau_+$ in the limit $t\to 0$ is given by the following lemma:
\begin{lemma}\la{tauW4}
Suppose that the Torelli marking $\Tor_0$ of $\CC_0$ and contours $\g_i^0$ on $\CC_0$  are obtained  from the Torelli marking $\Tor$ on $\CC$ and the contours $\g_i$ 
on $\CC$ by a continuous deformation in the limit as $t\to 0$.
Then the following asymptotics holds as $t\to 0$:
\be
\tau_+(\CC,\qd,\Tor,\{\g_i\})=const\times t^{1/72} (1+o(1))\tau_+(\CC_0,\qd_0,\Tor_0,\{\g^0_i\})
\la{astaupl}
\ee
and 
\be
{\rm arg}\,\tau_+(\CC,\qd,\Tor,\{\g_i\})={\rm arg}\,\tau_+ +const+ (\CC_0,\qd_0,\Tor_0,\{\g^0_i\})+ o(1)
\la{asargtaup}
\ee
where the constants are independent of a point of $(\CC_0,\qd_0)$ of the facet.
\end{lemma}
{\it Proof.} 
The proof of asymptotical behaviour (\ref{astaupl}) is analogous to the proof of asymptotics of the tau-function on Hurwitz spaces given in Sec.3.2 of \cite{Advances}
and Sec.5.1 of \cite{contemp}.
Namely, one can use   equations  (\ref{deftaupm}) for the tau-functions $\tau_+(\CC,\qd)$ and  $\tau_+(\CC_0,\qd_0)$
and the homogeneity properties of these two tau-functions.
The bidifferential $B(x,y)$ on $\CC$ tends in the limit $t\to 0$ to the bidifferential $B_0(x,y)$ on $\CC_0$ due to our assumption about Torelli markings on
$\CC$ and $\CC_0$. Moreover, $\qd\to \qd_0$; thus also the abelian 
differential $w^+_v$ on $\Ch$ tends to the corresponding Abelian differential 
$(w_v^+)^0$ on $\Ch_0$ (as long as one stays away from the neighbourhood of the degenerating edge).
From  (\ref{deftaupm}) we see that   partial derivatives of  $\ln \tau_+(\CC,\qd)$ with respect to all homological coordinates except $t$ 
tend to the defining equations for 
 $\ln \tau_+(\CC_0,\qd_0)$. It follows that as $t\to 0$ the function $\tau_+(\CC,\qd)$ behaves as $f(t)\tau_+(\CC_0,\qd_0)$. 
 Since $\tau_+$ is algebraic in moduli, $f(t)=t^\alpha (1+o(1))$ for some rational degree $\alpha$,  i.e.
 \be
 \tau_+(\CC,\qd)=const\times t^\alpha(1+o(1)) \tau_+(\CC_0,\qd_0)\;.
 \ee
Under rescaling $\qd\to\epsilon\qd$, $\qd_0\to \epsilon\qd_0$ we have $t\to \epsilon^{1/2}t$. According to (\ref{kapl}) the difference of homogeneity coefficients of $\tau_+(\CC,\qd)$ and $\tau_+(\CC_0,\qd_0)$ equals $1/48(10/3-3)= 1/144$ 
and, therefore, $\alpha=1/72$.
 Notice that this asymptotics agrees with Lemma 9 of \cite{contemp} (in 
 \cite{contemp} one uses the $48$th power of $\tau_+$).
 
 Taking the argument of  (\ref{astaupl})  we get rid of the real-valued factor which diverges at facets of  $W$ and get the formula (\ref{asargtaup}) which allows to continue ${\rm arg}\tau_+$ from one cell of $W$ to another by appropriate choice of Torelli marking and contours 
 $\gamma_i$ inside of the cells and at their facets.
 \QED
 
 A similar statement holds for the Prym tau-function $\tau_-$, which  depends on the choice of  symplectic basis in the space $H_-$.

 While the cycles from the  symplectic basis in $H_+$  can always be chosen to avoid  the degenerating edge of $\Gamma$,  this is not the case  for  $H_-$.  
  We have ${\rm dim}H_-^0={\rm dim}H_--2$.
 Under the choice of canonical basis in $H_-$ shown in Fig.\ref{Wh_prop2},
 in the limit $t\to 0$ the cycle $a_1^-$ disappears while $b_1^-$ becomes non-closed. The remaining elements of the symplectic basis of $H_-$ naturally provide a symplectic basis in $H_-^0$.

\begin{lemma}
\label{lemmaarg}
Assume that the symplectic basis in $H_-^0$ is induced by   symplectic basis in $H_-$ as described above; assume also that the integration contours $\g^0_i$ on $\CC_0$ are induced by these contours on $\CC$. Then 
the asymptotics of the Prym tau-function $\tau_-(\CC,\qd,\Tor,\{\g_i\})$ in the limit $t\to 0$ is:
\be
\tau_-(\CC,\qd,\Tor,\{\g_i\})=const\times t^{13/72} (1+o(1))\tau_-(\CC_0,\qd_0,\Tor_0,\{\g^0_i\})
\la{astaum}
\ee
and 
\be
{\rm arg}\,\tau_-(\CC,\qd,\Tor,\{\g_i\})={\rm arg}\,\tau_-(\CC_0,\qd_0,\Tor_0,\{\g^0_i\})+ const +o(1)
\la{asargtaum}
\ee
where the constants are independent of a point of the facet.
\end{lemma}
{\it Proof.}
The proof is completely parallel to the proof of the previous lemma. 
Under our assumption about Torelli marking of $\Ch$ the Bergman bidifferential
$\Bh(x,y)$ on $\Ch$ tends to the Bergman bidifferential
$\Bh_0(x,y)$ on $\Ch_0$. Therefore, the abelian differential $w_v^-$ on $\Ch$ also tends to
the abelian differential  $(w_v^-)^0$ on $\Ch_0$; the same statement holds for 
all of periods of   $w_v^-$   over 
canonical cycles surviving in the limit. Therefore, since contours $\gamma_i^0$ 
on $\Ch_0$ are obtained by continuous deformation of contours $\gamma_i$
( we assume that the initial point of these contours does not coincide with any 
of zeros coalescing in the limit $t\to 0$), equations (\ref{deftaupm}) imply
Thus, again we have 
\be
 \tau_-(\CC,\qd,\Tor,\{\g_i\})=const\times t^\alpha(1+o(1)) \tau_-(\CC_0,\qd_0,\Tor^0,\{\g_i^0\})
 \ee
for some value of $\alpha$. Again, this value can be computed looking at rescaling
$\qd\to \epsilon \qd$,  $\qd_0\to \epsilon \qd_0$ using homogeneity coefficients (\ref{kam}) of both
tau-functions (recall that then $t\to \epsilon^{1/2}t$). This gives $\a=\frac{1}{72}+\f{1}{16}$ implying  (\ref{astaum}).

\QED

The constant in (\ref{asargtaum}) can be different when approaching $F$ from $K_1$ or $K_2$. Therefore,
from Lemmas \ref{tauW4} and \ref{lemmaarg} we  obtain the following 
\begin{proposition}
\la{propagation}
Let $F$ be the  facet between two maximal cells  $K_1,K_2$ of $\Mcal_{g,n}[\pb]$. Then 
\begin{enumerate}
\item  The arguments of $\tau_\pm$ have well--defined  limit at any point of $F$ while  approaching from the interior of  either $K_{1}$ or $K_2$.

\item Suppose that  in $K_{1,2}$ the Torelli markings $\Tor_{1,2}$ of $\CC$ and 
$\hat{\Tor}_{1,2}$  
of $\Ch$  are chosen such that  all pairs of canonical cycles except a pair 
shown in Fig.\ref{Wh_prop2} remain outside of a neighbourhood of the vanishing edge and continue smoothly from $K_1$ to $K_2$. The remaining pair is assumed to transform as shown in Fig. \ref{Wh_prop2}.
(In particular the Torelli marking 
$\Tor_0$ of $\CC_0$ and $\hat{\Tor}_0$ of $\Ch_0$ is the same whether it is induced from $K_1$ or $K_2$). Assume also that that the  contours $\g_i^0$ on $\CC_0$  are obtained  from the contours $\g_i^{1,2}$ on $\CC$  by continuity (both from $K_1$ and  $K_2$).  Then the limits of $\Phi_\pm = \arg(\tau_\pm)$ as $t\to 0$ taken from $K_1$ and $K_2$  differ by a constant $\phi_\pm$  independent of a point of $F$.
\end{enumerate}
\end{proposition}

\subsection{Monodromy  of $\Phi_+$ around $\Wfive $ and $\Wbdr $}

\subsubsection{Variation of  $\Phi_+$ under a  pentagon move}

The canonical basis of cycles in $\CC$  can always be chosen so  that a  neighbourhood of the two edges involved  in the pentagon move lies inside the fundamental polygon; thus the Torelli marking of $\CC$ remains invariant under the move.

\begin{proposition}
\la{propmon0}
Under a positively oriented  pentagon move (Prop. \ref{proppent}, Fig. \ref{Wh_Penta}),   the increment of the continuous extension of  $\Phi_+=\arg\,\tau_+$  
equals $\pi/72$. 
\end{proposition}
{\it Proof.}
 First, notice that the variation of $\Phi_+={\rm arg}\,\tau_+$ along a closed path depends only on the free homotopy class 
of the path.

Therefore, to compute the variation of ${\rm arg}\;\tau_+$ under the pentagon move involving three selected three-valent vertices 
 (denote these vertices by $x_1$, $x_2$ and $x_3$) we  assume that the lengths of both edges connecting 
$x_2$ with $x_1$ and $x_3$ are small in comparison with lengths of the other edges, and, moreover, that these lengths remain small during the whole pentagon move. 

\begin{figure}[htb]
\centering
\includegraphics[width=0.6\textwidth]{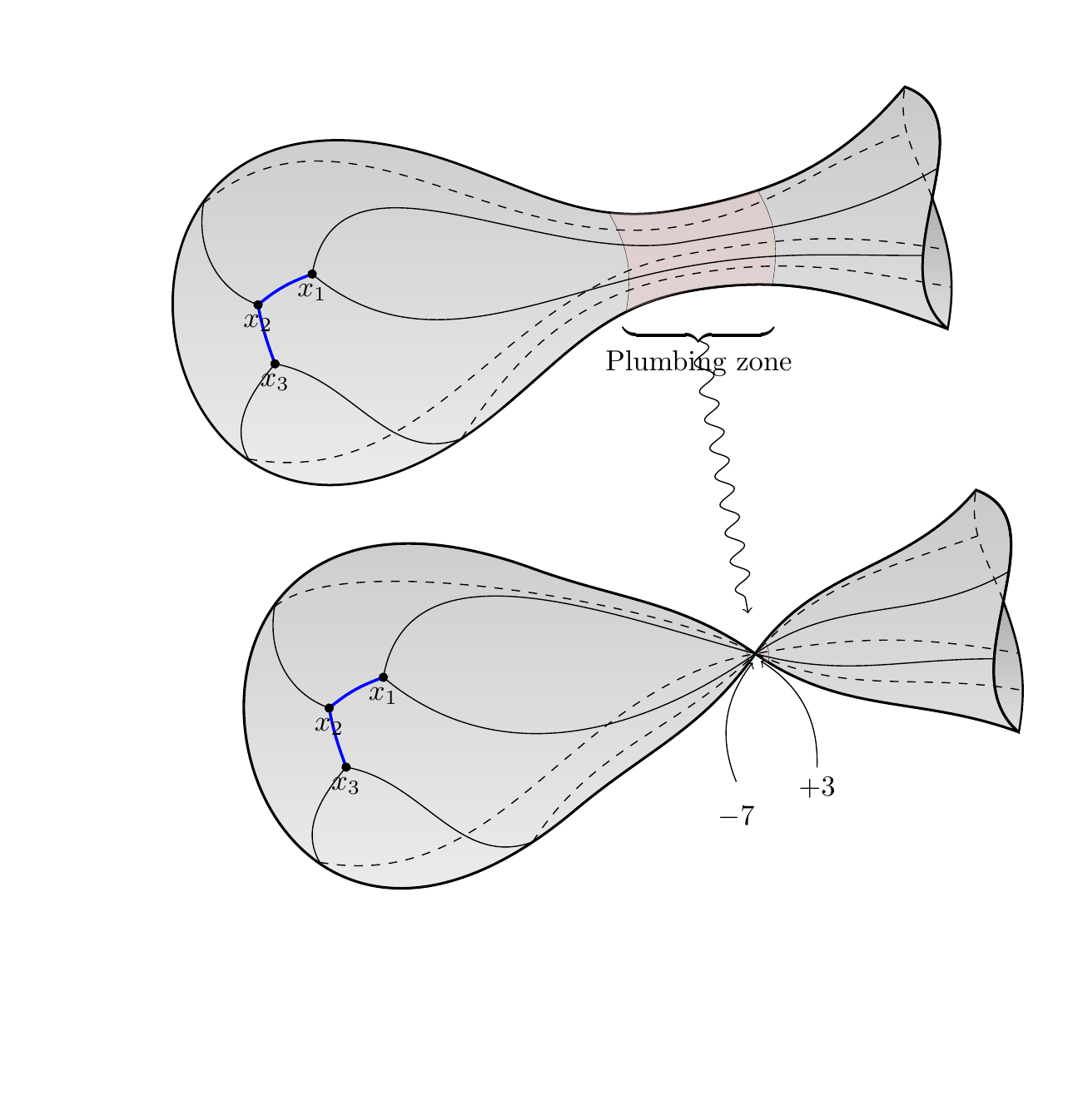}
\vspace{-2cm}
\caption{Separating of the Riemann sphere with three simple zeros and a pole of order 7 
of $Q$ in a neighbourhood of $\Wfive  $}
\la{splitW5}
\end{figure}

Denote the lengths of the edges connecting the three selected zeros by $tA$ and $tB$ where $A+B=1$, $A, B\in [0,1]$ and $t$ is a small parameter
which is kept fixed during  the pentagon move.
The  three zeros $x_i$ are close to each other in the metric $|\qd|$ and represent a small resolution of the triple zero.  Let us rescale the flat coordinate $z(x)= t(A+B)+\int_{x_2}^x v $ in a neighbourhood of these three zeros by a factor of $1/t$ (here $x_2$ is the "central zero"; see Section 3.1 of \cite{BK1} for justification 
of the choice of constant when defining $z(x)$). 
Then the surface  $\CC$ can be represented via a plumbing construction as follows.  Let  $\CC_1$ be the surface with the non-resolved zero $\tilde{x}$ of order 3  (five valent vertex of the corresponding ribbon graph) of
the JS differential $\qd_1$ such that the perimeters  and all lengths of all edges except the  edges connected to zeros  $x_{1,2,3}$ are the same on $\CC$ as on $\CC_1$. 
Let $\CC_0$  be the Riemann sphere equipped with the quadratic differential of the form (\ref{qex1}):
\be
Q_0=(x-x_1)(x-x_2)(x-x_3)(dx)^2\;.
\la{Q1}
\ee 
As in Section \ref{Q01117}, we assume that $x_1+x_2+x_3=0$; the integrals of $\sqrt{Q_0}$ between $x_2$ and $x_1$ and $x_3$ are assumed now to be real and equal to
$\pm A$ and $\pm B$. % (see for details \cite{BK1}). 
The flat coordinate on $\CC_0$ is given by the integral $w(x)=1+\int_{x_2}^x \sqrt{(x-x_1)(x-x_2)(x-x_3)}dx$. 
The plumbing of the Riemann surfaces $\CC_0$ and $\CC_1$ is defined by the equation (see Section 3.2 of \cite{BK1})
\be
\xi(x)\zeta(y)=t^{4/5}
\ee
where $\xi(x)$ is the distinguished local coordinate near $x=\infty$ on $\CC_0$ given by $\xi(x)=[\frac{5}{2} w(x)]^{-2/5}$; 
$\zeta(y) = \left(\frac{5}{2} z(y)\right)^{2/5}$ is the distinguished local coordinate near the triple  pole $\tilde{x}$ of $\CC_1$;
$z(x)=\int_{\tilde{x}}^x v_1$ is the flat coordinate on  $\CC_1$ near $\tilde{x}$.

The gluing of $\CC_0$ and $\CC_1$ is illustrated in Fig. \ref{splitW5}.
As $t\to 0$ one gets a union of Riemann sphere equipped with quadratic differential of the form (\ref{Q1}),  and the Riemann surface $\CC_1$ which belongs to the Witten's stratum $\Wfive $.

The quadratic differentials of the form (\ref{Q1}) on the Riemann sphere have three simple zeros and one pole of order 7;
such differentials form the space $\Qcal_0([1]^3,-7)$ discussed in Sec. \ref{Q01117}. 

Denote a Torelli marking of $\CC$ by $\Tor$; it naturally induces a Torelli marking $\Tor_1$ on $\CC_1$ 
 in the limit $t\to 0$ by continuous deformation of the canonical basis of cycles. Define a system of cuts $\{\gamma_i\}_{i=1}^n$ on $\CC$ choosing $x_1$ as an initial point. In the limit $t\to 0$
one gets  a system of cuts $\{\gamma_i^1\}_{i=1}^n$ on $\CC_1$ which connect the triple zero $x_1$ with $n$ poles of $\qd_1$
(one also gets cuts on $\CC_0$ connecting the point at infinity with $x_{1,2,3}$ but these cuts are inessential for our purposes).

Denote the resulting tau-function $\tau_+$ on the Witten's stratum $\Wfive $ by
$\tau_+(\CC_1,\qd_1,\Tor_1,\{\gamma_i^1\}_{i=1}^n)$ and on the space  $\Qcal_0([1]^3,-7)$ by $\tau_+(\CC_0,Q_0)$.  
The tau-functions 
$\tau_+(\CC_0,Q_0) $ is given by  (\ref{tau111m7}).

\begin{lemma}\la{astaupen}
 Assuming  that the Torelli marking $\Tor_1$ and the cuts $\{\gamma_i^1\}$ on  $\CC_{1}$ are induced from $\CC$ in the limit $t\to 0$ as described above the following asymptotics
 holds:
 \be
  \la{taupfac}
  \tau_+(\CC,\qd,\Tor,\{\gamma_i\}_{i=1}^n)= const\times (1+ o(1))\, \tau_+(\CC_0,\qd_0)\,\tau_+(\CC_1,\qd_1,\Tor_1,\{\gamma_i^1\}_{i=1}^n)\;.
 \ee
\end{lemma}
{\it Proof.} In the limit $t\to 0$ we have 
\be
\frac{\tau_+(\CC,\qd)}{\tau_+(\CC_0,\qd_0)\,\tau_+(\CC_1,\qd_1)}=f(t)(1+o(1))
\la{ratiotau}
\ee
 where $f$ is some function independent of the  moduli of $(\CC_0,\qd_0)$ and $(\CC_1,\qd_1)$. Similarly to the proof of Lemma \ref{tauW4} the
 asymptotics (\ref{ratiotau}) 
follows from  equations (\ref{deftaupm}) for three tau-functions from the r.h.s. of (\ref{ratiotau}) and the behaviour  of 
$B(x,y)$ in the limit $t\to 0$.
Namely, our assumption about the relationship between Torelli markings $\Tor$ and $\Tor_1$ implies  that  $B(x,y)$ tends
to $B_0(x,y)$ or $B_1(x,y)$ if both points $x$ and $y$ remain in the limit $t\to 0$ on $\CC_0$ or
$\CC_1$, respectively, see \cite{Fay73}, Cor. 3.2. In turn, this implies that   the meromorphic differential  $w_v^+$ (\ref{Qvpm}) on $\CC$ tends to the corresponding meromorphic differentials on $\CC_1$ and $\CC_0$. Using our assumptions about cuts
 $\gamma_i$ we conclude than that equations (\ref{deftaupm}) for $\tau_+(\CC,\qd)$ with respect to all homological coordinates except $t$
tend in the limit $t\to 0$ to the  equations for the product $\tau_+(\CC_0,\qd_0)\tau_+(\CC_1,\qd_1)$ which implies (\ref{ratiotau}).

To show that function $f$ is a constant up to terms vanishing in the limit $t\to 0$ (this fact will not be used here since we are  interested only in the
phase   of tau-functions) one makes use of homogeneity property of the tau-functions
$\tau_+$  on spaces $W$, $\Wfive $ and $\Qcal_0([1]^3,-7)$. 
Namely, the homogeneity coefficient (\ref{kapl}) of the function $\tau_+(\CC)$ equals $\f{5}{3\times 48}(4g-4+2n)$ which exactly coincides with the sum of homogeneity coefficients of $\tau_+(\CC_0)$
and $\tau_+(\CC_1)$. Thus the ratio (\ref{ratiotau}) tends to a constant as $t\to 0$.  \QED

Returning to the proof of Prop. \ref{propmon0}, we observe that the tau-function  $\tau_+(\CC_1,Q_1,\Tor_1,\{\gamma_i^1\}_{i=1}^n)$ does not change under the pentagon move, and, therefore,  the
variation of its argument  is only due to the variation of the argument of
\be
\tau_+(\CC_0,Q_0)= [(x_1-x_2)(x_2-x_3)(x_3-x_1)]^{1/36}
\la{tauplus1}
\ee
which  equals $\pi/72$ according to   Th. 2 of \cite{BK1} (it is an elementary although  technically non-trivial  result).\QED

\subsubsection{Variation of $\Phi_+$ under combinatorial Dehn's twist}

To compute the variation of the argument of $\tau_+(\CC,\qd,\Tor,\{\gamma_i\}_{i=1}^n)$ 
under the combinatorial Dehn's twist defined in Section \ref{combDehnsect}  we  assume that the Torelli marking $\Tor$ is chosen such that
the Lagrangian subspace of $a$-cycles remains invariant under the twist. Namely, if the twist goes around $W_{1,1}^r$ (i.e. corresponding loop separates  $\CC$
into two components) the  Torelli marking $\Tor$  of $\CC$ is obtained by taking the 
union of Torelli markings of these components.
 If the Dehn's twist is performed around $W_{1,1}^{irr}$ i.e. along a  non-separating loop then one of $a$-cycles should be chosen to follow this loop while 
 all other $a$-cycles 
 remain outside of the zone of the twist.

 In the rest of this section we prove the following proposition.
 
 \begin{proposition}\la{rmon}
 Let $e\cup \wt e$ be a closed loop on $\CC$ formed by the edges $e$ and $\wt e$ as in Prop. \ref{PDehn}. Let the  Torelli marking $\Tor$ of $\CC$ be chosen as discussed above. Then
the variation of the argument of the tau-function $\tau_+(\CC,\qd,\Tor,\{\gamma_i\}_{i=1}^n)$ under the combinatorial Dehn's twist 
along $e\cup \wt e$ equals $13\pi/72$.
\end{proposition}
 
 {\it Proof.}
 Under our assumption about the choice of Torelli marking on $\CC$ the Lagrangian subspace of $a$-cycles remains invariant 
  after the combinatorial Dehn's twist; thus
   the variation of ${\rm arg}\tau_+$ 
under the twist is independent of the lengths of edges within a given cell of $W$. Therefore, we can assume that 
the lengths of the edges $e$ and $\wt e$ are "small" i.e. we denote them by $tA$ and $tB$ with $A+B=1$ and $A,B,t\in \R_+$ and compute 
the variation of ${\rm arg}\tau_+$ in the limit $t\to 0$.

Consider the  twists around reducible and irreducible components of Kontsevich's boundary separately.
\vskip0.3cm
{\it 1. Combinatorial  Dehn's twist around  $\Wbdr ^r$.}

In the limit $t\to 0$ when the lengths of both edges tend to $0$ while the lengths of other edges are adjusted accordingly to preserve the 
perimeters of all faces, $\CC$ splits into two Riemann surfaces, $\CC_1$ (of genus  $g_1$) and $\CC_2$ (or genus $g_2$).
Corresponding  JS differentials We denote by $\qd_{1,2}$ the corresponding JS differential, by $n_{1,2}$ the  number of poles and by $\Gamma_{1,2}$ the  ribbon graphs.
The differentials $\qd_{1,2}$ have simple poles at the points of resolution of the node and the ribbon graphs $\Gamma_{1,2}$ 
have one-valent vertices at these points.

According to our assumption the Torelli marking of  $\CC$ is chosen in such a way that in the limit $t\to 0$  it gives rise to Torelli markings $\Tor_{1,2}$ of  $\CC_{1,2}$. 

The "first" zero $x_1$ and the cuts $\g_i$ are chosen as follows:
the "first" zero $x_1$ is chosen to be  one of the zeros involved in the Dehn's twist. Starting from this zero, we 
choose $n_1$ cuts $\{\g_i\}_{i=1}^{n_1}$ going towards the poles which remain on the   $\CC_1$ side, and $n_2$ cuts $\{\g_i\}_{i=n_1+1}^n$ going towards the poles on $\CC_2$ side. Such a system of cuts naturally splits under degeneration into two systems: $\{\g_i^{1}\}_{i=1}^{n_1}$ on $\CC_1$ and $\{\g_i^{2}\}_{i=1}^{n_2}$ 
on $\CC_2$. The initial point of integration then becomes the one-valent vertex of the corresponding ribbon graphs $\Gamma_{1,2}$. 

Introduce now the tau-functions 
$\tau_+(\CC,\qd,\Tor_1,\{\gamma_i\}_{i=1}^n)$, $\tau_+(\CC_1,\qd_1,\Tor_1,\{\gamma_i^1\}_{i=1}^{n_1})$ and 
$\tau_+(\CC_2,\qd_2,\Tor_2,\{\gamma_i^2\}_{i=1}^{n_2})$.

Consider also the following 
quadratic differential on the Riemann sphere $\CC_0$:
\be
\qdt_0(x)=\frac{(x-x_1)(x-x_2)(dx)^2}{x^3}\;
\la{Q1-2}
\ee
The differential $\qdt_0$ has third order poles
at the points $x=0$ and $x=\infty$.
The homological coordinates of  ${\qdt_0}$ are equal to $\pm A$ and $\pm B$.

The differential (\ref{Q1-2}) with real homological coordinates defines a point of the space $\Qcal_0^{[1,1,-3,-3]}$ and the corresponding tau-function $\tau_+({\bf CP}^1,\qdt_0)$ is given by (\ref{taup1133}).

\begin{figure}[t]
\begin{center}
\includegraphics[width=0.6\textwidth]{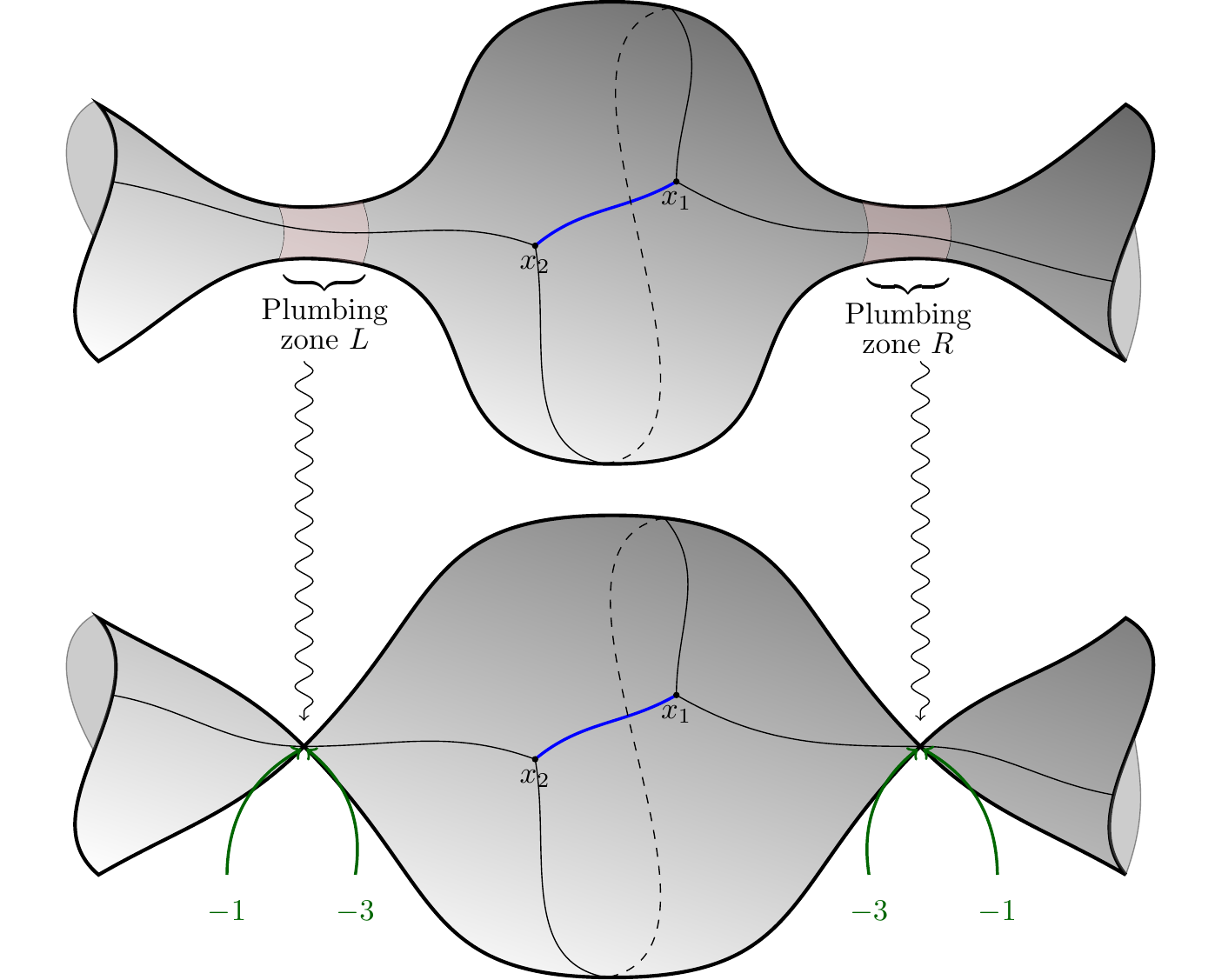}
\end{center}
 \caption{Plumbing construction interpretation of resolution of Kontsevich boundary: a Riemann sphere with two three-valent vertices is glued 
 between one-valent vertices via two plumbing zones}
 \label{W11-blowup-Fig}
\end{figure}

The Riemann surface  $\CC$ equipped with the differential $\qd$ can be obtained via a "plumbing" construction by inserting the
Riemann sphere equipped with the differential $\qdt_0$ between Riemann surfaces 
$(\CC_1,\qd_1)$ and $(\CC_2,\qd_2)$ (see for details \cite{BK1}).
A neighbourhood of the point $x=0$ of the Riemann sphere is "plumbed" with a neighbourhood of the one-valent vertex
(i.e. simple pole of $\qd_1$) on $\CC_1$ while a neighbourhood of the point $x=\infty$ of the Riemann sphere is "plumbed" with a neighbourhood of the one-valent vertex
(i.e. simple pole of $\qd_2$) on $\CC_2$. This procedure is illustrated in Fig.\ref{W11-blowup-Fig}.

%\vskip3.0cm

The following analog of Lemma \ref{astaupen} holds in this case:
\begin{lemma}\la{astaupen2}
 Assuming  that the Torelli marking on  $\CC_{1,2}$ and the cuts $\g^{(1,2)}_i$ are induced from $\CC$ in the limit $t\to 0$ the following asymptotics
 holds:
 \be
  \la{taupfac1}
  \tau_+(\CC,\qd,\Tor,\{\gamma_i\})= const\times 
  (1+ o(1))\, \tau_+({|bf CP}^1,\qdt_0)\,\tau_+(\CC_1,\qd_1,\Tor_1, \{\gamma_i^1\})\,\tau_+(\CC_2,\qd_2,\Tor_2,\{\gamma_i^2\})
 \ee
\end{lemma}
{\it Proof.} As in the proof of Lemma \ref{astaupen},  consider the expression
\be
\frac{\tau_+(\CC,\qd,\Tor,\{\gamma_i\})}{\tau_+({|bf CP}^1,\qdt_0)\tau_+(\CC_1,\qd_1,\Tor_1, \{\gamma_i^1\})\tau_+(\CC_2,\qd_2,\Tor_2,\{\gamma_i^2\})}\;.
\la{ratiotau1}
\ee
Under our assumptions about Torelli markings and cuts $\g_i$ on $\CC$ and $\CC_{1,2}$ the expression (\ref{ratiotau1}) behaves as 
$f(t)(1+o(1))$ as $t\to 0$. This follows again from the asymptotics  of the bidifferential $B(x,y)$ as $t\to 0$ which in particular 
implies (under our choice of Torelli markings on $\CC$ and $\CC_{1,2}$) that the Bergman projective connection $S_B$ on $\CC$ 
tends to the Bergman projective connections at the corresponding points of $\CC_1$, $\CC_2$ and $\CC_0$
(and also the fact that the quadratic differential $\qd$ tends to   $\qd_{1,2}$ and $\qdt_0$ in the limit $t\to 0$
on corresponding components). 

Using (\ref{kapl}) we find that  the expression (\ref{ratiotau1}) remains invariant under simultaneous rescaling of  $\qd$ and $\qd_{0,1,2}$.
Therefore, $f=const$. 
\QED

In the  limit $t\to 0$  the Whitehead move is performed entirely on the Riemann sphere equipped with $\qdt_0$;
under this Whitehead move only the  first term in the r.h.s. of (\ref{taupfac}) contributes to monodromy of ${\rm arg}\tau_+$.
The  monodromy of (\ref{taup1133}) under the Whitehead's move was computed in Theorem 4 of  \cite{BK1}; it equals  $13\pi/72$ implying the statement of the proposition in the reducible case.

\vskip0.3cm
{\it 2. Combinatorial  Dehn's twist around the irreducible component $\Wbdr ^{irr}$.}
In this case   the edges, $e_1$ and $e_2$ connecting two  vertices of the ribbon graph
bound a loop which does not separate $\CC$ into two components.
 
As stated before Prop.\ref{rmon}, we choose the Torelli marking  $\Tor$ such  that the Lagrangian subspace of $a$-cycles remains invariant under the move.
Namely,  the "first" $a$-cycle $a_1$ goes along the loop formed by the edges $(e_1,e_2)$.  The cycle $b_1$ then goes "along" the corresponding cylinder (Fig. \ref{Dehnplus}). 
The other cycles forming the canonical basis can be placed far from the zone of the Dehn's twist.
Then under the combinatorial Dehn's twist shown in Fig. \ref{Dehnplus}  the pair of cycles $(a_1,b_1)$ transforms as
\be
(a_1,b_1)\to (a_1, b_1 -  a_1)
\la{transab}
\ee
while all other cycles do not vary. 
\begin{figure}
\centering
\includegraphics[width=0.8\textwidth]{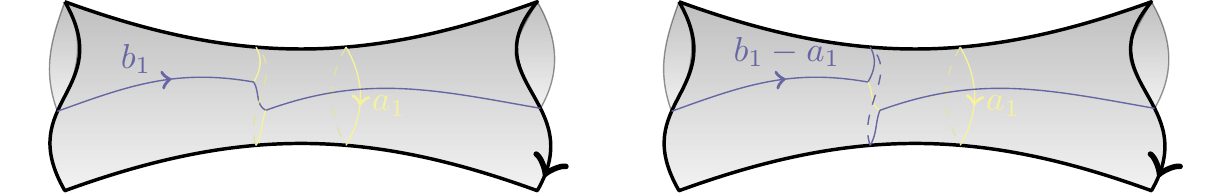}
\caption{Action of combinatorial Dehn twist along non-separating loop on a pair of canonical cycles from $H_1(\CC)$. }
\la{Dehnplus}
\end{figure}

Denoting the lengths of the edges $e_1$ and $e_2$ by $tA$ and $tB$, we compute the variation of ${\rm arg}\tau_+$ 
as $t\to 0$. For that we assume that the basic cycles on $\CC$ are chosen as discussed above and that the
basic cycles on $\CC$ (except $(a_1,b_1)$) are used to Torelli mark the Riemann surface $\CC_1$ of genus $g-1$.
Moreover, we assume that the cuts $\g_i$ are chosen on $\CC$ to start from, say,  zero $x_1$ and all go from $x_1$ to the "right" inducing   in the limit $t\to 0$ the system of cuts $\{\gamma_i^1\}_{i=1}^n$ (then on $\CC_1$ they start at one of simple poles of $\qd_1$).

The pair $(\CC,\qd)$ can be then restored via plumbing construction by gluing the 
Riemann surface $(\CC_1,\qd_1)$ to the Riemann sphere equipped with the differential $\qdt_0$. The  plumbing zones
 are placed on $\CC_1$ near the simple poles of $\qd_1$ and on the Riemann sphere near the points $x=0$ and $x=\infty$
(see \cite{BK1}).

The following lemma is and analog of lemmas \ref{astaupen}, \ref{astaupen2} and is proved in the same way.
\begin{lemma}\la{astaupen3}
 Assuming  that the Torelli marking and the cuts $\{\g_i^1\}_{i=1}^n$ on  $\CC_{1}$  are induced from cuts
  $\{\g_i\}_{i=1}^n$ on
 $\CC$ in the limit $t\to 0$ the following asymptotics
 holds:
 \be
  \la{taupfac2}
  \tau_+(\CC,\qd,\Tor,\{\gamma_i\})= const\times (1+ o(1)) \tau_+(\qdt_0)\,\tau_+(\CC_1,\qd_1,\Tor_1,\{\gamma_i^1\})
 \ee
\end{lemma}

The variation of the argument of  $\tau_+(\CC,\qd,\Tor,\{\gamma_i\})$ under the Dehn's twist is again only due to variation of the 
argument of $\tau_+({|bf CP}^1,\qdt_0)$ which gives $13\pi/72$ as before.
This ends the proof of Proposition \ref{rmon}.

\QED

%\vskip3.0cm

\subsection{Monodromy  of $\Phi_-$ around $\Wfive $ and $\Wbdr $}

When computing the  monodromy of  ${\rm arg}\tau_-$ around $W_5$ and $W_{1,1}$    a pair of  cycles
forming a canonical basis in $H_-(\Ch)$ always passes through  the zone of the corresponding move.
Therefore, when computing the monodromy of $\arg\tau_-$ with respect to these moves one needs to take into account
the contribution from the change of Torelli marking  $\Tor_-$. Let us call this pair of cycles $(a_1^-,b_1^-)$.

Under an elementary Whitehead move the cycle $a_1^-$ vanishes  at the facet  between two cells
of $W$ and then re-opens again. 
This requirement does not uniquely define the propagation of Torelli marking  $\Tor_-$ from one cell to another;
Fig.\ref{Wh_prop2} gives just one of many possible ways of such propagation.

After a pentagon move or a combinatorial Dehn's twist the Torelli marking in $\Tor_-$ does not necessarily return back to the 
original one, which leads to an extra contribution to variation of $\arg\tau_-$ in comparison with $\tau_+$-case.

\subsubsection{Variation of  $\Phi_-$ under pentagon move}

Assuming that the number of zeros of $\qd$ is greater than 3 (the remaining case of $g=1$ and $n=1$ can be treated similarly) we 
 choose the initial point  $x_1$ of cuts  $\{\gamma_i\}_{i=1}^n$ to not participate in the pentagon move; the cuts themselves are assumed to stay away from the zone of the move. Then the cuts $\{\gamma_i\}_{i=1}^n$ naturally propagate from cell to cell, 
and  return to their initial positions  after the pentagon move.

Indeed, if the evolution of cycles $(a_i^-,b_i^-)$ under five Whitehead moves is chosen as shown in Fig. \ref{Wh_Penta}
(in particular, we choose three  branch cuts outgoing from the zone of the pentagon move as in Fig.\ref{Wh_Penta};
tilde marks the edge on which we perform the Whitehead move on each step)
then the pair of cycles $(a_1^-,b_1^-)$ evolves to $(\tilde{a}_1^-,\tilde{b}_1^-)$ such that  
\be
\tilde{a}^-=a^-\;\hskip0.7cm \tilde{b}^-= b^--a^-
\la{transmin}
\ee
Although the Torelli marking $\Tor_-$ has monodromy under the pentagon move, this monodromy does not impact the
 $\arg\tau_-$ since the Lagrangian subspace of $a$-cycles in $H_-$ (and, therefore, the Prym bidifferential $B_-$) are invariant under such symplectic transformations.

\begin{proposition}
\la{propmon1}
Let the pentagon move 
induce  a symplectic transformation  in $H_-$
given  by a matrix   $\sigma_-=\left(\ba{cc} C_- & D_-\\ A_- & B_- \ea\right)\in Sp(6g-6+2n,\Z)$.  Denote the Prym matrix of the canonical cover by $\Pi$.
Then the variation of ${\rm arg}\,\tau_-$  under the pentagon move  is given by:
\be
{\rm var}\,\arg\,\tau_-=\frac{13}{72}\pi+ \arg\,\det\,(C_-\Pi+ D_-) \;.
\la{vartaum}
\ee
\end{proposition}
{\it Proof.}  It is sufficient to prove (\ref{vartaum}) choosing the canonical basis $\Tor_-$ in $H_-$ as described above;
all canonical pairs $(a_i^-,b_i^-)$ except $(a_1^-,b_1^-)$ stay outside of the zone of the pentagon move,
while the pair $(a_1^-,b_1^-)$ evolves as shown in Fig.\ref{Wh_Penta} according to (\ref{transmin}).
Then the second term in the r.h.s. of (\ref{vartaum}) is absent and the variation of $\arg\,\tau_-(\CC,\qd,\Tor_-,\{\gamma_i\}_{i=1}^n)$ should be proven to equal 
 $13\pi/72$.

To compute this variation we follow the logic of the proof of Prop.\ref{propmon0}. Namely, we represent the canonical cover $\Ch$ of the
Riemann surface $\CC$ by  plumbing of the canonical cover $\Ch_1$ of $\CC_1$  and the canonical cover $\Ch_0$ of the Riemann sphere equipped with quadratic differential $\qd_0$
(\ref{Q1}) ($\qd_0$ is an element of the space $\Qcal_0([1]^3,-7)$).  In the limit as the plumbing parameter $t$ tends to $0$ the  cover $\Ch$ decomposes in the
disjoint union of $\Ch_0$ and $\Ch_1$. The canonical bases $\Tor_0^-$ in $H_-(\Ch_0)$ and $\Tor_1^-$ in $H_-(\Ch_1)$ are naturally induced from
the canonical base $\Tor^-$ in $H_-(\Ch)$ chosen according to the above convention.

In the limit $t\to 0$ the tau-function $\tau_-(\CC,\qd,\Tor^-)$ decomposes as
\be
 \tau_-(\CC,Q,\Tor^-,\{\gamma_i\})= const\times (1+ o(1)) \tau_-(\CC_0,\qd_0,\Tor^-_0)\,\tau_-(\CC_1,\qd_1,\Tor^-_1,\{\gamma_i^1\})
 \la{taumfac}
 \ee
 Similarly to the proof of Lemma \ref{astaupen}, the proof of the asymptotics (\ref{taumfac}) is based on the fact that
 our choice of Torelli markings $\Tor^-$ and $\Tor^-_{0,1}$ implies that the Prym projective connection $S_B^-(x)$ on
 $\Ch$ tends to the Prym projective connections ${S_B^-}_{0,1}$   
  if the argument $x$ remains in the limit on $\Ch_0$ 
 or $\Ch_1$, respectively. Therefore, the same statement applies to the meromorphic differentials $w_v^-$ 
 (\ref{Qvpm}) which enter the equations 
 (\ref{deftaupm}) for $\tau_-$.
 The constant factor $const\times(1+o(1))$ is obtained by comparing the homogeneity properties of the three tau-functions entering (\ref{taumfac})  (this coefficient equals $\f{5}{3\times 48}(4g-4+2n)$ on both sides).

The tau-function  $\tau_-(\CC_1,\qd_1,\Tor^-_1)$ is invariant under the pentagon move as $t\to 0$ i.e. in this limit the
move is localized to $\CC_0$ part and the variation of  $arg\,\tau_-(\CC,Q,\Tor^-,\{\gamma_i\})$ is equal to the variation of the argument of 
 $\tau_-(\CC_0,\qd_0,\Tor^-_0)$.
The latter tau-function
is  given by the formula (\ref{taum33}):
\be
\tau_-= \o_1\,[(x_1-x_2)(x_1-x_3)(x_2-x_3)]^{13/36}
\la{tauminus}
\ee
where $\o_1$ is the Abelian integral of the first kind over the cycle $a_1^-$ on $\Ch_0$.

 Therefore, under the pentagon move around $\Wfive $ the variation of $\arg\,\tau_-(\CC,\qd,\Tor_-)$ is given by the sum of 
variation of the argument of $[(x_1-x_2)(x_1-x_3)(x_2-x_3)]^{13/36}$ and variation of the argument of the abelian integral 
$\omega_1$ over cycle $a_1^-$. 
According to  Th.2 of \cite{BK1} the variation of ${\rm arg}[(x_1-x_2)(x_1-x_3)(x_2-x_3)]^{13/36}$ under the pentagon move equals $13/72 \pi$. On the other hand, the variation of $\arg\, \omega_1$  is zero since the cycle $a_1^-$ transforms to itself.

The second term in (\ref{vartaum}) is non-trivial if $\Tor_-$  on the initial cell is chosen differently, or if
Torelli's marking suitable for description of the pentagon move on one set of three vertices is propagated
along pentagon move on another set of three vertices (i.e. around another cell of $\Wfive $).
\QED

\subsubsection{Variation of $\Phi_-$ under combinatorial Dehn's twist}

To study the variation of $\arg\,\tau_-(\CC,\qd,\Tor^-,\{\gamma_i\})$ it is again convenient to choose a basis $\Tor_-$ in $H_-(\Ch)$ 
and the set of cuts $\{\gamma_i\}$
which transform
in the "best"  way under the combinatorial Dehn's twist both around reducible and irreducible components of $W_{1,1}$. As before, we assume that all pairs of canonical cycles in $H_-$ stay away from the twist zone. Two selected cycles, which we call  $a_1^-$ and $b_1^-$ are chosen as shown in Fig.\ref{Dehnminus1}.

 The zone of the Dehn's twist is also shown as an annulus in Fig.\ref{Dehnminus2}. Under the action of the Dehn's twist the cycles $(a^-,b^-)$ transform as follows:
\be
\tilde{a}^-= a^-,\hskip0.7cm
\tilde{b}^-= b^- + 2 a_-
\ee
therefore, under this choice of the basis in $H_-(\Ch)$ the Lagrangian subspace formed by $a$-cycles remains invariant 
under the Dehn's wist.

\begin{proposition}
\la{propDehn}

Let the combinatorial Dehn twist  induce  a symplectic transformation given 
by a matrix   $\sigma_-=\left(\ba{cc} C & D\\ A & B \ea\right)\in Sp(6g-6+2n,\Z)$ in the space $H_-$; denote the Prym matrix by $\Pi$.
Then the variation of ${\rm arg}\,\tau_-$  under the Dehn twist is given by:
\be
{\rm var}\,\arg\,\tau_-=\frac{25}{72}\pi+ \arg\,\det\,(C_-\Pi+ D_-) \;.
\la{vartaumdehn}
\ee
\end{proposition}

{\it Proof.}  The proof is parallel to the proof of the corresponding statement for $\tau_+$. Namely, in both reducible and irreducible case one can "localize" the Dehn's twist. 

In the irreducible case we represent the canonical cover 
$\Ch$ by plumbing of the canonical cover $\Ch_1$ of the Riemann surface $\CC_1$ of genus $g-1$ with the elliptic curve
$\Ch_0$ which is a canonical cover of the Riemann sphere with branch points at $0,\infty, x_1,x_2$.
Denoting the real plumbing parameter  by $t$, and the canonical bases in $H_-(\Ch_{0,1})$ inherited from
$H_-(\Ch)$ by $\Tor_{0,1}$, we have the asymptotics as $t\to 0$:
\be
\tau_-(\CC,\qd,\Tor_-,\{\gamma_i\})=const\times (1+o(1))\tau_-(\CC_1,\qd_1,\Tor^-_1,\{\gamma_i^1\})\tau_-(\CC_0,\qd_0,\Tor^-_0)\;.
\la{asympDehnm}
\ee
It is a assumed as before that the cuts $\{\gamma_i^1\}$ are induced on $\CC_1$ from cuts 
 $\{\gamma_i\}$ on
$\CC$ in the limit $t\to 0$.
In the limit $t\to 0$ the function $\tau_-(\CC_1,\qd_1,\Tor^-_1)$ does not change under the Dehn's twist.

The function $\tau_-(\CC_0,\qd_0,\Tor^-_0)$ is given explicitly by (\ref{tauh1134}):
\be
\tau_-= \o_1\, (x_1 x_2)^{1/12} (x_1-x_2)^{13/36}
\la{tauh11new}
\ee
where $\o_1$ is the elliptic integral of first kind over over cycle $a_1^-$.
According to our assumption about transformation of the $a$-cycle under the Dehn's twist the integral $\o_1$ remains invariant.
The variation of the argument of the remaining explicit expression equals $\frac{25}{72}\pi$ according to Theorem 4 of \cite{BK1}.
This gives (\ref{vartaumdehn}) in the irreducible case.

The case of reducible component $\Wbdr ^{r}$, when the Riemann surface $\CC$ is separated into the union of two
Riemann surfaces, $\CC_1$ and $\CC_2$, with the Riemann sphere equipped with a differential of the form (\ref{q2z2p}) is
glued between them, is treated similarly. Namely, he canonical cover $\Ch$ can be obtained via plumbing of $\Ch_0$ 
with both $\Ch_1$ and $\Ch_2$. Then, under the usual assumption that the set of contours $\{\gamma_i^1\}_{i=1}^n$ on $\Ch_1$ and $\Ch_2$
is inherited from $\Ch$ and that the Torelli markings in $H_-(\Ch_1)$, $H_-(\Ch_2)$ and $H_-(\Ch_0)$ are inherited from
$H_-(\Ch)$, we get the asymptotics as $t\to 0$:
\be
\tau_-(\CC,\qd,\Tor_-,\{\gamma_i\})=const\times (1+o(1))\,\tau_-(\CC_1,\qd_1,\Tor^-_1,\{\gamma_i^1\})\,
\tau_-(\CC_1,\qd_1,\Tor^-_1,\{\gamma_i^2\})\,\tau_-(\CC_0,\qd_0,\Tor^-_0)\;
%\la{asympDehnm}
\ee
where the last term is the tau-function (\ref{tauh11new}) and the other two terms do not change under the Dehn's twist.
This leads to  (\ref{vartaumdehn}) in the irreducible case.

Again, the second term in  (\ref{vartaumdehn})  appears if we either start from a different Torelli marking in $H_-(\Ch)$ or propagate the
given Torelli marking along a different Dehn's twist.
\QED

\subsection{Expressing Hodge and Prym classes via $\Wfive $ and $\Wbdr $}

The Prym vector bundle $\Lambda_P$ over $\Mcomb$ is defined by restriction of the Prym bundle over
$\Qcal_{g,n} [\pb]$ to  $\Mcomb$, as in Section \ref{PrymonQ}.

The theorem \ref{PrymQuad} is applicable also in the context of $\Mcomb$ and provides the isomorphism between 
the Prym vector bundle $\Lambda_P$ with the vector bundle $\Lambda_2^{(n)}$ of quadratic differentials with simple poles at the punctures, therefore the determinant line bundle ${\rm det}\,\Lambda_P$ is the isomorphic to the  determinant line bundle ${\rm det}\,\Lambda_2^{(n)}$. 
We denote by $\lambda_P$ the corresponding class: $\lambda_P=c_1({\rm det}\,\Lambda_P)$.

According to Proposition \ref{propTpm} the tau functions $\tau_\pm^{48}$ are sections of the line bundles $\lambda^{48}\otimes \prod_{i=1}^n \Lcal _i^4$ and $\lambda_P^{48}\otimes \prod_{i=1}^n \Lcal _i^4$;  furthermore, according to the discussion in Section \ref{sectcircle}  their phases provide  sections of the corresponding circle bundles over the combinatorial model $\Mcal_{g,n}[{\bf p}]$. Therefore, from the computations of the increment of ${\rm arg} \,\tau_\pm$ around $\Wfive $ and $\Wbdr $ in Propositions \ref{propmon0}, \ref{rmon}, \ref{propmon1}, \ref{propDehn},  we get the following relations between the first Chern classes of the determinants of the Hodge and Prym vector bundles ($\lambda$ and $\lambda_P$, respectively), $\psi$-classes, Witten's cycle $\Wfive $ and Kontsevich's boundary $\Wbdr $ of the combinatorial model:
\begin{theorem}
\label{thmlambda1}
The following relations hold in ${\rm Pic}(\Mcomb,\Q)$:
\be
\lambda+\f{1}{12}\sum\psi_i= \f{1}{144}\Wfive  +\f{13}{144} \Wbdr 
\la{lplus}
\ee
\be
\lambda_P+\f{1}{12}\sum\psi_i= \f{13}{144}\Wfive  +\f{25}{144}\Wbdr 
\la{lminus}
\ee
\end{theorem}

The formula (\ref{lplus}) is an analog of  formulas for the $\lambda$-class 
derived in various complex-analytic frameworks: Hurwitz spaces \cite{Advances}, spaces of Abelian \cite{MRL} and
quadratic \cite{contemp} differentials.  The formula (\ref{lminus}) is an analog of the expression of Prym class on spaces of
quadratic differentials \cite{contemp,KSZ}.

\subsection{The $\kappa_1$ circle bundle $S[\chi_\kappa]$ and $\tau_\pm$}

Recall that the Mumford-Morita-Miller class $\kappa_1$ can be expressed as follows (\ref{kall}):
\be
\kappa_1=\l_2^{(n)}-\l_1
\la{kall1}
\ee

Using (\ref{kall1}) and taking into account that $\l_2^{(n)}=\l_P$ we have the following corollary of
(\ref{lplus}), (\ref{lminus}):
\begin{corollary}
The following formula holds in ${\rm Pic}(\Mcomb,\Q)$:
\be
12\kappa_1=\Wfive +\Wbdr 
\la{kaWW1}
\ee
\end{corollary}
The formula (\ref{kaWW1}) was originally derived in \cite{ArbCor} and then reproved in \cite{Mondello1}.
This paper  gives an alternative proof of fact.

Observe now that   $\kappa_1=c_1(\chi_\kappa)$ where the line bundle $\chi_\kappa$ over $\Mcal_{g,n}$ is defined by
\be
\chi_\kappa=\frac{{\rm det}\Lambda_2^{(n)}}{{\rm det}\Lambda_H}
\la{Lkappa1}\ee

As before, given any line bundle $\chi$  we denote the associated circle bundle  by  $S[\chi]$.
As another corollary of (\ref{lplus}), (\ref{lminus}) we get
\begin{corollary}
\label{corsection}
  A section of  the circle bundle $ S[(\chi_{\ka})^{48}]$  over $\Mcomb$ is given by $ \frac {\Theta_-}{\Theta_+}$ where $ \Theta_\pm = 
\left(\f{\tau_\pm} {|\tau_\pm|} \right)^{48}$.
\end{corollary}

To get another corollary of formulas 
(\ref{lplus}) and    (\ref{lminus}) one can eliminate $\Wfive $.
Then
we get the  formula (using that $\lambda_P=\lambda^{(n)}_2$):
%\label{corlambda2}
\be
\lambda_2-13\lambda_1=\sum_{i=1}^n \psi_i  - \Wbdr 
\la{Mummain}\ee
valid in $\Mcomb$.  This formula is the combinatorial analog of the Mumford's formula \eqref{Mumint} which is valid in the Deligne-Mumford compactification of $\Mcal_{g,n}$.

\subsection{Open problems}

After the pioneering work \cite{Kontsevich} the approach to the study of 
of various tautological  classes and their intersection numbers via the JS combinatorial model remained under-developed in spite of progress summarized in \cite{Mondello2}. 

We mention only few 
 natural questions which remain to be answered:
\begin{enumerate}
\item
What is the combinatorial cycle which is Poincar\'e dual  to the psi-class $\psi_i$? This seems to be a very 
elementary question whose answer we were not able to find in the literature. 
\item
How to complete the Kontsevich's boundary $W_{1,1}$ of the JS combinatorial model to get the one-to-one correspondence with the Deligne-Mumford compactification $\overline{\Mcal}_{g,n}$ of the moduli space?
\item
 How to express all $\lambda$-classes directly via cycles of the JS combinatorial model without relying on
relationship with $\kappa$-classes and expressions for $\kappa$-classes obtained in \cite{ArbCor,Igusa1,Mondello1}?

\item
 How to get a self-contained  proof, based on "flat" combinatorial model, of the theorem of \cite{Kazarian} about generating function 
of linear Hodge integrals and KP hierarchy, without the use of inversion of the ELSV formula for Hurwitz numbers?
Furthermore,   intersection numbers of an arbitrary set of tautological classes ($\kappa$-, $\lambda$- and $\psi$-) should admit a complete description based on the  combinatorial model in the spirit of \cite{Kontsevich}.
\end{enumerate}

{\bf Acknowledgements.}  The authors thank Peter Zograf for numerous illuminating discussions.
We thank Sam Grushevsky and Martin M\"oller for comments on the formula (\ref{Mumint}).
The work of M.B. is supported in part by the Natural Sciences and Engineering Research Council of Canada (NSERC) grant RGPIN-2016-06660.
The work of D.K. was supported in part by the Natural Sciences and Engineering Research Council of Canada grant
RGPIN/3827-2015 and by Alexander von Humboldt Stiftung.
Both authors are partially supported by the  FQRNT grant "Matrices Al\'eatoires, Processus Stochastiques et Syst\`emes Int\'egrables" (2013--PR--166790). 
Both authors thank the {\it Institut Mittag--Leffler} for  hospitality during the workshop "Moduli Integrability and Dynamics", where parts of the paper where written. 

D.K. thanks   Max-Planck Institute for Gravitational Physics in Golm (Albert Einstein Institute) and SISSA (Trieste) for  hospitality and support during  preparation of this work.

\end{document}